\documentclass{jfm}

\usepackage{graphicx}
\usepackage{newtxtext}
\usepackage{newtxmath}
\usepackage{natbib}
\usepackage{hyperref}
\usepackage{comment}
\hypersetup{
    colorlinks = true,
    urlcolor   = blue,
    citecolor  = black,
}

\newcommand{\RomanNumeralCaps}[1]
\linenumbers

\title{An experimental study on the multiscale properties of turbulence in bubble-laden flows} 

\author{
Tian Ma\aff{1,2},
Hendrik Hessenkemper\aff{1}\corresp{\email{h.hessenkemper@hzdr.de}},
Dirk Lucas\aff{1}
\and Andrew D. Bragg\aff{2}\corresp{\email{andrew.bragg@duke.edu}}
}

\affiliation{\aff{1}Helmholtz-Zentrum Dresden -- Rossendorf, Institute of Fluid Dynamics, 01328 Dresden, Germany
\aff{2}Department of Civil and Environmental Engineering, Duke University, Durham, NC 27708, USA}

\begin{document}
\maketitle

\begin{abstract}

The properties of bubble-laden turbulent flows at different scales are investigated experimentally, focusing on the flow kinetic energy, energy transfer, and extreme events. The experiments employed particle shadow velocimetry measurements to measure the flow in a column generated by a homogeneous bubble swarm rising in water, for two different bubble diameters ($2.7$ mm  $\&$ $3.9$ mm) and moderate gas volume fractions ($0.26\%\sim1.31\%$). The two velocity components were measured at high-resolution, and used to construct structure functions up to twelfth order for separations spanning the small to large scales in the flow. Concerning the flow anisotropy, the velocity structure functions are found to differ for separations in the vertical and horizontal directions of the flow, and the cases with smaller bubbles are the most anisotropic, with a dependence on void fraction. The degree of anisotropy is shown to increase as the order of the structure functions is increased, showing that extreme events in the flow are the most anisotropic. Our results show that the average energy transfer with the horizontal velocity component is downscale, just as for the three-dimensional single-phase turbulence. However, the energy transfer associated with the vertical component of the fluid velocity is upscale. The probability density functions of the velocity increments reveal that extreme values become more probable with decreasing Reynolds number, the opposite of the behaviour in single-phase turbulence. We visualize those extreme events and find that regions of intense small scale velocity increments occur near the turbulent/non-turbulent interface at the boundary of the bubble wake.  

\end{abstract}


\section{Introduction}\label{sec: introduction}

Particle-laden flows are a central topic in fluid mechanics and omnipresent in nature and technology \citep{2009_Prosperetti,2010_Balachandar}. While great attention has been given to investigating flows containing suspensions of small inertial, heavy particles \citep{1987_Maxey,2006_Bec,2007_Bec,2014_Fox,2016_Ireland_a,2016_Ireland_b,2016_Gustavsson,2018_Hogendoorn,2018_Dou_a,2018_Dou_b,2019_Tom,2019_Petersen,2021_Berk}, there has been less of a focus on bubbly turbulent flows, partly due to the increased complexity associated with performing experiments or simulations for such flows \citep{2018_Lohse}. Among the various topics relevant to bubbly flows, bubble-induced turbulence (BIT) is an important area for investigation, both for its own fundamental importance and also for understanding bubble motion \citep{2000_Magnaudet}, deformation \citep{2021_Masuk_a,2021_Perrard}, coalescence/breakup \citep{2010_Liao,2021_Riviere}, clustering mechanisms \citep{2001_Zenit,2021_Maeda}, and mixing processes \citep{2019_Almeras} in bubbly flows. 

Recently, attention has been given to investigating single-point turbulence statistics in bubble-laden flows. Readers are referred to \cite{2018_Risso} and \cite{2020_Mathai} for detailed reviews. For flows with low to moderate gas void fraction ($\alpha<5\%$), we summarize the following characteristics for homogeneous bubble swarms rising either within a background quiescent or weakly turbulent carrier liquid: (i) the liquid velocity fluctuations are highly anisotropic, with much larger fluctuations in the direction of the mean bubble motion (vertical direction in standard coordinates) \citep{2005_Mudde,2013_Lu,2020_Ma}; (ii) the probability density functions (PDFs) of all fluctuating velocity components are non-Gaussian, and the PDF of the vertical velocity fluctuation is strongly positively skewed, while the other two directions have symmetric PDFs \citep{2010_Riboux,2011_Roghair,2019_Lai}; (iii) bubbles wakes introduce additional turbulence and enhance turbulence dissipation rates in the vicinity of the bubble surface \citep{2016_Santarelli_b,2021_Masuk}; (iv) modulation of the liquid mean velocity profile due to interphase momentum transfer, resulting in modifications to the background shear-induced turbulence \citep{2008_Lu,2021_Bragg}.

Studies that have explored the multiscale properties of bubble-laden turbulence have mainly focused on investigating modifications to the energy spectra of the liquid velocity fluctuations due to the bubbles. \cite{1991_Lance} were the first to find a power law scaling with a slope of about $-3$ for the vertical velocity fluctuation in a bubble-laden turbulent channel flows using hot-film anemometry. This scaling that emerges in BIT-dominated flows is in contrast to the classical $-5/3$ scaling that appears in many single-phase turbulent flows \citep{2000_Pope}. This $-3$ scaling was also confirmed by subsequent studies for all the components of the fluctuating fluid velocity, and has been observed both in experiments and direct numerical simulations (DNS) \citep{2010_Riboux,2013_Mendez,2020_Pandey,2021_Innocenti} for varying bubble properties.

Beyond the behavior of the energy spectra, knowledge about the multiscale properties of bubble-laden turbulence is quite limited. A pioneering study on this topic is that by \cite{2005_Rensen}, who performed hot-film anemometry measurements in a bubble-laden water tunnel and found an increase of the second-order structure function for the two-phase case compared with the single-phase case for the same bulk Reynolds number, and that this increase was more pronounced at the small scales than the large scales. Moreover, they considered the probability density functions (PDFs) of the velocity increments and used extended self-similarity (ESS) \citep{1993_Benzi} to show that the flow intermittency is enhanced by the bubbles. Furthermore, they argued that once the bubbles are present in the flow, the dependence of the flow properties on the actual bubble concentration is weak for the range they investigated ($ 0.5\%\leq\alpha\leq 2.9\%$). Similar behaviour was also observed later in \cite{2012_Biferale} and \cite{2021_Ma} when comparing the small-scale properties of bubble-laden and unladen turbulent flows.

To gain more insight into the multiscale energetics of bubble-laden turbulence, \cite{2020_Pandey} computed the scale-by-scale average energy budget equation in Fourier space using results from an interface-resolved DNS with several tens of bubbles rising in a initially quiescent flow. They showed that on average there is a downscale energy transfer, just as occurs for the single-phase turbulence in three dimensions \citep{2018_Alexakis}. A similar finding was also reported by two more recent DNS studies \citep{2021_Innocenti,2021_Ma}. An issue with our previous study \citep{2021_Ma}, however, is that the conclusion was drawn based on a one-dimensional dataset, which only allowed us to construct the velocity increments for separations in the spanwise direction of the bubble-laden turbulent channel flow.

Another important point to be quantified is how the bubbles influence the anisotropy of the turbulent flow across the scales. Thanks to significant research efforts, single-phase multiscale anisotropy is understood in considerable detail \citep{1997_Sreenivasan,2005_Biferale}. While phenomenological turbulence theories postulate a return-to-isotropy at small scales \citep{1941_Kolmogorov_a,1995_Frisch}, experimental and numerical data have showed persistent small-scale anisotropy \citep{2000_Shen,2006_Ouellette,2016_Pumir}, especially when considering high-order structure functions \citep{2017_Carter,2000_Kurien}. In contrast to single-phase turbulence, where anisotropy is usually injected into the flow at the large scales \citep{2012_Chang}, bubbles inject anisotropy into the flow at the scale of their diameter/wake, and this often corresponds to the small-scales of the turbulence. As a result of this, bubble-laden turbulent flows can exhibit much stronger anisotropy at the small-scales of the flow compared to their single-phase counterparts with the same bulk Reynolds number \citep{2021_Ma}. This behavior was quantified in our recent study by developing a new method based on the barycentric map approach \citep{2007_Banerjee}, and the results also revealed that the bubble Reynolds number is the key factor responsible for governing the flow anisotropy, whereas the void fraction does not seem to play an important role, at least for the void fractions considered.

To advance the understanding of the multiscale properties of bubble-laden turbulence, in this paper we present an experimental study based on measurements of millimetre-sized air bubbles with (approximately) fixed shape/size rising in a vertical column of water, using high-resolution particle shadow velocimetry (PSV). The water is deliberately contaminated to allow for the no-slip condition to be satisfied on the bubble surface \citep{2019_Elghobashi}, similar to what is usually assumed in DNS studies. The bubbles are injected uniformly at a low to moderate gas void fractions to minimize the large scale velocity fluctuations in the flow \citep{2003_Harteveld}, and we focus on a region of the flow sufficiently far away from the walls of the column where the one-point statistics from both phases are approximately homogeneous across the measurement window. Unlike our previous study that used a one-dimensional DNS dataset \citep{2021_Ma}, PSV measures two-dimensional, two component (2D-2C) velocity fields, providing access to both longitudinal and transverse structure function associated with separations along two directions. Moreover, we record a large number of uncorrelated velocity fields, so that structure functions up to twelfth-order can be computed. One of the ways the present work advances that in \cite{2021_Ma} is that the resolution of the DNS used in \cite{2021_Ma} was insufficient to resolve extreme fluctuations in the flow, and therefore only structure functions up to order four were considered. Here, since experimental data is used there is no question regarding whether the physics of the flow is properly captured in the measured quantities.

The rest of this paper is organized as follows. In \S\,\ref{sec: Experimental setup}, we introduce the experimental set-up and the measurement techniques. We then first present the single-point statistics for both phases in \S\,\ref{sec: one-point}. The multipoint results are divided into three parts, namely, anisotropy in \S\,\ref{sec: Anisotropy}, energy transfer in \S\,\ref{sec: Energy transfer}, and intermittency in \S\,\ref{sec: Extreme velocity increment}.

\section{Experimental setup and measurement techniques}\label{sec: Experimental setup}

\subsection{Experimental facility}

\begin{figure}
	\centering
	\includegraphics[height=9.5cm]{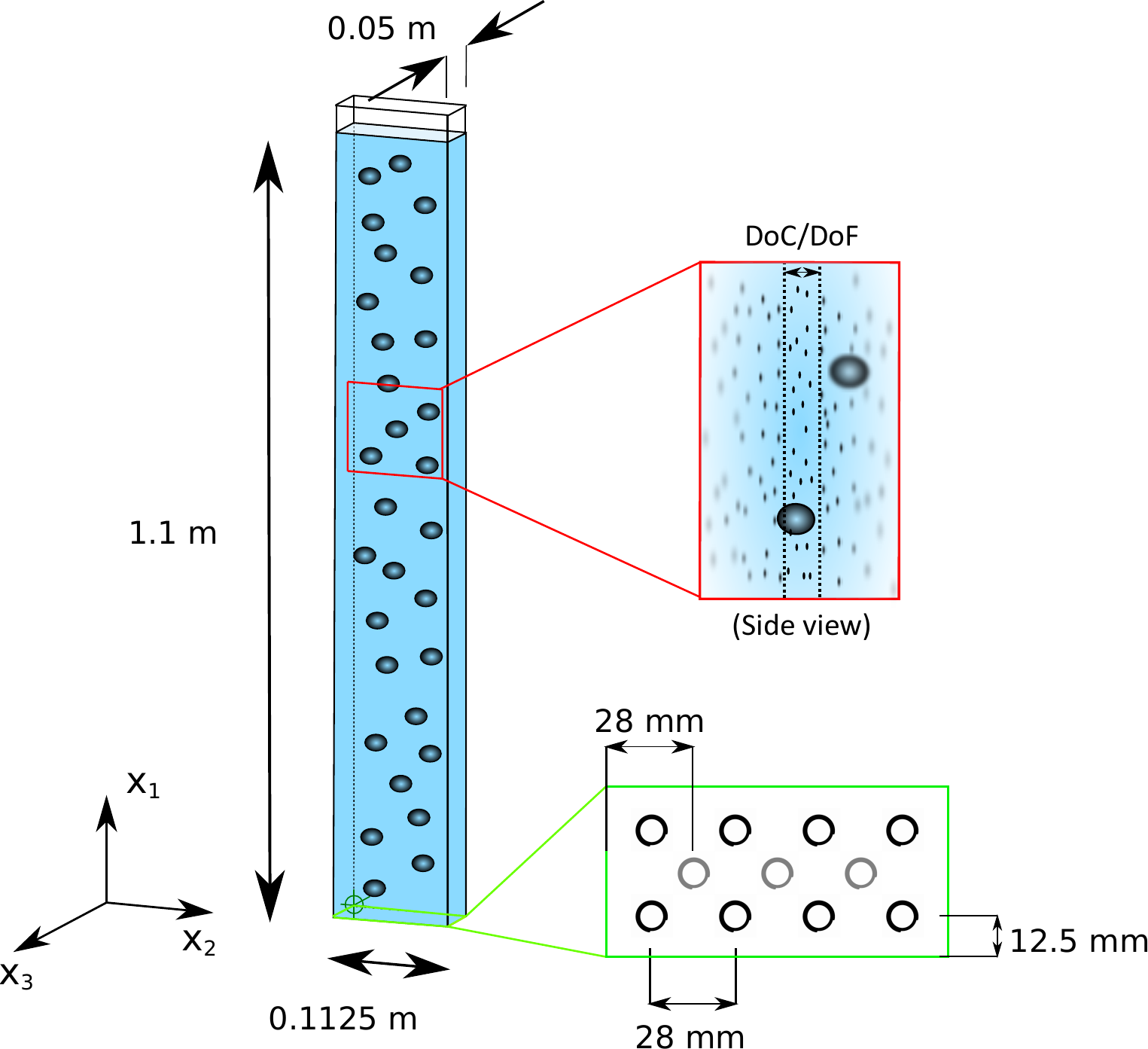}
	\caption{Sketch of the bubble column used in the experiments (note that in the actual experiment, the number of bubbles in the column is $O(10^3)$). Right top shows a representation of the shallow DoF seen from the side view and bottom right shows the sparger arrangement.} \label{fig: Bubble_column}
\end{figure}

The experimental data required for the present study has only recently become available thanks to the advance of the PSV measurement technique for two-phase flows \citep{2018_Hessenkemper}. We used this method in our experiments, which were conducted in Helmholtz-Zentrum Dresden -- Rossendorf, Germany. The experimental setup consists of a rectangular column (depth $50\,\mathrm{mm}$ and width $112.5\,\mathrm{mm}$) made of acrylic glass which is filled with tap water to a height of $1,100\,\mathrm{mm}$ (figure \ref{fig: Bubble_column}). The temperature in the lab was kept constant at $20\,^{\circ}\mathrm{C}$ and the density and kinematic viscosity of the water are assumed to be constant with values $\rho_l=998\,\mathrm{kg/m^3}$ and $\nu=1\times 10^{-6}\,\mathrm{m^2/s}$, respectively. 

To suppress bubble deformation, $1,000\,\mathrm{ppm}$ 1-Pentanol is added to the flow. As demonstrated by \cite{2014_Tagawa}, 1-Pentanol at such a high concentration leads to a rapid full contamination of the bubbles. The thereby induced immobilization of the otherwise mobile bubble surface results in a nearly no-slip condition at the gas-liquid interphase \citep{2011_Takagi,2020_Manikantan}. A reduction of the bubble rising velocity can also be clearly observed in the experiment by adding this amount of 1-Pentanol, which is explained by the Marangoni effect \citep{2011_Takagi}. Another feature of 1-Pentanol is that the adsorbed surfactant inhibits bubble coalescence and break-up, so that in each experiment the bubbles are mono-disperse with a fixed bubble size.

Air bubbles are injected through several spargers that are inserted into $11$ holes that have been drilled into the bottom of the column (see bottom right of figure \ref{fig: Bubble_column} for the sparger configuration). All the spargers are removable, hence, both the gas fraction and the bubble size can be varied by removing or replacing different spargers. In the present study, either all $11$ spargers or the $8$ outer spargers are used to ensure a homogenous bubble distribution at the measurement height. Here, we use two different spargers with the inner diameter $0.2\,\mathrm{mm}$ and $0.6\,\mathrm{mm}$, corresponding to a constant gas flow rate of $0.04\,\mathrm{l/min}$ and $0.1\,\mathrm{l/min}$ per sparger, respectively (our preliminary tests show that these flow rates lead to a stable production of bubbles of constant size). In summary, we consider two different bubble sizes, and for each bubble size we consider two gas void fractions. These four mono-dispersed cases are labelled as \textit{SmLess}, \textit{SmMore}, \textit{LaLess}, and \textit{LaMore} in table \ref{tab: bubble papa} (\textit{Sm}/\textit{La} for smaller/larger bubbles and \textit{More}/\textit{Less} for higher/lower gas void fraction), including some basic characteristic dimensionless numbers for the bubbles.

\subsection{Measurement techniques}

\subsubsection{Liquid velocity measurement}

Although it is possible to use conventional particle imagine velocimetry (PIV) with a laser-sheet to measure the liquid velocity in bubbly flows, such a side-wise high intensity illumination can result in an inhomogeneous illumination due to unwanted lateral shadows of the bubbles as well as strong light scattering and reflection at the gas-liquid interfaces \citep{2007_Broder,2011_Tropea,2016_Ziegenhein}. To circumvent these problems, the very recently developed PSV method for dispersed two-phase flows \citep{2018_Hessenkemper} is performed here. The feature of such a measurement is to use a volumetric direct in-line illumination with e.g. LED-backlights for the region of interest, whereby scattering effects are strongly reduced and no lateral bubble shadows occur. By using a shallow depth of field (DoF), sharp tracer particle shadows positioned inside the DoF region can be identified and the particle displacement is evaluated in a PIV-like manner. Hence sometimes this method is also called \textquoteleft PIV with LED\textquoteright\, or \textquoteleft particle shadow image velocimetry\textquoteright\, \citep{2005_Estevadeordal}. The top right of figure \ref{fig: Bubble_column} shows a representation of such a shallow DoF based on the side view of the column. This kind of thin DoF is also frequently used in laser-based $\mu$PIV measurements, since a laser-sheet thickness below $0.5\,\mathrm{mm}$ is hard to achieve, while the DoF can be adjusted to even smaller depth expansions \citep{2010_Wereley}. With the present setup, the effective DoF (i.e., depth of correlation (DoC) in the terminology of measurement) is calculated using eq.(2) in \cite{2006_Bourdon} as $\mathrm{DoC}\simeq370\,\mathrm{\mu m}$, which is approximately one seventh of the smaller bubble diameter (table \ref{tab: bubble papa}).

The liquid velocity measurements are performed along the $x_1$-$x_2$ symmetry plane at the centre of the depth ($x_3$). In the following, we refer to $x_1$ and $x_2$ as the vertical and horizontal direction, respectively. The measurement height is $x_1=0.65\,\mathrm{m}$ based on the centre of the field of view (FOV) (figure \ref{fig: Bubble_column}) to ensure that bubbles entering the measurement region have already lost any memory of the way they were injected. The flow is seeded with $10\,\mathrm{\mu m}$ hollow glass spheres (HGS) -- Dantec, and illuminated by a $200\,\mathrm{W}$ LED-lamp, which consists of $160$ small LED arrays arranged in a circular area of $100\,\mathrm{mm}$ diameter and located directly facing the high-speed camera in the $x_3$-direction. Since the HGS particles are hydrophilic, almost no flotation due to the added surfactant occurs. To fully capture the small-scale fluctuations of the flow, the Stokes number of the seeding particle, defined as $St=\tau_d⁄\tau_f$ must be much smaller than unity, where $\tau_d$ is the particle response time and $\tau_f$ is the characteristic time scale of the flow. In the current experiments the highest $St$ is $O(10^{-3})$ for $\tau_f$ based on the estimated Kolmogorov time scale (table \ref{tab: flow para}), whose estimate is explained in \S\,\ref{sec: one-point}. Image pairs for the velocity determination are acquired with a MotionPro Y3 high-speed CMOS camera, with a frame rate of $900$ fps for the smaller bubble cases and 1,100 fps for the larger bubble cases. 

Before the velocity interrogation is performed, the comparably large bubble shadows are masked and only interrogation areas outside the mask are considered. Velocity fields are processed using a multipass-refinement procedure, with a final interrogation window size of $32 \times 32$ pixels and $50\,\%$ overlap. Measurements are obtained by mounting a macro lens (Samyang) with a focal length of $100\,\mathrm{mm}$ and an f-stop of $2.8$. This provides a FOV of $13.9\,\mathrm{mm}\,(\mathrm{H_1}) \times 19.5\,\mathrm{mm}\,(\mathrm{H_2})$, with a pixel size of $14.7$ $\mu$m. Since this pixel size is even slightly larger than the seeding particle size, the so-called particle sharpening step \citep{2018_Hessenkemper} is used, which artificially increases the particle size due to a swelling to the neighbouring pixels. Hence, an effective particle size of about $3$-$4$ pixels is achieved. In the resulting liquid velocity fields, the spatial resolution $\Delta$ (PSV grid) is approximately $0.23\,\mathrm{mm}$ ($d_p/\Delta\approx11\sim17$, see table \ref{tab: bubble papa} for different cases). For more details on the PSV method used in the present measurements, including the whole PIV-like post-processing steps, we refer the readers to \cite{2018_Hessenkemper}.

\begin{figure}
\centering
\includegraphics[height=6.5cm]{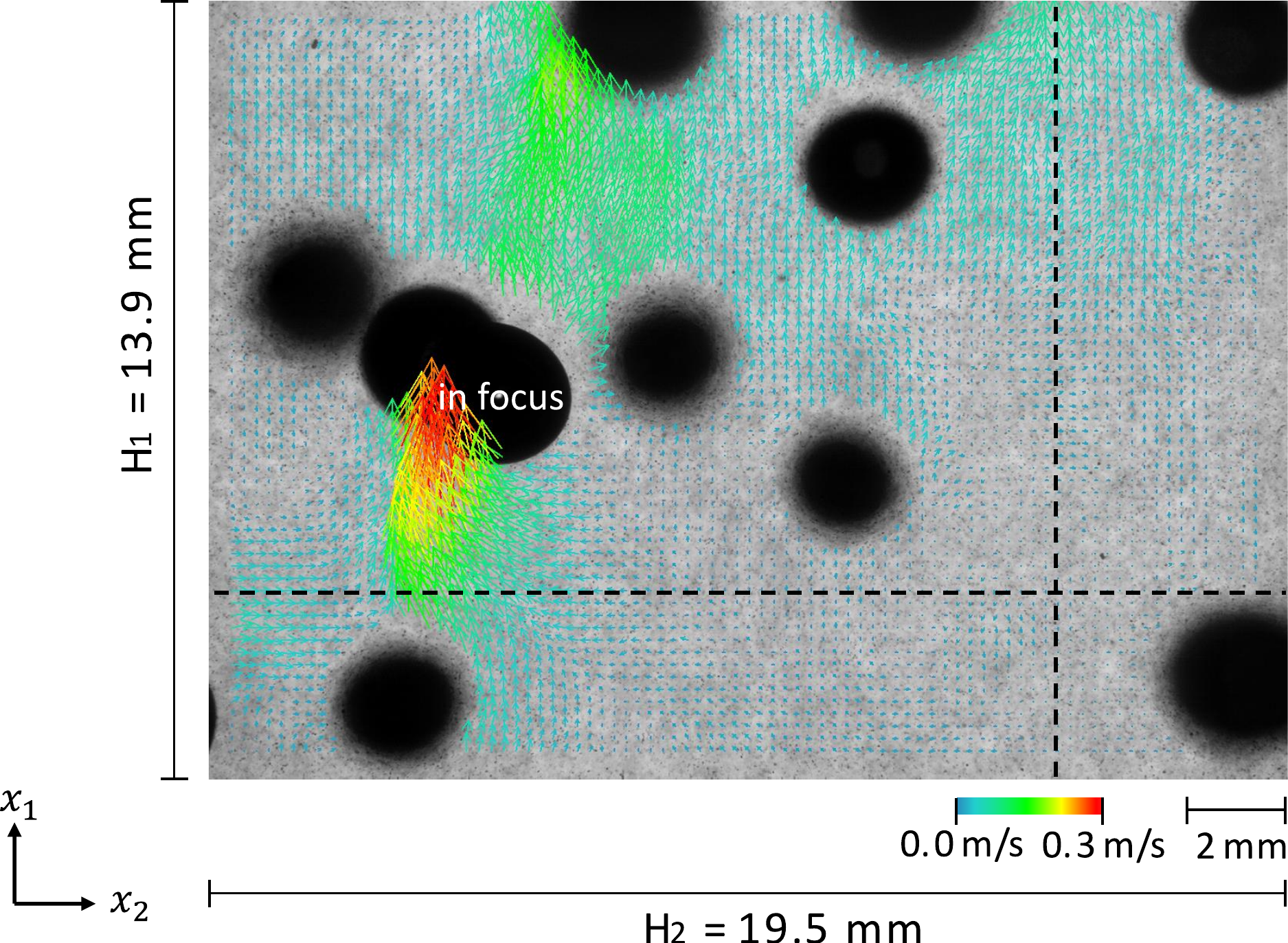}
\caption{Instantaneous realization of velocity vector over the FOV in the case \textit{SmMore} with a detected in-focus bubble (sharp interface) in the middle left. The horizontal/vertical dashed lines are examples, indicating the representative points along the lines where the structure functions were computed.} \label{fig: PSV FOV}
\end{figure}

In figure \ref{fig: PSV FOV} a typical instantaneous FOV with the bubbles and the resulting liquid velocity vector field are shown for the case \textit{SmMore}. In this snapshot, while there is only one in-focus bubble (identified by the sharp interface), with strong wake entrainment identified in the velocity vector field, the other bubbles are blurred, i.e. out-of-focus. Tracer particles at these locations cannot be detected, so that no liquid velocity fields are available there. 

To obtain converged high-order statistics, for each case considered $60,000$ velocity fields were recorded with an image pair acquisition rate of $1\,\mathrm{Hz}$. With $60\times84$ interrogation windows in each velocity field, this yields in total $\approx3\times10^9$ data points for each case.

\subsubsection{Bubble statistics measurement}

\begin{table}
	\begin{center}
		\def~{\hphantom{0}}
		\begin{tabular}{ccccc}
			Parameter  
			&\textsl{SmLess}&\textsl{SmMore}&\textsl{LaLess}&\textsl{LaMore}\\	
			\hline
			$\alpha_p$   &0.26\%&0.54\%&0.93\%&1.31\%\\
			$d_p \;(\mathrm{mm})$    &2.7&2.7&3.8&3.9\\
			$\chi$     &1.12&1.12&1.3&1.3\\
			$Ga$       &437&437&730&759\\
			$E_o$      &1.05&1.05&2.08&2.19\\
			$Re_p$     &524.4&522&828.5&809.5\\
			$C_D$      &0.925&0.931&1.027&1.158\\
			$d_p/\Delta$      &11.6&11.6&16.4&16.8\\
		\end{tabular}
\caption{Selected basic statistics of the gas phase for the four investigated cases. Here, $\alpha_p$ is the averaged gas void fraction, $d_p$ the equivalent bubble diameter, $\chi$ the aspect ratio, $Ga\equiv\sqrt{\left|\pi_\rho-1\right|gd_{p}^{3}}/\nu$ the Galileo number. The values of $Re_p$, the bubble Reynolds number and $C_D$, the drag coefficient are based on $d_p$ and the bubble to fluid relative velocity obtained from the experiment. $\Delta$ is one PSV grid, i.e. the smallest spatial resolution of the present experiment.} \label{tab: bubble papa}
	\end{center}
\end{table}

The bubble statistics and properties are evaluated with a separate set of measurements. For these measurements, the flow configurations are kept exactly the same with the exception that no tracer particles are added to the flow and that a larger FOV of $40\,\mathrm{mm} \times 80\,\mathrm{mm}$ (width $\times$ height) than with PSV for the liquid phase is investigated to obtain sufficient bubble statistics. Here, we have evaluated $500$ such image pairs for each case considered. 

\begin{figure}
	\begin{minipage}[b]{1.0\linewidth}
		\begin{minipage}[b]{0.5\linewidth}
			\centering
			\makebox[2em][l]{\raisebox{-\height}{(\textit{a})}}%
			\raisebox{-\height}{\includegraphics[height=5cm]{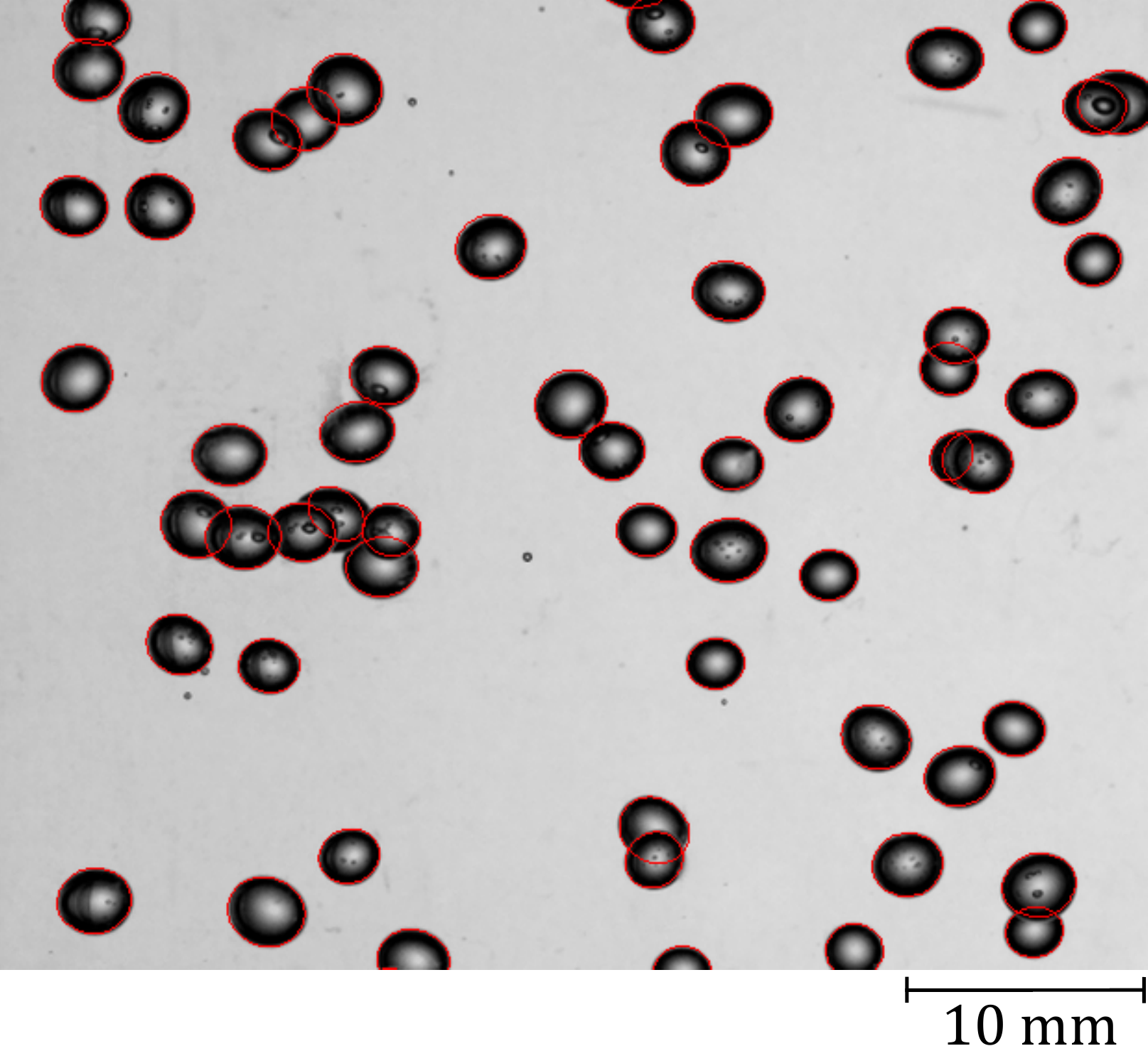}}
		\end{minipage}
		\begin{minipage}[b]{0.5\linewidth}
			\centering
			\makebox[2em][l]{\raisebox{-\height}{(\textit{b})}}%
			\raisebox{-\height}{\includegraphics[height=5cm]{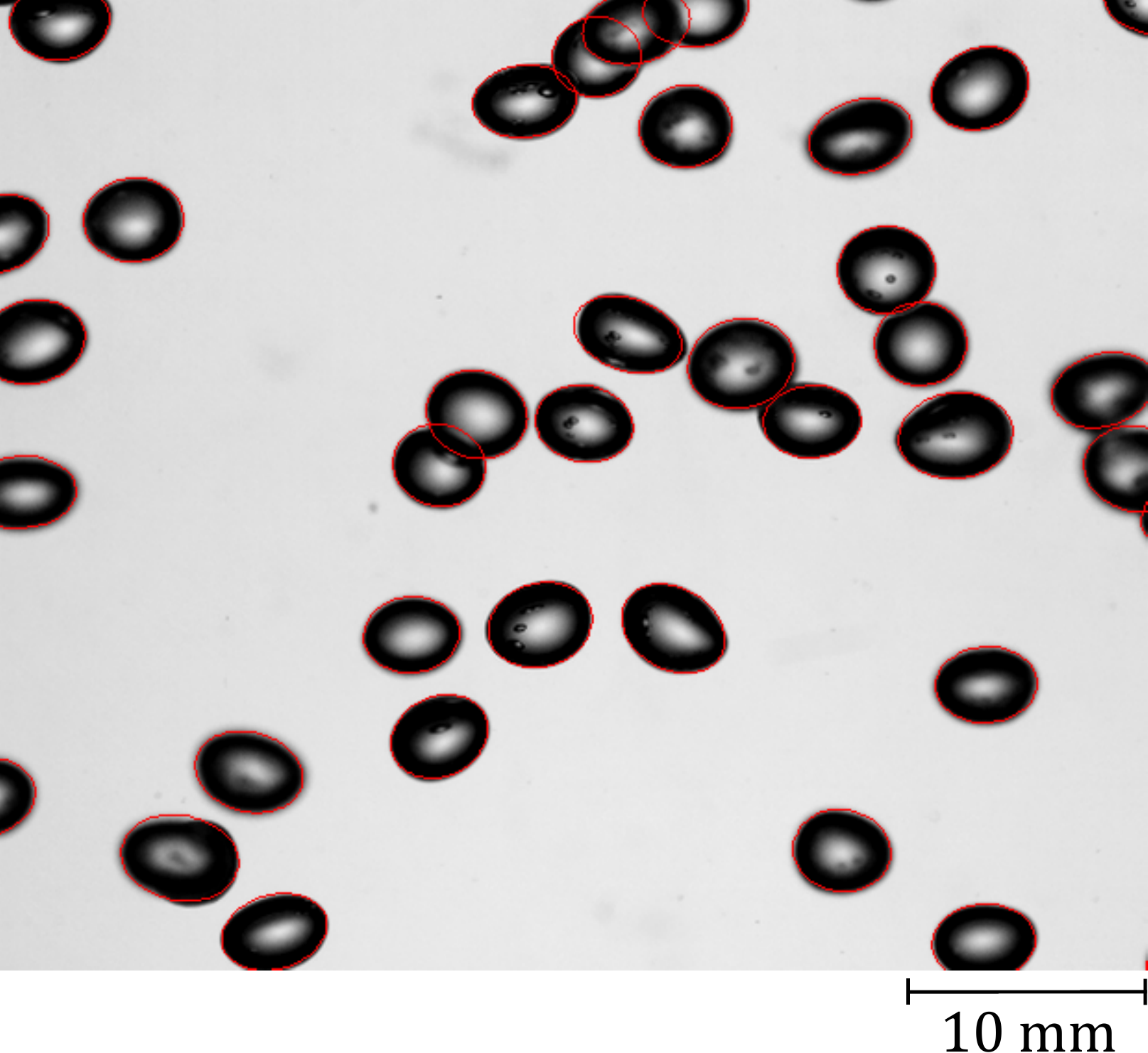}}
		\end{minipage}
	\end{minipage}
	\caption{Example images of the bubbles with fitted ellipses for an arbitrary instant: (\textit{a}) smaller bubbles from the case \textit{SmMore} and (\textit{b}) larger bubbles from the case \textit{LaLess}. (Figures are cut from a region of the FOV for the gas phase for the corresponding case.)} \label{fig: real bubbles}
\end{figure}

A machine learning procedure is used to detect and intersect overlapping bubbles in the images \citep{2021_Hessenkemper}. The underlying neural network uses a U-Net architecture \citep{2015_Ronneberger} and is trained to find the intersection boundary of a bubble in front of another bubble. Using these intersections, the bubbles are split and an ellipse is fitted to the contour of each bubble. The ellipse parameters are then used to calculate the volume of an axisymmetric spheroid. Since the gas fraction is comparably small, the number of overlapping bubbles is also small and up to now unevaluated errors connected to this are estimated to about $5\%$. Due to the contamination of the bubbles, deviations from an ellipse due to irregular (wobbling) bubble surfaces are only minor. Example images for the two different bubble sizes with fitted ellipses are given in figure \ref{fig: real bubbles} for the cases \textit{SmMore} and \textit{LaLess}, respectively. The bubble size is then calculated using the volume-equivalent bubble diameter of a spheroid as $d_p=(d_{maj}^2d_{min})^{1/3}$, where $d_{maj}$ and $d_{min}$ are the lengths of the major and minor axes of the fitted ellipse, respectively. For the smaller bubbles, the shape is close to a sphere with a small aspect ratio $\chi=d_{maj}/d_{min}=1.12$ and we obtain $d_p=2.7\,\mathrm{mm}$ corresponding to an E\"{o}tv\"{o}s number, $Eo=\Delta\rho gd_p^2/\sigma\approx1$ in our system. In contrast with the larger bubbles, we have $d_p=3.8\sim3.9\,\mathrm{mm}$, $\chi=1.3$, and $Eo\approx2$. Here, $\Delta\rho$ is the density difference between the carrier phase and the dispersed phase and $\sigma$ is the surface tensor coefficient. The range of the bubble Reynolds number, $Re_p$, in our experiments are $500\sim900$ based on $d_p$ and the bubble to liquid relative velocity. Note that although no net liquid flow through a cross-section ($x_2$-$x_3$ plane) is present (i.e. the bulk velocity of the liquid is zero), for bubble columns, the flow exhibits a large-scale circulation in a time-averaged sense with upward flow in the centre and downward flow close to the walls \citep{2005_Mudde}. This results in a non-negligible upward flow in the measurement area of the present PSV (shown later in \S\,\ref{sec: one-point}), so that the relative velocity in the FOV is in general not equal to the bubble terminal rising velocity. Table \ref{tab: bubble papa} lists all the relevant parameters for the bubble properties. 

A shallow DoF is used for the measurements conducted to obtain bubble statistics, but it is an order of magnitude larger than that for the liquid velocity measurements due to the different optical system for these two sets of experiments. To measure the bubbles in the $x_3$-centre region where the liquid velocity measurements were also taken, the average grey value derivative along the bubble contour is calculated. A threshold for this average grey value derivative is used so that only bubbles in the $x_3$-centre region are considered, and the depth of this $x_3$-centre region is about $5\,\mathrm{mm}$. Hence, the evaluated depth for the gas phase is an order of magnitude larger than for the liquid velocity, which is however necessary due to the size of the bubbles. The number of bubbles that fall into this region ($40\,\mathrm{mm} \times 80\,\mathrm{mm} \times 5\,\mathrm{mm}$) is between $2,500$ and $5,000$ summed up over the considered $500$ image pairs depending on the case being considered, and this corresponds to an average void fraction range of $0.26\%$ to $1.31\%$ (table \ref{tab: bubble papa}). The bubble velocity was determined by calculating the distance translated by individual bubble centroids, using a standard particle tracking velocimetry algorithm that matches the closest neighbouring bubble in a small search window in the presumed direction of bubble motion \citep{2007_Broder}.

\section{Basic flow characterisation}\label{sec: one-point}

\begin{figure}	
	\begin{minipage}[b]{1.0\linewidth}
		\begin{minipage}[b]{0.5\linewidth}
			\centering
			\makebox[0.5em][l]{\raisebox{-\height}{(\textit{a})}}%
			\raisebox{-\height}{\includegraphics[height=4.1cm]{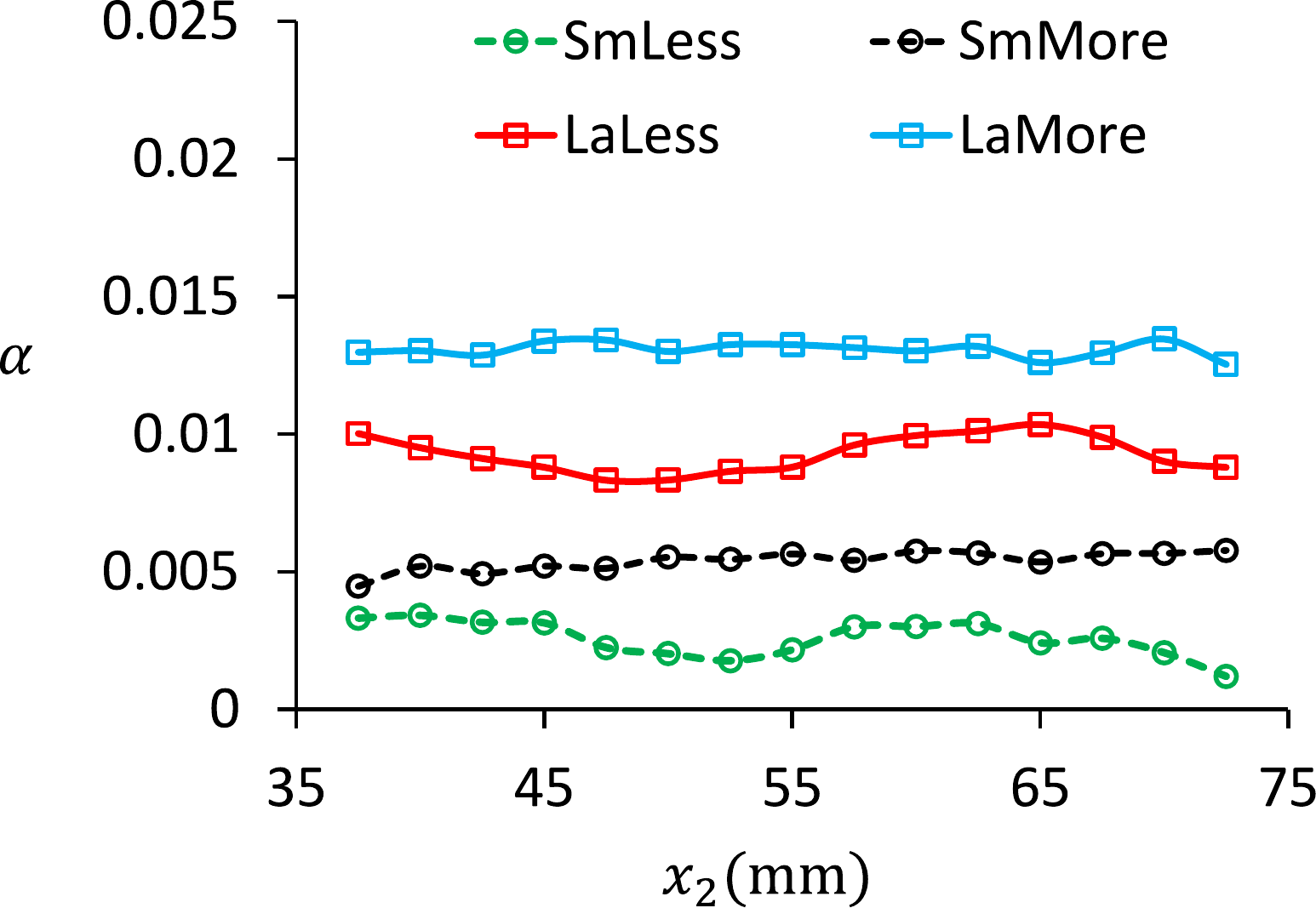}}
		\end{minipage}
		\begin{minipage}[b]{0.5\linewidth}
			\centering
			\makebox[1.2em][l]{\raisebox{-\height}{(\textit{b})}}%
			\raisebox{-\height}{\includegraphics[height=4.1cm]{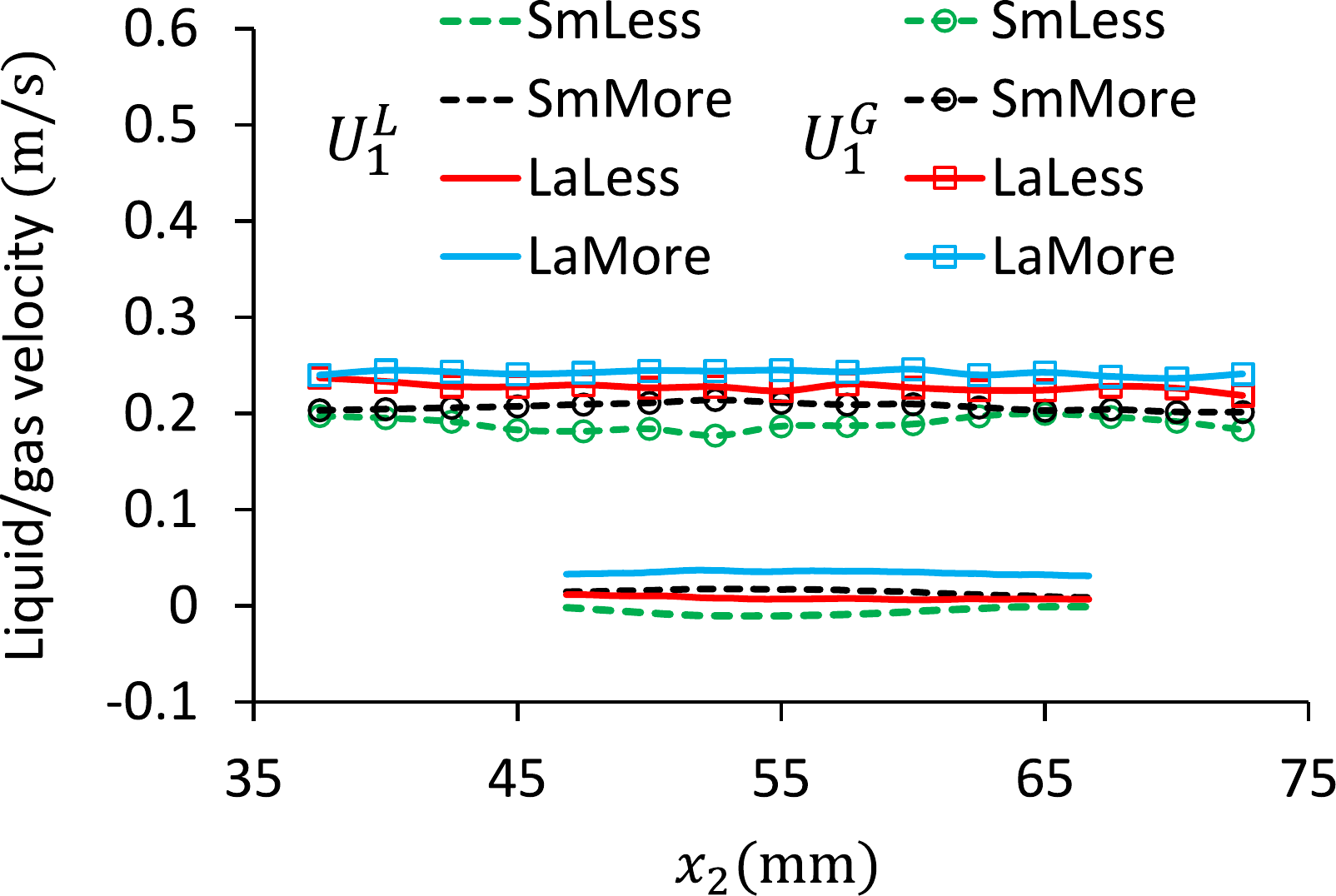}}
		\end{minipage}
	\end{minipage}	
	\begin{minipage}[b]{1.0\linewidth}
		\vspace{3mm}
		\begin{minipage}[b]{0.5\linewidth}
			\centering
			\makebox[0.5em][l]{\raisebox{-\height}{(\textit{c})}}%
			\raisebox{-\height}{\includegraphics[height=4.1cm]{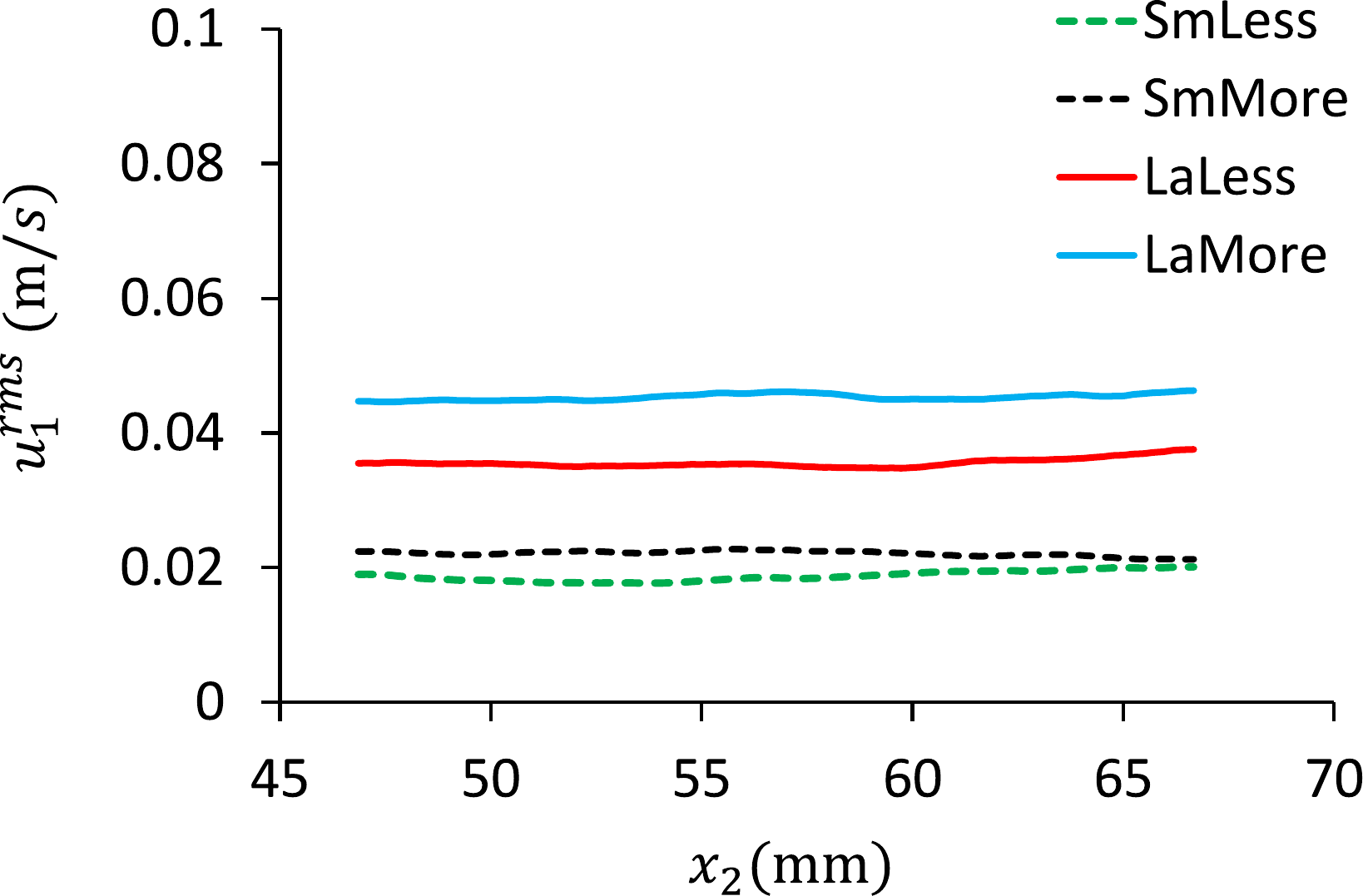}}
		\end{minipage}
		\begin{minipage}[b]{0.5\linewidth}
			\centering
			\makebox[0.5em][l]{\raisebox{-\height}{(\textit{d})}}%
			\raisebox{-\height}{\includegraphics[height=4.1cm]{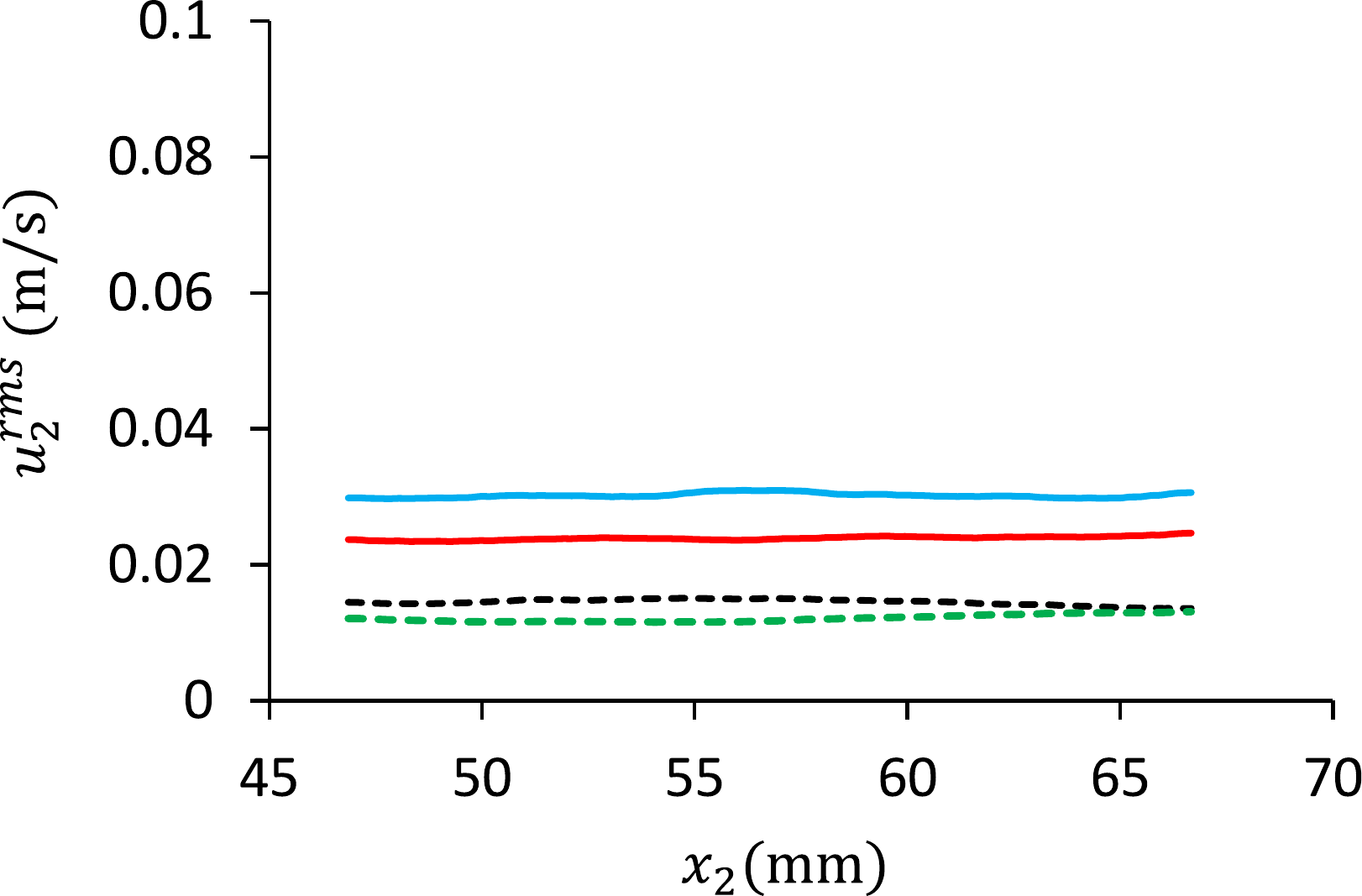}}
		\end{minipage}
	\end{minipage}	
	\caption{One-point statistics along the horizontal axis of FOV for the four considered cases: (\textit{a}) gas void fraction, (\textit{b}) liquid/gas vertical velocity, and liquid fluctuating velocity in (\textit{c}) vertical and (\textit{d}) horizontal component.} \label{fig: One-point statistics}
\end{figure}

Before exploring the properties of the flow at different scales, we begin with some basic characteristics of the flow as quantified by single-point statistics of both phases. The instantaneous velocity $\tilde{\boldsymbol{u}}(\boldsymbol{x},t)$ is decomposed into a ensemble-averaged part $\boldsymbol{U}(\boldsymbol{x},t)$ and a fluctuating part $\boldsymbol{u}(\boldsymbol{x},t)$, with associated components $\tilde{u}_i=U_i+u_i$.

For all the cases investigated, it is observed that the local void fraction (figure \ref{fig: One-point statistics}\textit{a}), the vertical gas/liquid mean velocity (figure \ref{fig: One-point statistics}\textit{b}), and the liquid fluctuating velocity profiles (figure \ref{fig: One-point statistics}\textit{c,d}) in the observation region (FOV of either liquid or gas) of the column are constant to a very good approximation, impllying statistical homogeneity of the flow in the observation region. As mentioned earlier, for most cases there is a slight upwards mean flow for the liquid phase in the FOV.

According to figure \ref{fig: One-point statistics}, the r.m.s. value of the vertical velocity fluctuation $u_1^{rms}$ is significantly larger than the horizontal component $u_2^{rms}$ for the each case, indicating the large scale anisotropy in the flow due to the preferential direction of the bubble motion. The ratio of the vertical to horizontal velocity fluctuation $u_1^{rms}/u_2^{rms}$ is in the range $1.4$ to $1.6$ for the four cases investigated, which is close to the previous studies of \cite{2010_Riboux} and \cite{2021_Ma} for bubbles with sizes in the range $1$ to $3\,\mathrm{mm}$. For both components, we find the velocity fluctuations increase with increasing $\alpha$ -- from \textit{SmLess} to \textit{LaMore}.  

\begin{figure}
	\centering
	\includegraphics[height=4.2cm]{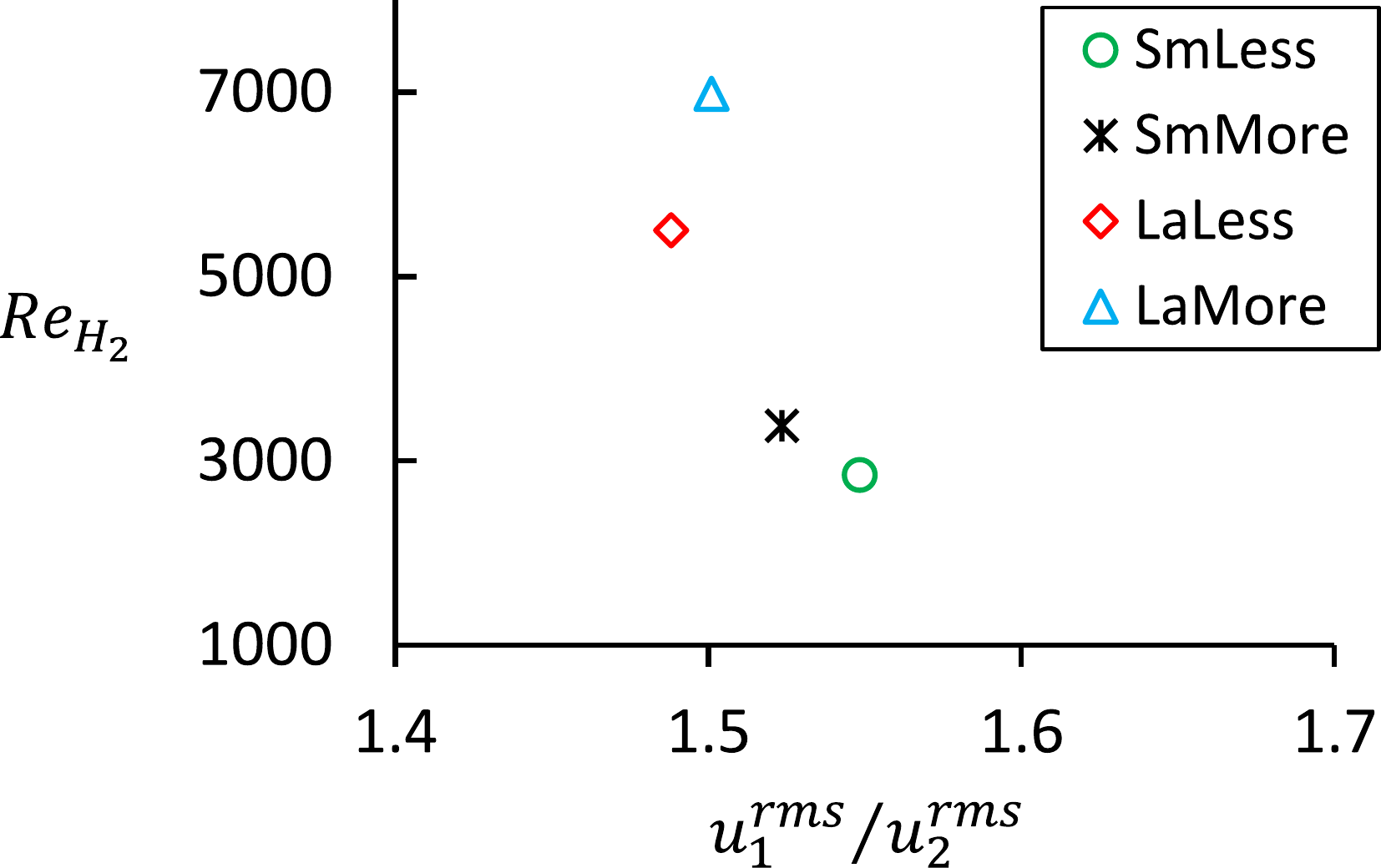}
	\caption{Reynolds number, $Re_{H_2}$ plotted versus large-scale anisotropy ratio, $u_1^{rms}/u_2^{rms}$.} \label{fig: Re_H2}
\end{figure}

Following \cite{2021_Ma}, we define a Reynolds number $Re_{H_2}\equiv u^\ast H_2/\nu$ which indicates the range of scales in the turbulent bubbly flows. Here,
$u^\ast\equiv \sqrt{(2/3)k_{\mathrm{FOV}}}$, and $k_{\mathrm{FOV}}$ is the weighted turbulent kinetic energy (TKE), $k_{\mathrm{FOV}}=((u_1^{rms})^2+2(u_2^{rms})^2)/2$, assuming axisymmetry of the flow in the FOV about the vertical direction, and averaged over the FOV of the liquid phase. Note that for our experiments the bulk Reynolds number is zero, since there is no averaged net liquid flow when averaged over the entire flow cross section. Moreover, we are interested in the properties of the fluctuating velocity field, and hence it is more appropriate to consider a Reynolds number based on $u^\ast$.

In figure \ref{fig: Re_H2} we plot $Re_{H_2}$ versus large-scale anisotropy ratio $u_1^{rms}/u_2^{rms}$ (averaged over the FOV of liquid). The figure shows that $Re_{H_2}$ increases in the order of \textit{SmLess}, \textit{SmMore}, \textit{LaLess} to \textit{LaMore}, which corresponds to larger bubbles and higher void fraction. Since the Reynolds number of the flow is usually understood to be related to the range of excited scales of motion in the system, this result implies that in the aforementioned sequence the range of excited scales also increases in the flow, hence, the flow becomes increasingly multiscale. Furthermore, we find that for large scales whose velocities are characterized by $u_1^{rms}$ and $u_2^{rms}$, the smaller bubbles produce more anisotropy in the flow than the larger bubbles, as indicated by a larger ratio of $u_1^{rms}/u_2^{rms}$ for the cases \textit{SmLess} and \textit{SmMore}. This is in very close agreement with our previous study based on DNS data of bubble-laden turbulent channel flow driven by a vertical pressure gradient \citep{2021_Ma}.

To quantify the PSV resolution with respect to the Kolmogorov scale $\eta$, we estimate the mean dissipation $\epsilon$ based on an algebraic relation derived by \cite{2020_Ma_a} for BIT dominated flows
\begin{equation}
\epsilon=S_k/(1-\alpha) \;,
\end{equation}
where the interfacial term $S_k$ for the TKE transport equation is adopted from \cite{2017_Ma}
\begin{equation}
S_{k}=\min(0.18Re_{p}^{0.23},1)\boldsymbol{F}_{D}\boldsymbol{\cdot}\boldsymbol{u}_r \;.
\end{equation}
Here, $\boldsymbol{u_r}$ is the mean slip velocity between the bubble and the liquid, and $\boldsymbol{F}_{D}=(3/4d_p)C_D\alpha\|\boldsymbol{u}_r\|\boldsymbol{u}_r$ is the drag force on the bubbles averaged over the FOV of the liquid. We can then compute the Kolmogorov length scale $\eta\equiv(\nu^3/\epsilon)^{1/4}$, as well the time scale $\tau_\eta\equiv(\nu/\epsilon)^{1/2}$. Both parameters and the ratio of $\Delta/\eta$ (ranges between $1.8$ and $2.8$) are given in table \ref{tab: flow para} for all the cases. The current PSV spatial resolution and the size of the liquid FOV are comparable to the recent experimental study of \cite{2017_Carter} on single-phase turbulence using standard PIV  (see their small FOV/high resolution measurement). 
\begin{table}
	\begin{center}
		\def~{\hphantom{0}}
		\begin{tabular}{ccccc}
			Parameter  &\textsl{SmLess}&\textsl{SmMore}&\textsl{LaLess}&\textsl{LaMore}\\	
			\hline
			$Re_{H_2}$     &2849&3383&5502&6982\\
			$u^\ast\;(\mathrm{m/s})$   &0.0147&0.0174&0.0283&0.0359\\
			$u^{\mathrm{rms}}_1/u^{\mathrm{rms}}_2$   &1.54&1.52&1.49&1.50\\
			$\epsilon\;(\mathrm{m^2/s^3})$    &0.0049&0.0103&0.0199&0.0268\\
			$\eta\;(\mathrm{mm})$             &0.13&0.11&0.088&0.082\\
			$\tau_\eta\;(\mathrm{ms})$        &16.4&11.4&7.7&6.7\\
			$\Delta/\eta$                     &1.8&2.2&2.6&2.8\\
		\end{tabular}
		\caption{Selected Basic statistics of the liquid phase for the four investigated flow configurations.}
		\label{tab: flow para}
	\end{center}
\end{table}

\begin{figure}
	\begin{minipage}[b]{1.0\linewidth}
		\begin{minipage}[b]{0.5\linewidth}
			\centering
			\makebox[0.5em][l]{\raisebox{-\height}{(\textit{a})}}%
			\raisebox{-\height}{\includegraphics[height=4.2cm]{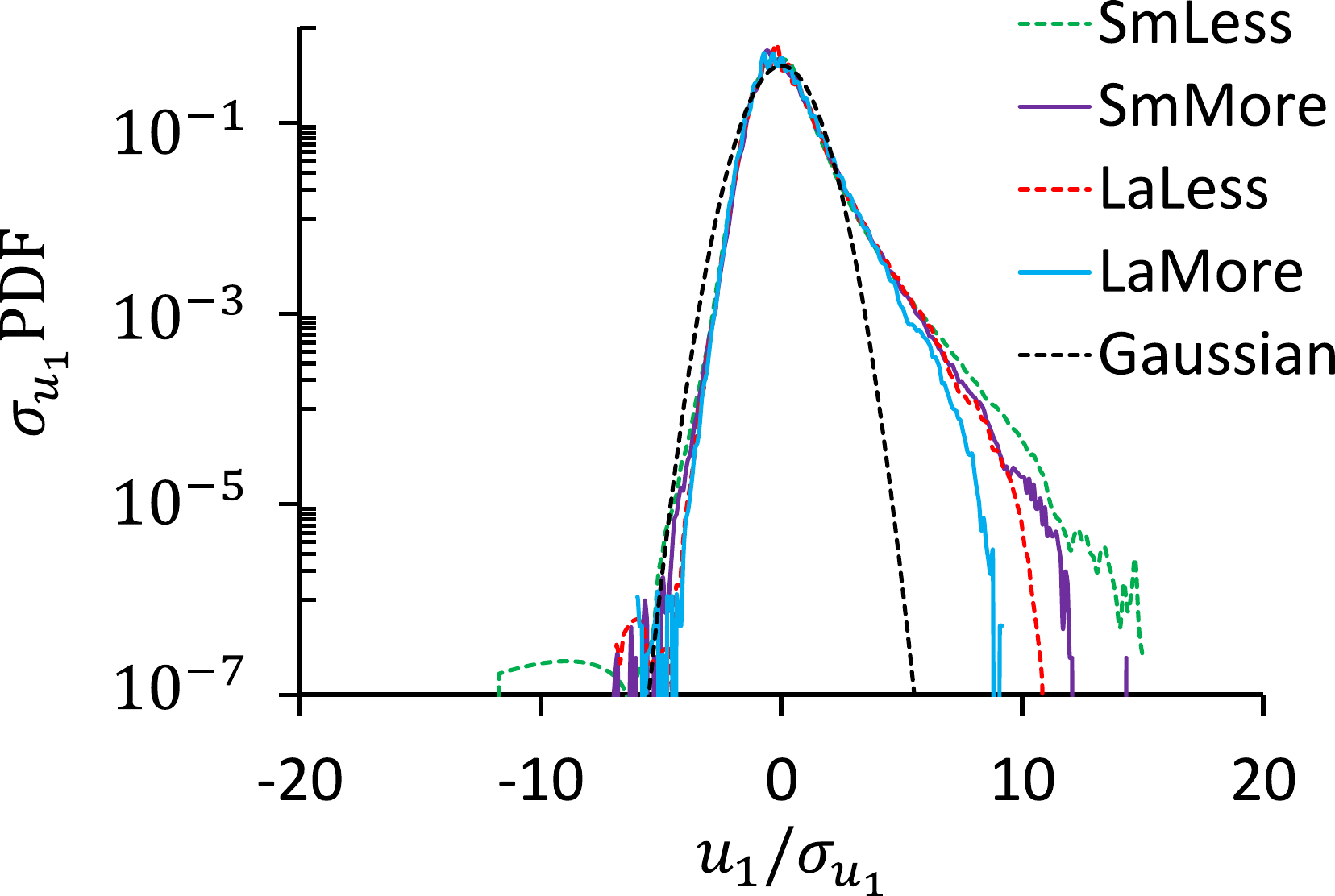}}
		\end{minipage}
		\begin{minipage}[b]{0.5\linewidth}
			\centering
			\makebox[1.2em][l]{\raisebox{-\height}{(\textit{b})}}%
			\raisebox{-\height}{\includegraphics[height=4.2cm]{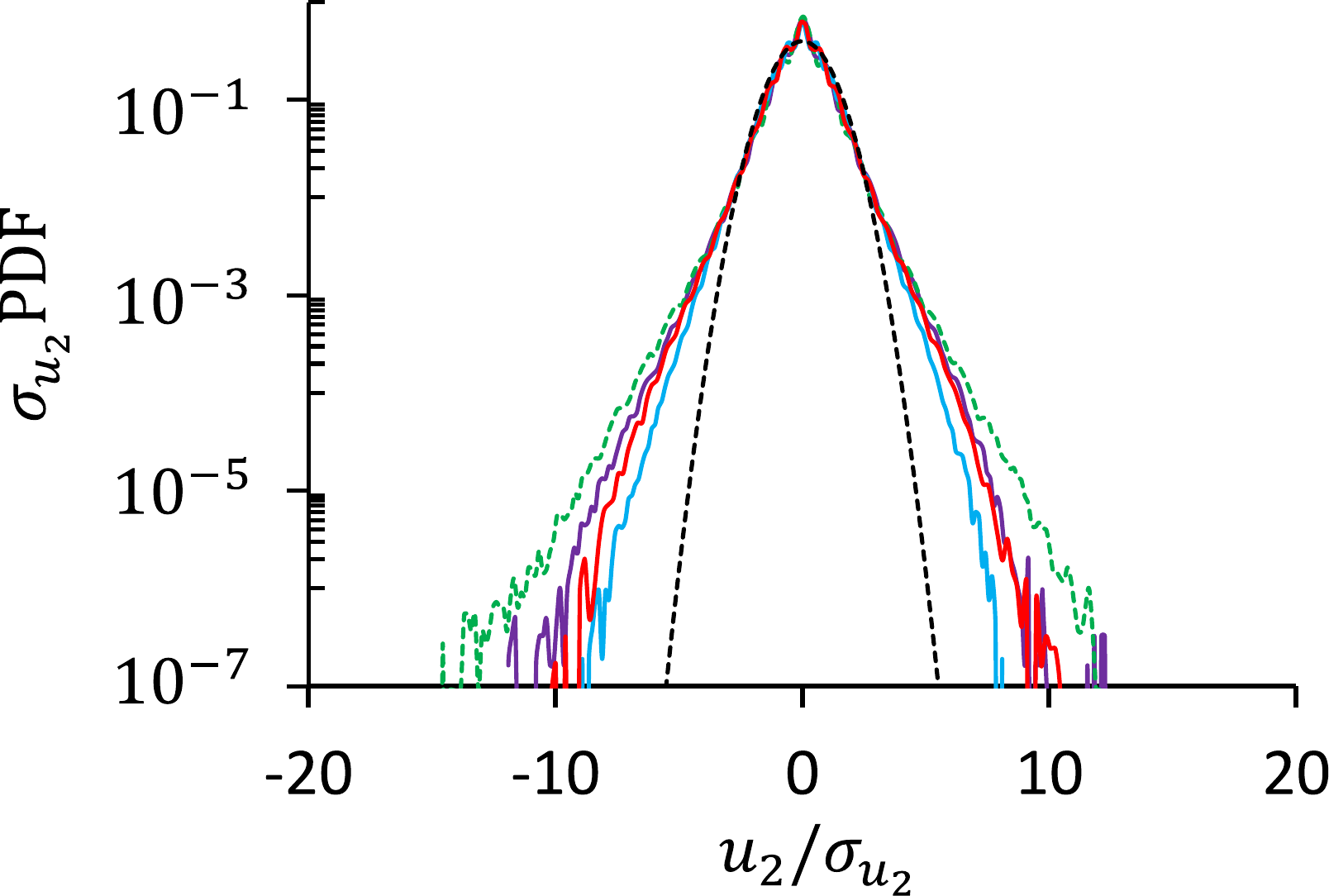}}
		\end{minipage}
	\end{minipage}
	\caption{Normalized probability density functions of liquid velocity fluctuations: (\textit{a}) the vertical component and (\textit{b}) the horizontal component.} \label{fig: pdf u and v}
\end{figure}
In figure \ref{fig: pdf u and v} we plot the PDFs of the liquid velocity fluctuations for both directions, normalized by their standard deviations. Due to the large quantity of velocity fields recorded from the experiment, the PDFs are well converged with tails extending to extreme values, and values of the PDF spanning six orders of magnitude, which is much greater than previous experiments for bubbly flows \citep{2010_Riboux,2017_Almeras,2019_Lai}. Both the vertical and horizontal velocity PDFs are in good quantitative agreement with the previous experimental studies just mentioned. While PDFs of the horizontal velocity fluctuations are symmetric and non-Gaussian, for the vertical component, the PDFs are strongly positively skewed for all the cases. This positive asymmetry originates from the wake entrainment as visualized in figure \ref{fig: PSV FOV} for the region directly behind the in-focus bubble, which leads to a larger probability of upward fluctuations. The asymmetry of the PDFs is highest for the case \textit{SmLess} with the the lowest gas void fraction and gradually reduces at larger $Re_{H_2}$. In our four cases, $Re_{H_2}$ increases with increasing $\alpha$, and a similar trend was also reported in figure 5(\textit{a}) of \cite{2016_Prakash}. Furthermore, for the PDF of the horizontal velocity component, \textit{SmLess} is the furthest from Gaussian, associated with stronger large-scale intermittency than the other cases. In many aspects, those results are very different from those of homogeneous isotropic turbulence (HIT) for single-phase flows, for which the PDFs of the velocity fluctuations are almost Gaussian \citep{2002_Gotoh}.

\section{Turbulence anisotropy quantified using structure functions}\label{sec: Anisotropy}

While Kolmogorov's theory assumes local isotropy at the small-scales of turbulent flows \citep[][K41 for brevity]{1941_Kolmogorov_a}, many experimental and DNS results reveal persistent anisotropy because the process of a return to isotropy at progressively smaller scales can be very slow \citep{1995_Pumir,2000_Shen,2006_Ouellette,2017_Carter}. The most systematic approach for characterizing anisotropy at different scales is based on the use of irreducible representations of the SO(3) group \citep{1999_Arad,2005_Biferale}. However, this method requires information on from the complete 3D flow fields, which is often not available. Due to this practical difficulty, many studies on turbulence anisotropy are based on the velocity structure function tensor, and then characterizing anisotropy based on how this quantity varies in different flow directions. The $n$-th order structure function is defined as
\begin{equation}
\boldsymbol{D}_n(\boldsymbol{r},t)\equiv\langle \Delta \boldsymbol{u}^n(\boldsymbol{x},\boldsymbol{r},t) \rangle  \;, \label{eq: Dij}
\end{equation}
where $\Delta \boldsymbol{u}(\boldsymbol{x},\boldsymbol{r},t)\equiv \boldsymbol{u}(\boldsymbol{x}+\boldsymbol{r},t)-\boldsymbol{u}(\boldsymbol{x},t)$ is the fluid velocity increment, and $\left\langle \cdot \right\rangle$ denotes an ensemble average. The calculation of liquid velocity increments in a bubbly flow is, however, delicate, since the liquid velocity is not defined at points occupied by a bubble. To overcome this non-continuous velocity signal challenge, we use the method proposed in \citep{2017_Ma} -- store only liquid velocity data along the horizontal/vertical PSV grid lines whenever the entire line was free from bubbles in the considered FOV (see the two dashed lines in figure \ref{fig: PSV FOV} as example). Based on the $60,000$ velocity fields that were recorded for each case, we were able to extract data along $500,000$ to $1,500,000$ such vertical/horizontal lines, depending on the case.

\subsection{Second-order structure function}\label{subsec: 2-order SF}

We first consider the second-order structure function, whose components are
\begin{equation}
	D_2^{ij}(\boldsymbol{x},\boldsymbol{r},t)\equiv\langle \Delta u_i(\boldsymbol{x},\boldsymbol{r},t)\Delta u_j(\boldsymbol{x},\boldsymbol{r},t) \rangle  \;, \label{eq: Dij}
\end{equation}
Hereafter, we suppress the space $\boldsymbol{x}$ and time $t$ arguments since we are considering a flow which is statistically homogeneous and stationary flows over the FOV. The PSV technique provides access to data associated with separations along two directions, namely, the vertical separation $\boldsymbol{r}=r_1\boldsymbol{e}_1$ ($r_1\equiv\|\boldsymbol{r}\|$) and the horizontal separation $\boldsymbol{r}=r_2\boldsymbol{e}_2$ ($r_2\equiv\|\boldsymbol{r}\|$). Hence, we are able to compute the four contributions 
\begin{equation}
D^L_2(r_1)=D_2^{11}(r_1)  \;, 
\end{equation}
\begin{equation}
D^L_2(r_2)=D_2^{22}(r_2)  \;, 
\end{equation}
\begin{equation} 
D^T_2(r_1)=D_2^{22}(r_1)  \;, 
\end{equation}
\begin{equation}
D^T_2(r_2)=D_2^{11}(r_2)  \;,
\end{equation}
based on the Cartesian coordinate system depicted in figure \ref{fig: Bubble_column}. For an incompressible and isotropic flow, the following relation holds for the transverse structure function
\begin{equation}
D^{T}_{2,iso}(r_\gamma)=D^{L}_{2}(r_\gamma)+\frac{r_\gamma}{2}\frac{\partial}{\partial{r_\gamma}}D^{L}_{2}(r_\gamma) \;, \label{eq: D_T-iso}
\end{equation}
where no summation over $\gamma$ is implied. 

\begin{figure}
	\begin{minipage}[b]{1.0\linewidth}
		\begin{minipage}[b]{0.5\linewidth}
			\centering
			\makebox[0.5em][l]{\raisebox{-\height}{(\textit{a})}}%
			\raisebox{-\height}{\includegraphics[height=3.6cm]{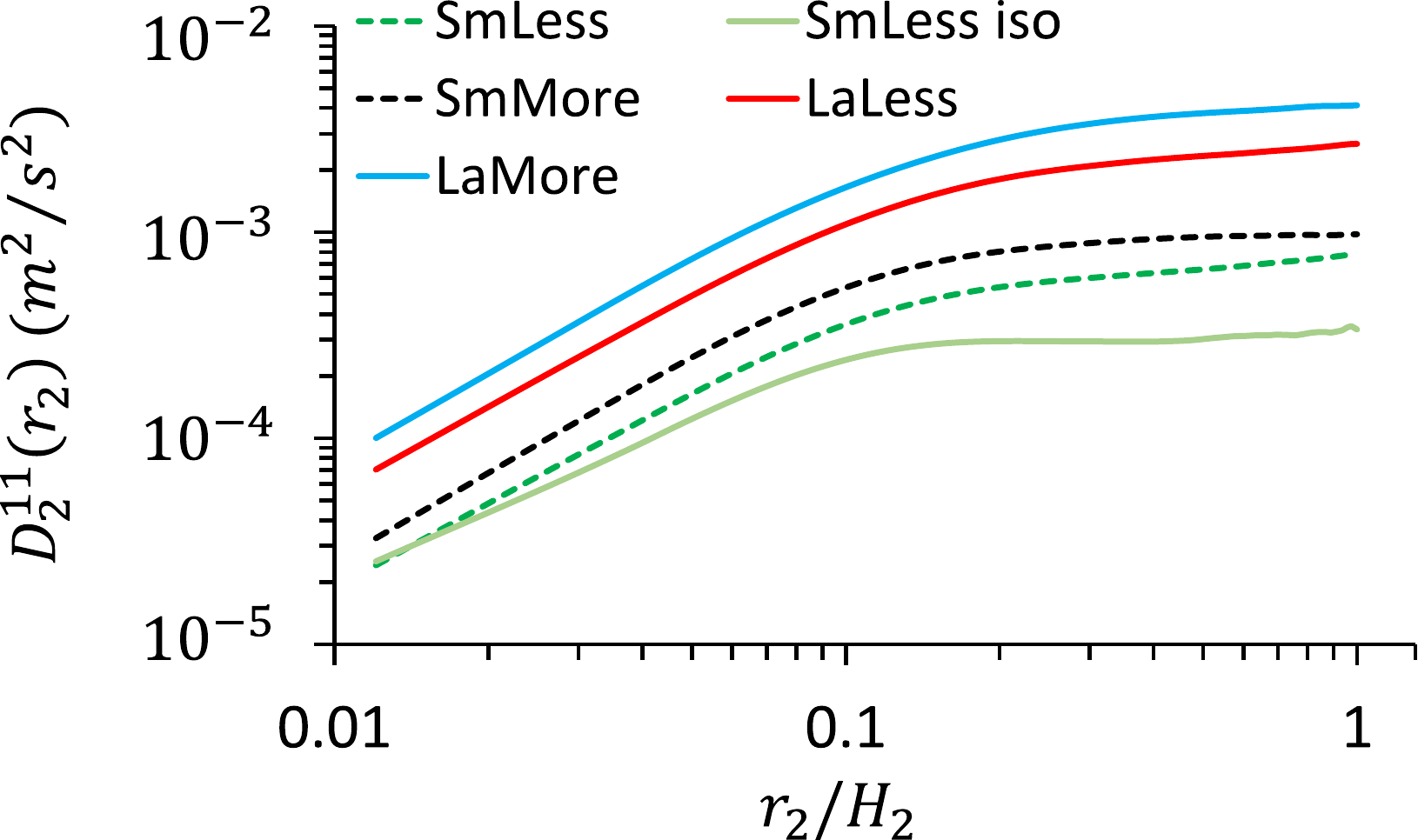}}
		\end{minipage}
		\begin{minipage}[b]{0.5\linewidth}
			\centering
			\makebox[0.5em][l]{\raisebox{-\height}{(\textit{b})}}%
			\raisebox{-\height}{\includegraphics[height=3.6cm]{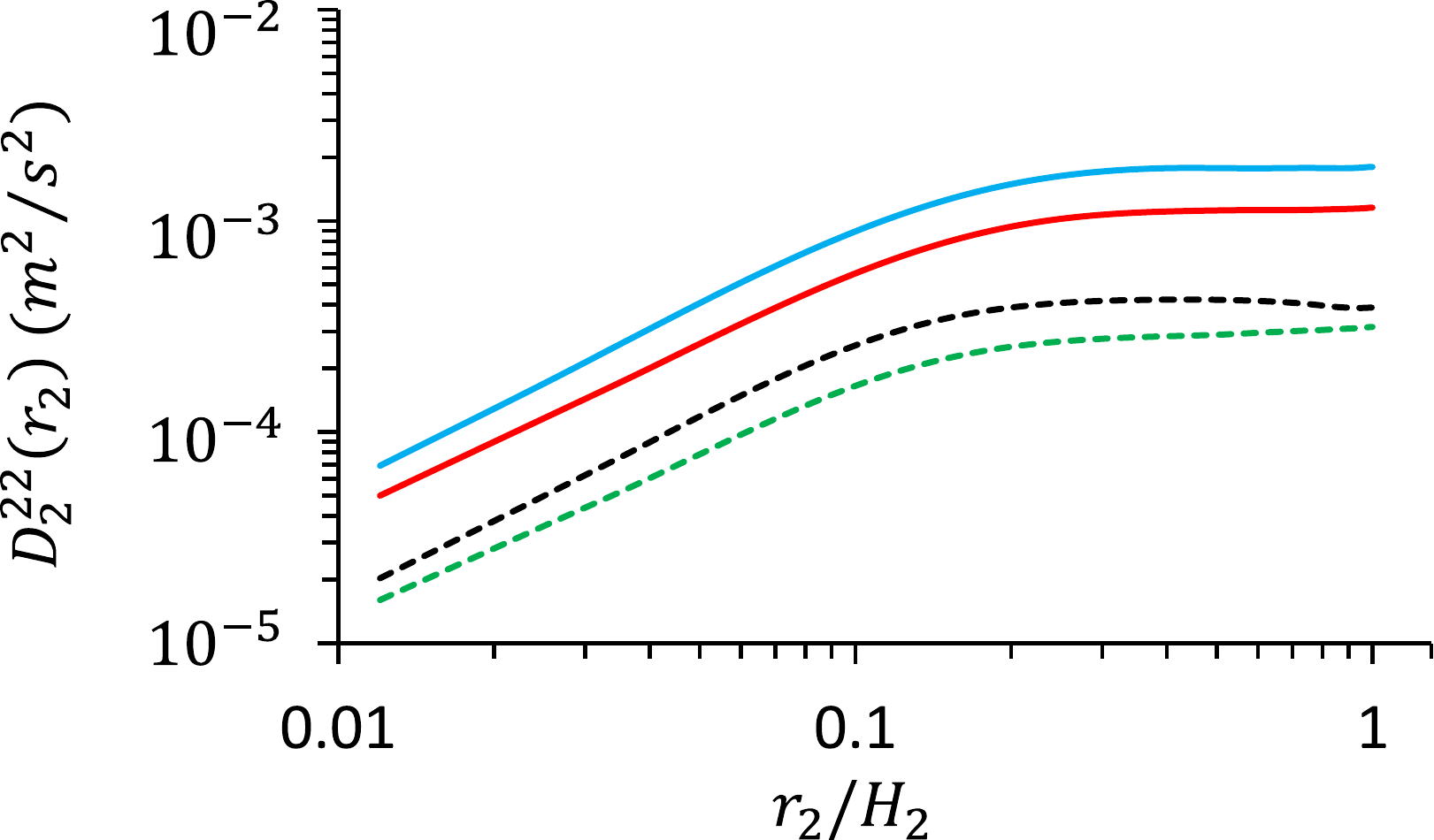}}
		\end{minipage}
	\end{minipage}
	\begin{minipage}[b]{1.0\linewidth}
		\vspace{3mm}
		\begin{minipage}[b]{0.5\linewidth}
			\centering
			\makebox[0.5em][l]{\raisebox{-\height}{(\textit{c})}}%
			\raisebox{-\height}{\includegraphics[height=3.6cm]{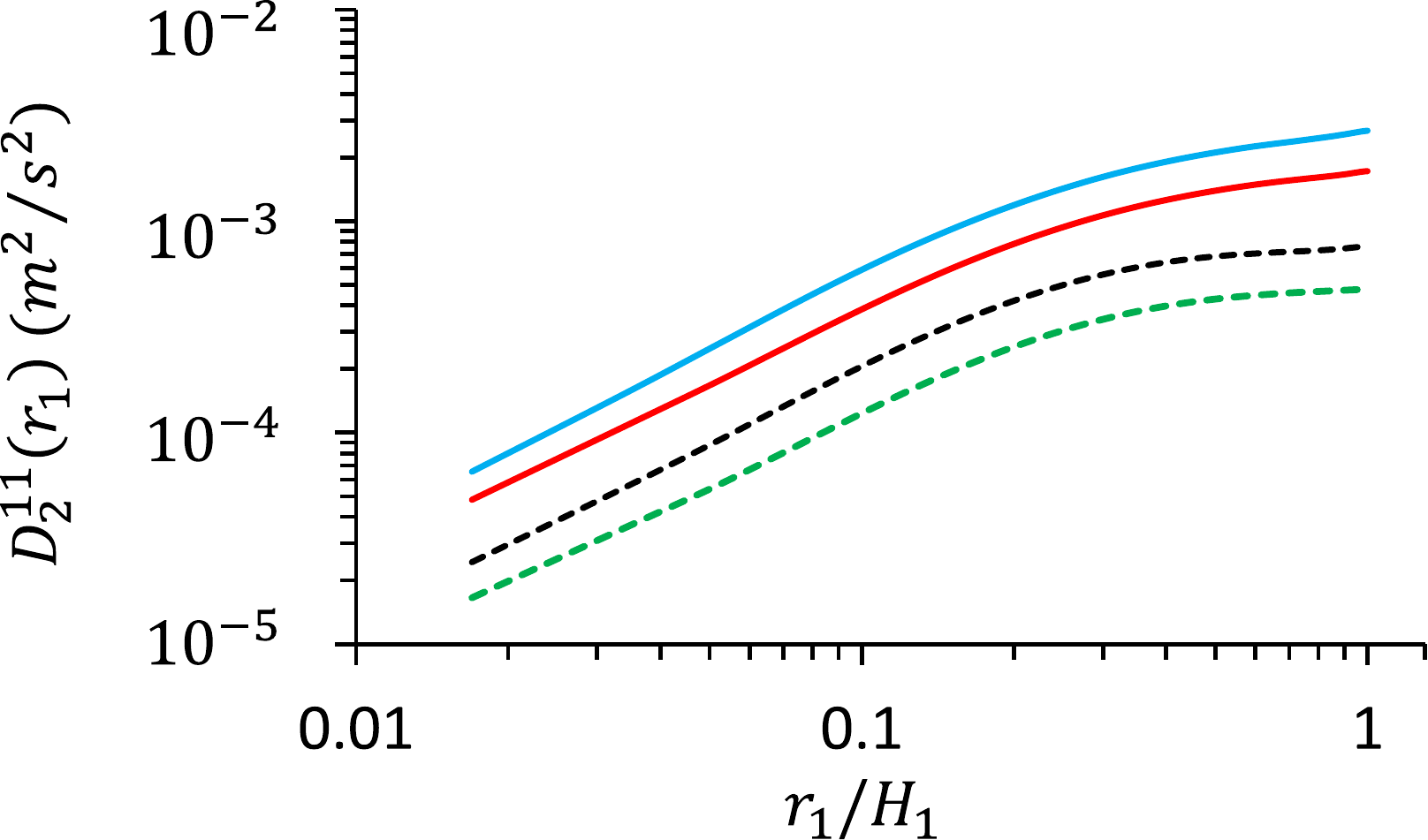}}
		\end{minipage}
		\begin{minipage}[b]{0.5\linewidth}
			\centering
			\makebox[0.5em][l]{\raisebox{-\height}{(\textit{d})}}%
			\raisebox{-\height}{\includegraphics[height=3.6cm]{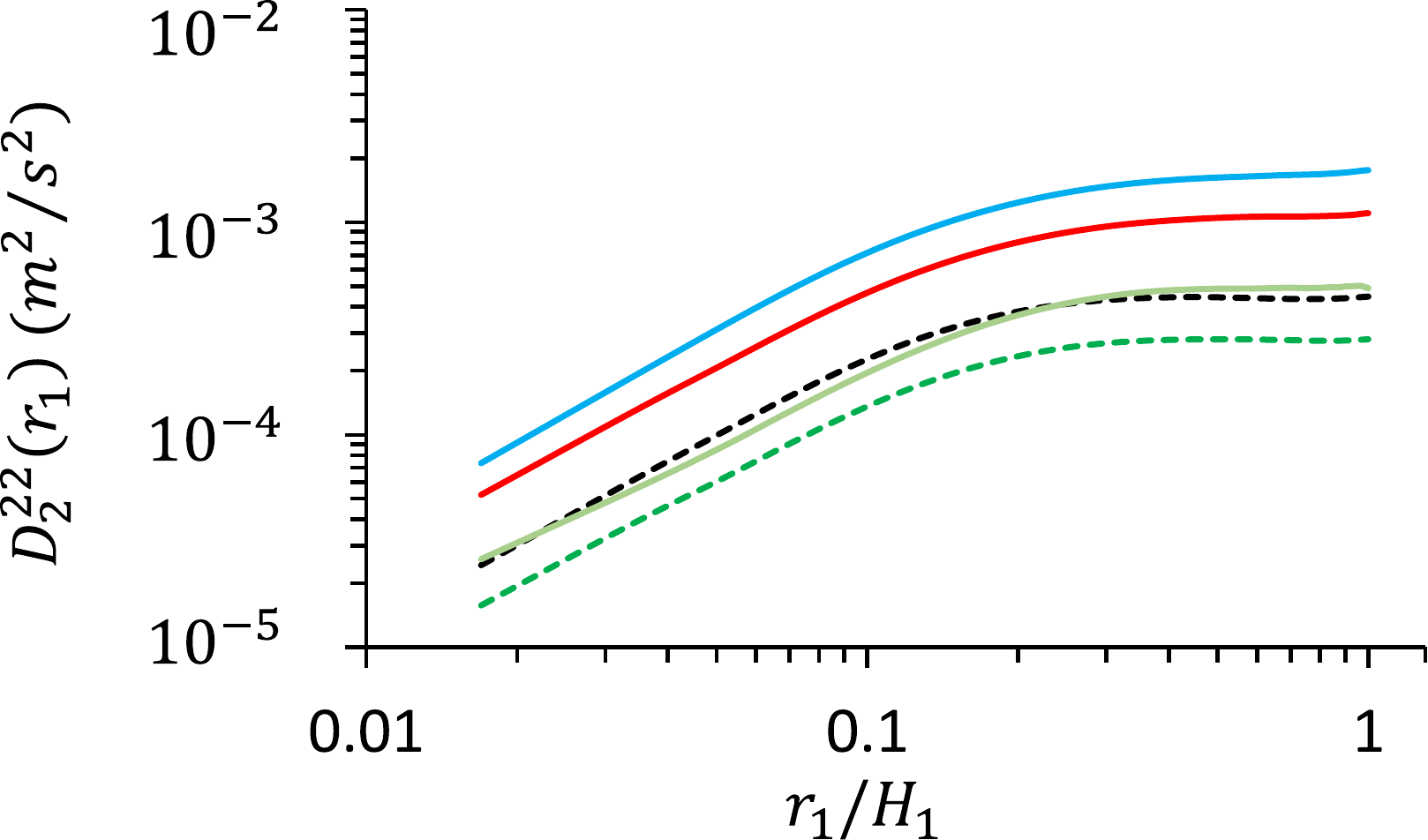}}
		\end{minipage}
	\end{minipage}
\caption{Second-order structure functions, with separations along the horizontal (\textit{a,b}) and the vertical (\textit{c,d}) directions. Note that $D^{T}_{2,iso}(r_i)$ is shown for \textit{SmLess} in (\textit{a,d}).} \label{fig: Dii}
\end{figure}

Figure \ref{fig: Dii} shows all the measured components of the second-order structure function, as well as $D^{T}_{2,iso}(r_\gamma)$ obtained from (\ref{eq: D_T-iso}) for the representative case \textit{SmLess} in figure \ref{fig: Dii}(\textit{a,d}). The results show that the values of the structure functions increase in the order \textit{SmLess, SmMore, LaLess}, and \textit{LaMore}, which corresponds to increasing $Re_{H_2}$, gas void fraction, and/or bubble Reynolds number. This relationship holds for all of the components computed and across all scales. A similar trend was also reported for all three diagonal components of the second-order structure function based on separations in the spanwise direction of a bubble-laden turbulent channel flow in our recent study \citep{2021_Ma}. At the large-scales of a homogeneous flow both $(1/2)D^{11}_2(r_\gamma)/\langle u_1u_1\rangle$ and $(1/2)D^{22}_2(r_\gamma)/\langle u_2u_2\rangle$ approach unity. Our data for these normalized quantities show that along the separation $r_2$ the quantities converge to unity when $r_2\rightarrow H_2$. However, while $(1/2)D^{22}_2(r_1\rightarrow H_1)/\langle u_2u_2\rangle$ approaches unity, $(1/2)D^{11}_2(r_1\rightarrow H_1)/\langle u_1u_1\rangle\approx0.8$ for all the cases. This implies that the FOV is not large enough in the vertical direction to resolve the integral lengthscale of the flow in this direction, while it is for the horizontal direction. However, this may be simply due to the fact that $H_1<H_2$ for the FOV. 

For the \textit{SmLess} case, departures from $D^{11}_2(r_2)=D^{11,iso}_2(r_2)$ are larger at the large scales, reflecting the large scale anisotropy characterized by e.g. $u^{\mathrm{rms}}_1/u^{\mathrm{rms}}_2$ in table \ref{tab: flow para}. At for smallest values of $r_2$, $D^{11}_2(r_2)\approx D^{11,iso}_2(r_2)$. However, for separations along the vertical direction there are strong departures from $D^{22}_2(r_1)=D^{22,iso}_2(r_1)$ at all scales (figure \ref{fig: Dii}\textit{d}). We obtained similar results (not shown) for the other three cases. Interestingly, such behavior was also observed in \cite{2017_Carter} for single-phase turbulence generated by jet-stirring (with zero-mean-flow) at a similar range of $u^{\mathrm{rms}}_1/u^{\mathrm{rms}}_2$.   

\begin{figure}
	\begin{minipage}[b]{1.0\linewidth}
		\begin{minipage}[b]{0.5\linewidth}
			\centering
			\makebox[1em][l]{\raisebox{-\height}{(\textit{a})}}%
			\raisebox{-\height}{\includegraphics[height=3.8cm]{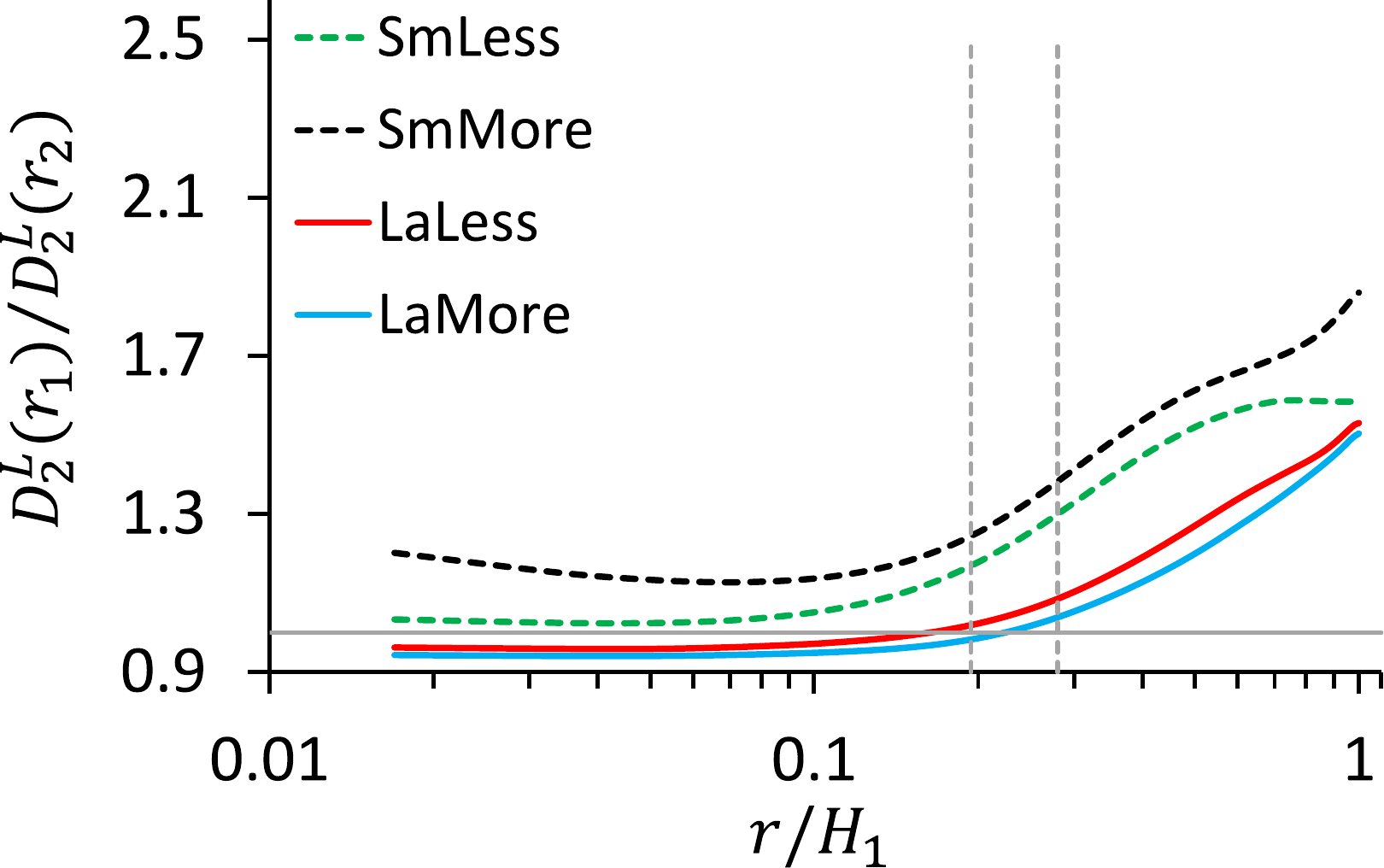}}
		\end{minipage}
		\begin{minipage}[b]{0.5\linewidth}
			\centering
			\makebox[1em][l]{\raisebox{-\height}{(\textit{b})}}%
			\raisebox{-\height}{\includegraphics[height=3.8cm]{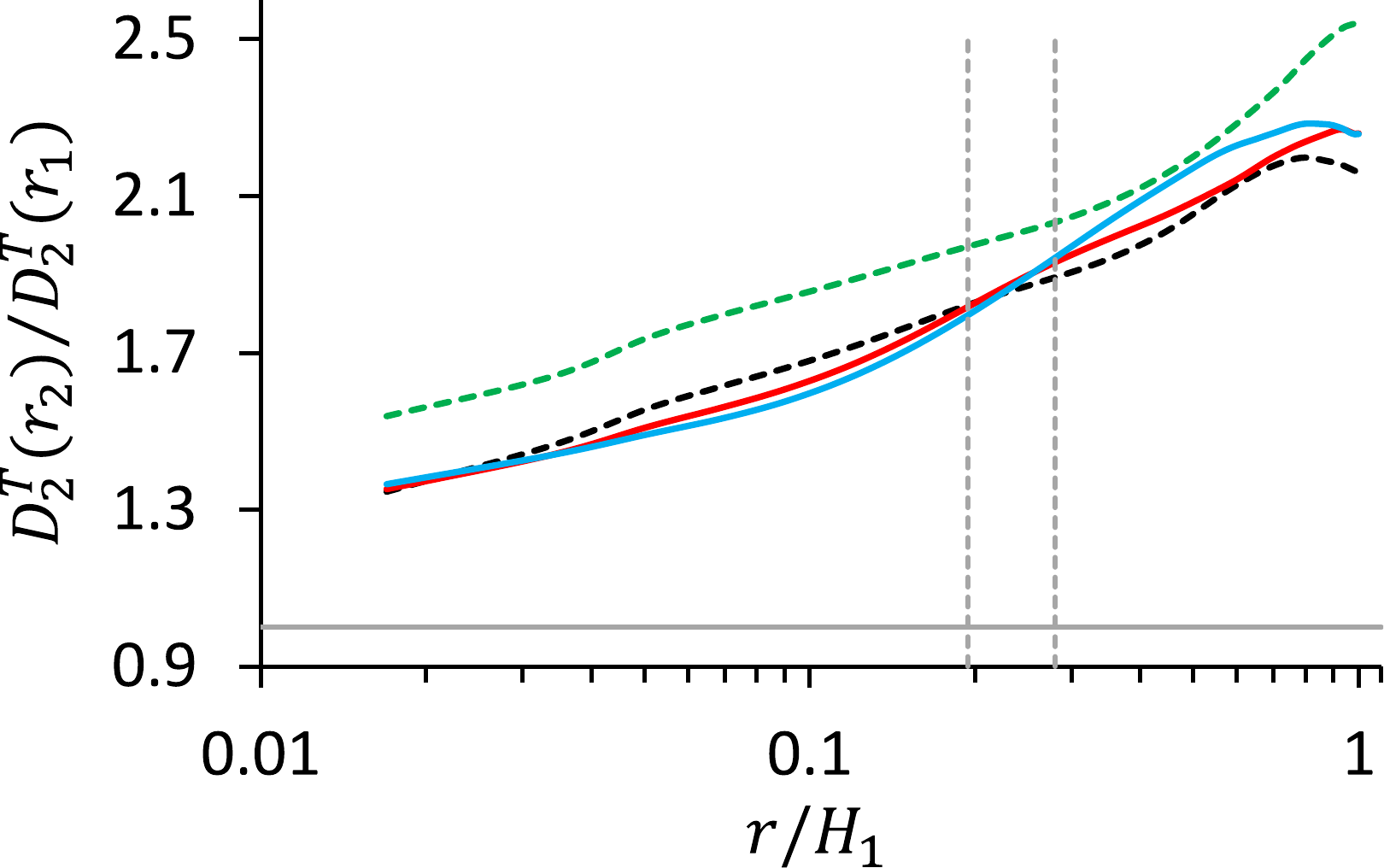}}
		\end{minipage}
	\end{minipage}
\caption{Ratio of longitudinal (\textit{a}) and transverse (\textit{b}) structure functions in different separation directions for all the cases. In (\textit{a,b}) the horizontal lines indicate the value of unity and the two vertical dashed lines show $r=d_p$ for smaller and larger bubbles, respectively.} \label{fig: D_LT_ratio}
\end{figure}

For the second-order structure functions, one of other ways to quantify anisotropy is by the ratios of components such as $D^L_2(r_1)/D^L_2(r_2)$ and $D^T_2(r_2)/D^T_2(r_1)$, which would be equal to unity for an isotropic flow. The results in figure \ref{fig: D_LT_ratio} show that both ratios monotonically decrease for decreasing separation. While the ratio of the longitudinal structure functions is very close to unity at small scales, the ratio of transverse structure functions departs more strongly from unity. By comparing the different cases, the results indicate that the smaller bubbles generate stronger anisotropy in the flow than the larger bubbles across the scales, which is in agreement with the DNS results in \cite{2021_Ma} and also the results of large-scale anisotropy in \S\,\ref{sec: one-point}. 

In general, the differing behaviour of the second-order structure function in the longitudinal and transverse directions shows that both need to be considered in order to fully consider the turbulence anisotropy across scales. This conclusion based on our bubble-laden flow is in agreement with that of \cite{2017_Carter} for single-phase turbulence.

\subsection{High-order structure function}\label{subsec: high-order SF}

\begin{figure}
	\begin{minipage}[b]{1.0\linewidth}
		\begin{minipage}[b]{0.33\linewidth}
			\centering
			\makebox[-2.2em][l]{\raisebox{-\height}{(\textit{a})}}%
			\raisebox{-\height}{\includegraphics[height=3cm]{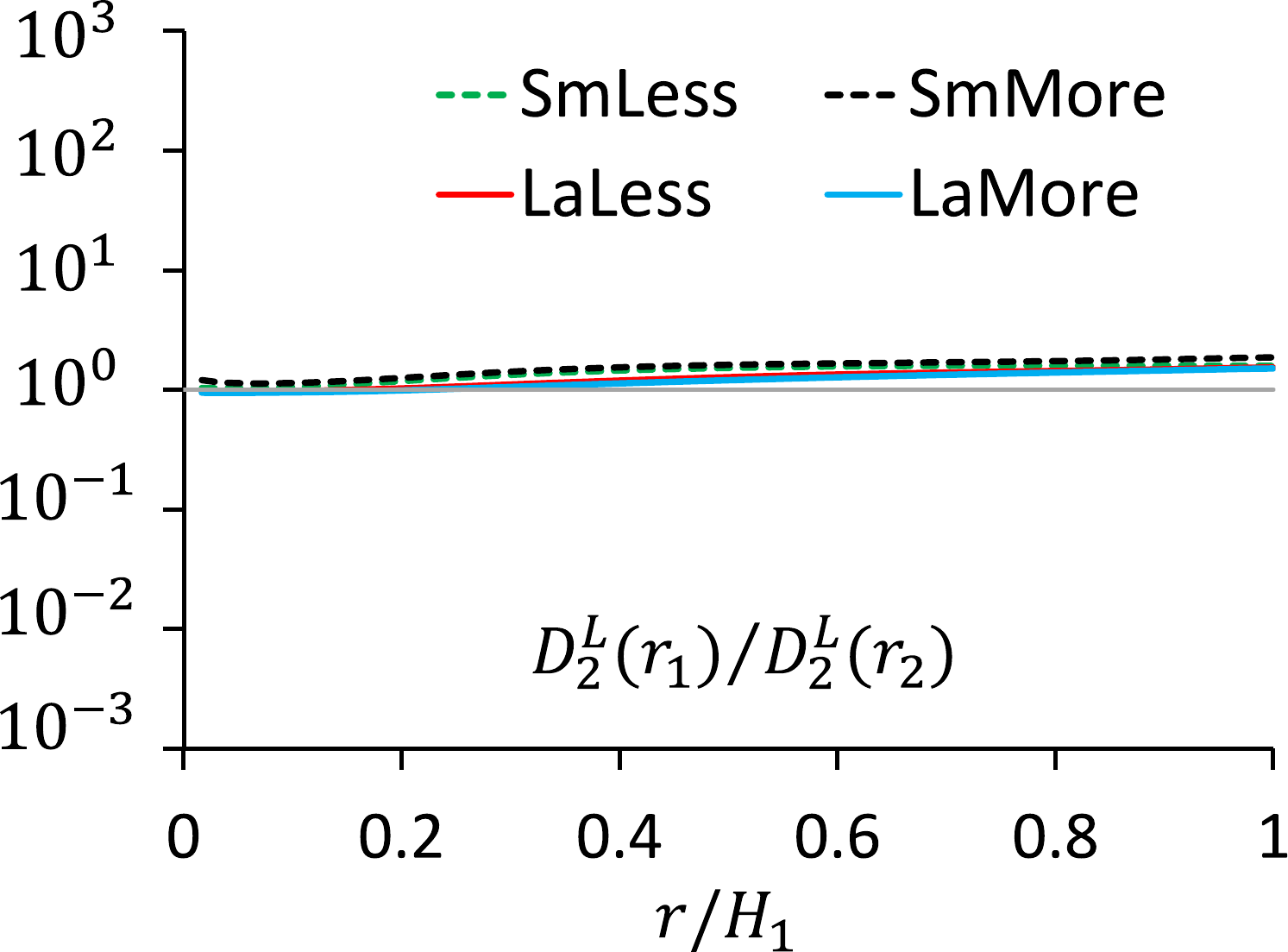}}
		\end{minipage}
		\begin{minipage}[b]{0.33\linewidth}
			\centering
			\makebox[-2.2em][l]{\raisebox{-\height}{(\textit{b})}}%
			\raisebox{-\height}{\includegraphics[height=3cm]{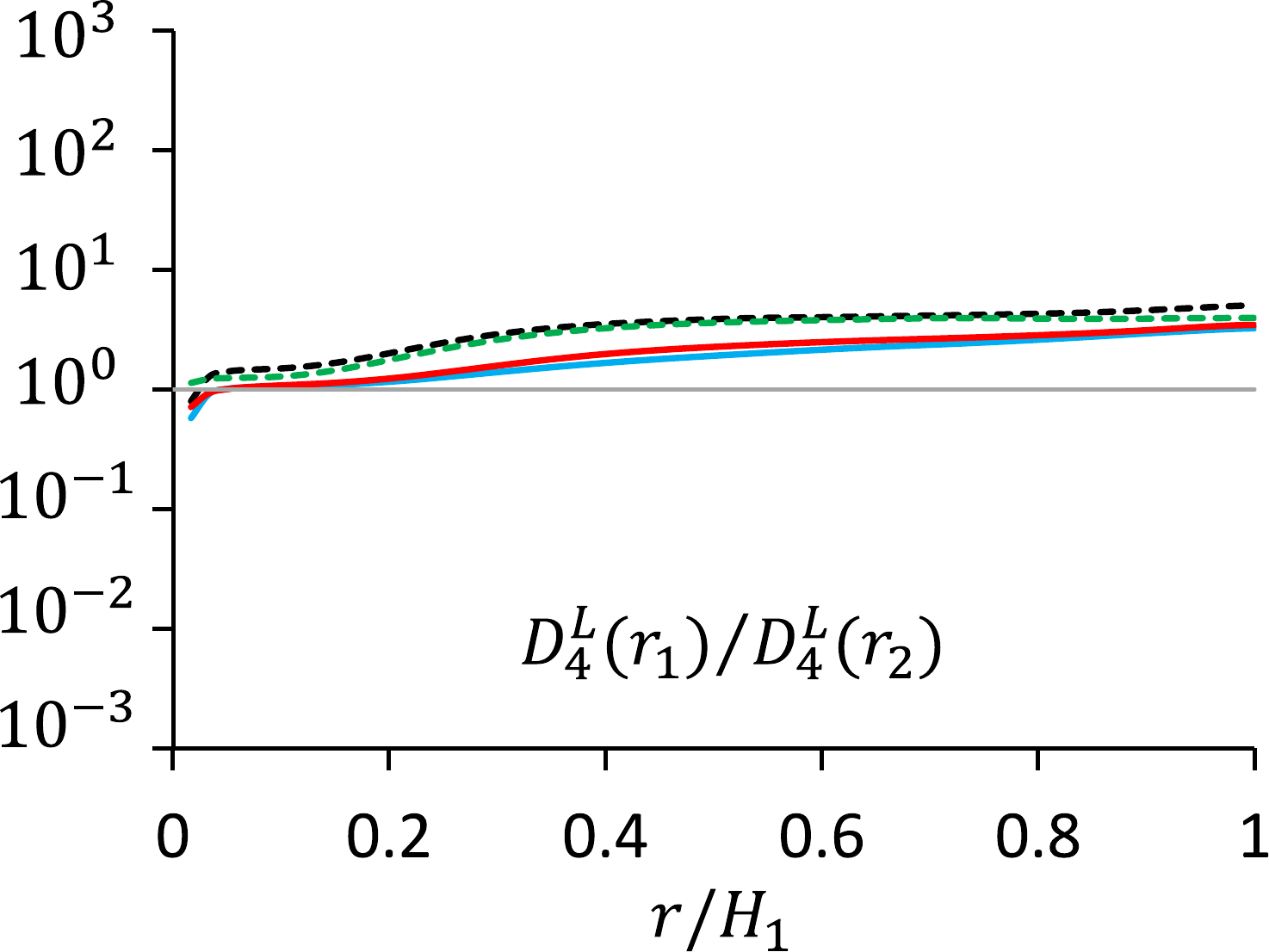}}
		\end{minipage}
		\begin{minipage}[b]{0.33\linewidth}
		    \centering
		    \makebox[-2.3em][l]{\raisebox{-\height}{(\textit{c})}}%
		    \raisebox{-\height}{\includegraphics[height=3cm]{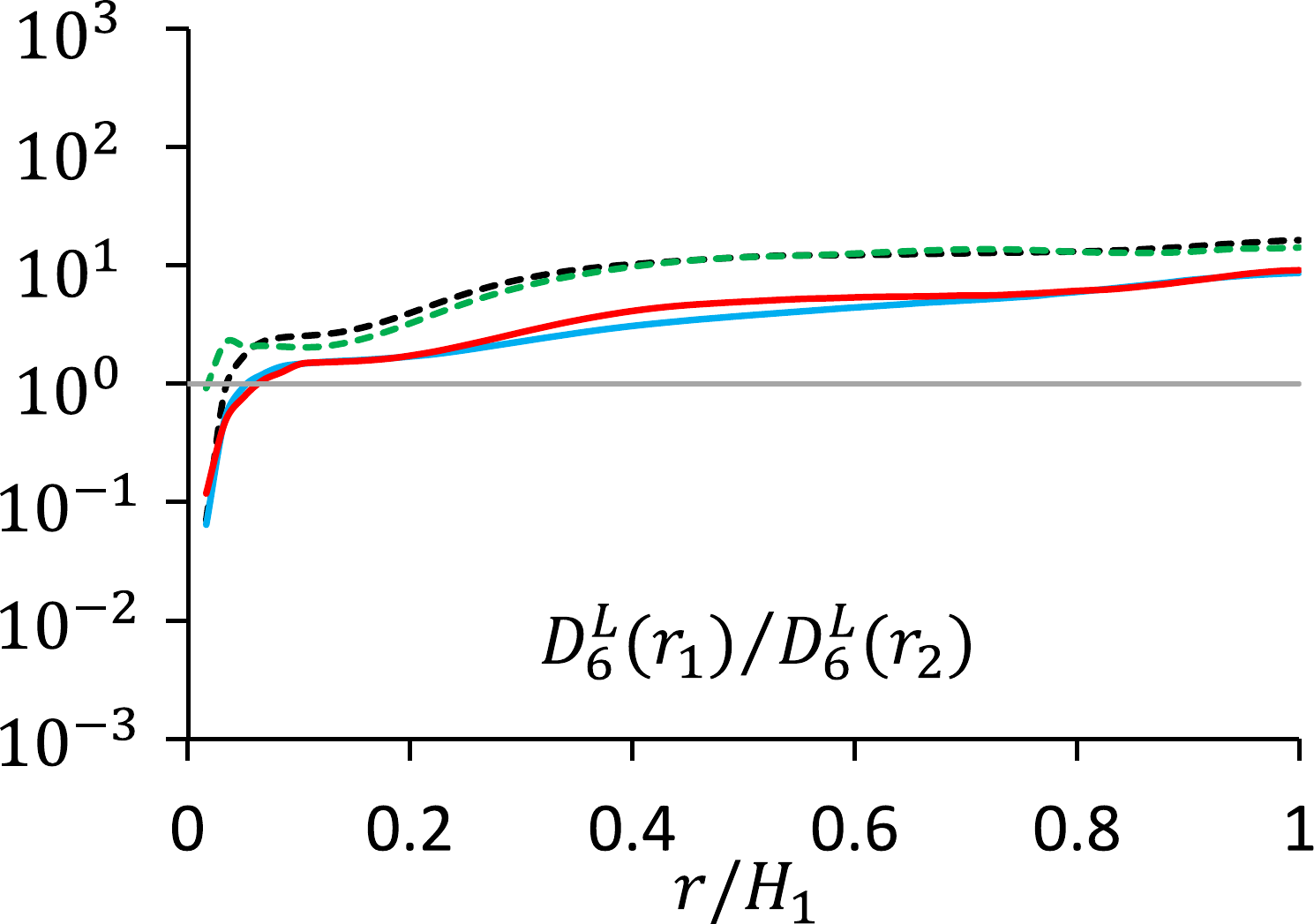}}
	    \end{minipage}
    \end{minipage}
	\begin{minipage}[b]{1.0\linewidth}
		\vspace{2mm}
		\begin{minipage}[b]{0.33\linewidth}
			\centering
			\makebox[-2.3em][l]{\raisebox{-\height}{(\textit{d})}}%
			\raisebox{-\height}{\includegraphics[height=3cm]{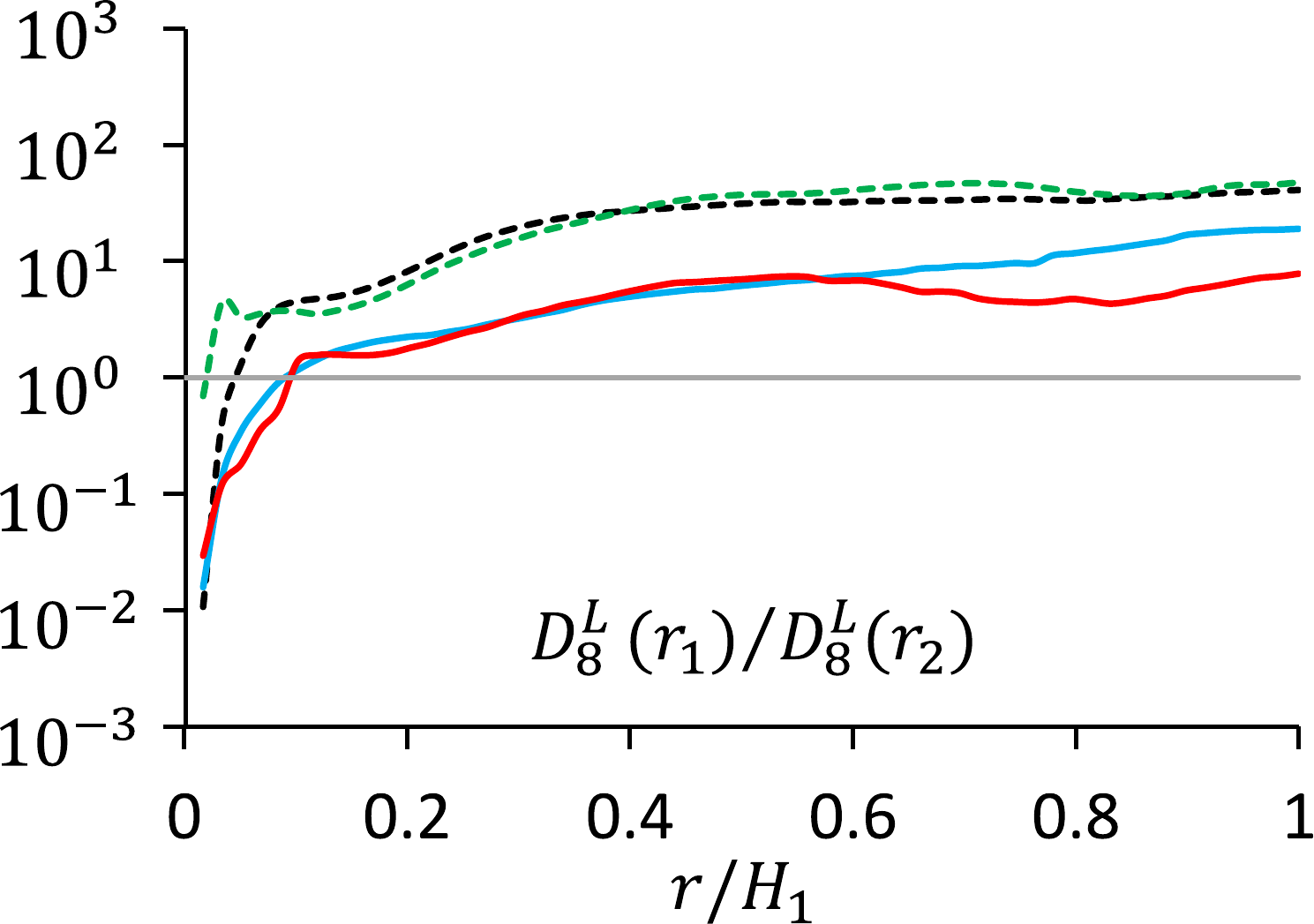}}
		\end{minipage}
		\begin{minipage}[b]{0.33\linewidth}
			\centering
			\makebox[-2.3em][l]{\raisebox{-\height}{(\textit{e})}}%
			\raisebox{-\height}{\includegraphics[height=3cm]{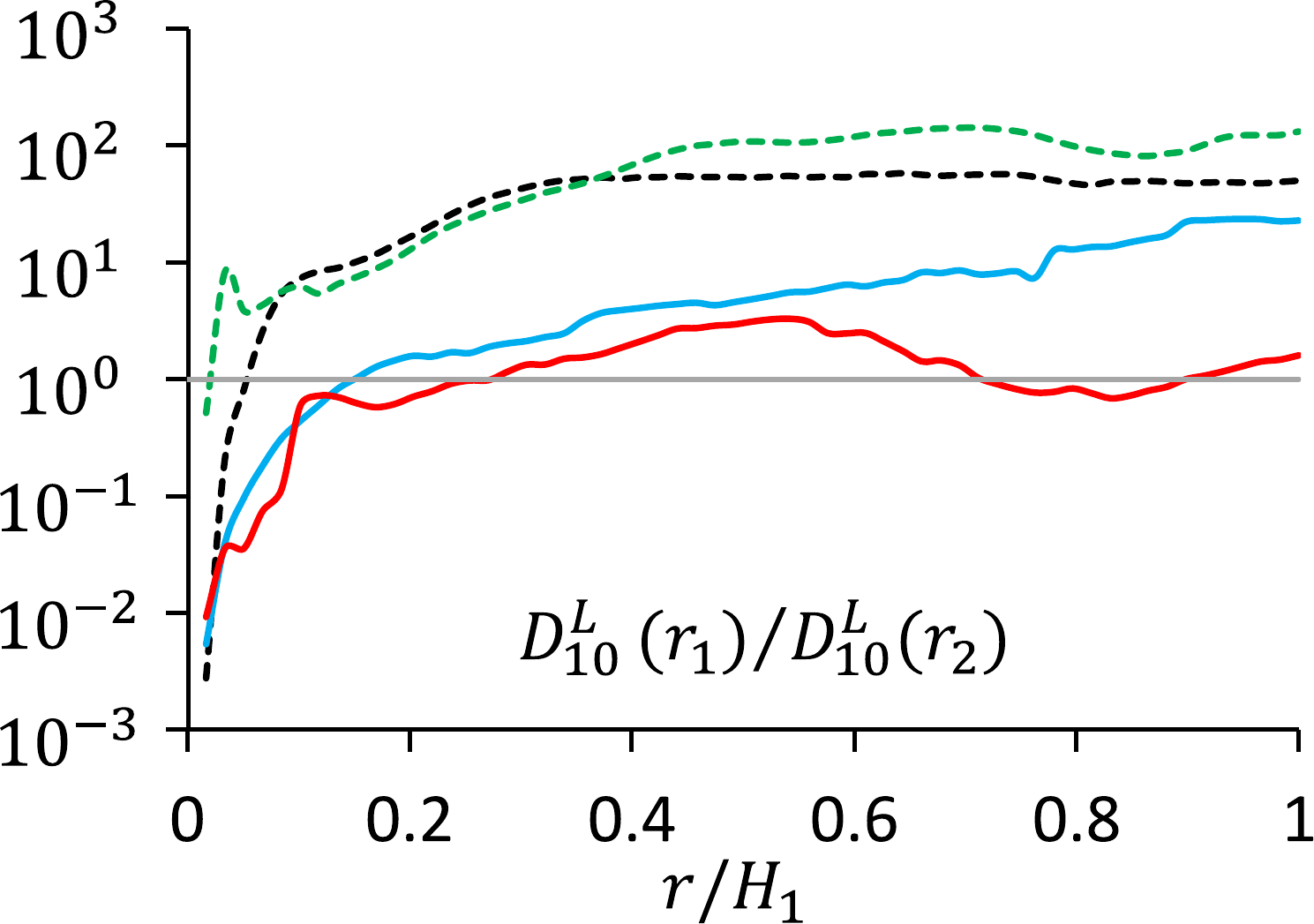}}
		\end{minipage}
		\begin{minipage}[b]{0.33\linewidth}
	        \centering
	        \makebox[-2.3em][l]{\raisebox{-\height}{(\textit{f})}}%
	        \raisebox{-\height}{\includegraphics[height=3cm]{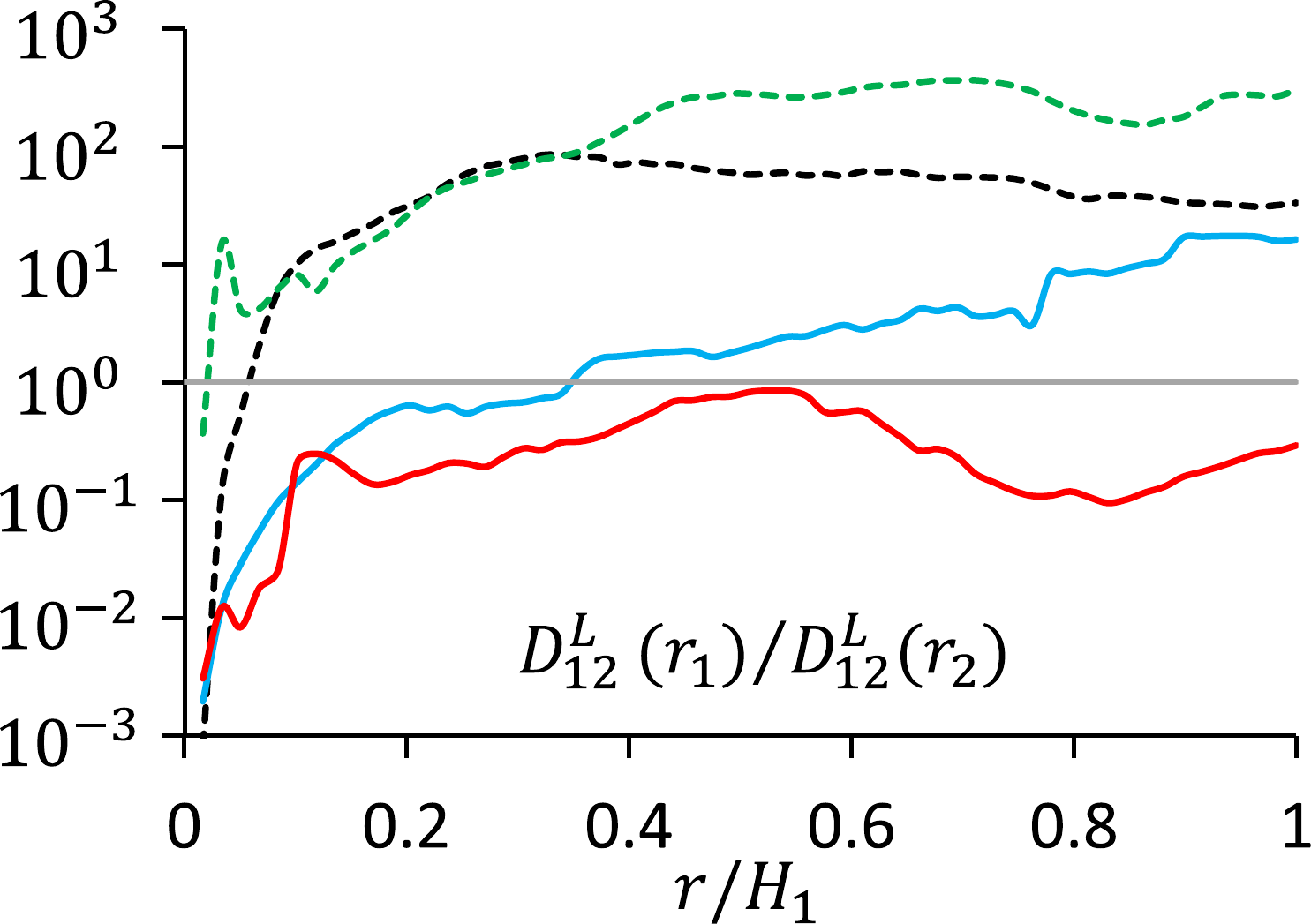}}
        \end{minipage}			
     \end{minipage}
\caption{Ratio of $n$th-even-order longitudinal structure functions in different separation directions for all the cases ($n=2, 4, 6, 8, 10, 12$). The grey horizontal line in each plot indicates the isotropic value of unity. Note that (\textit{a}) presents the same value as figure \ref{fig: D_LT_ratio}(\textit{a}), just with a different label of y-axis.} \label{fig: DL_ratio}
\end{figure}

\begin{figure}
	\begin{minipage}[b]{1.0\linewidth}
		\begin{minipage}[b]{0.33\linewidth}
			\centering
			\makebox[-2.1em][l]{\raisebox{-\height}{(\textit{a})}}%
			\raisebox{-\height}{\includegraphics[height=3cm]{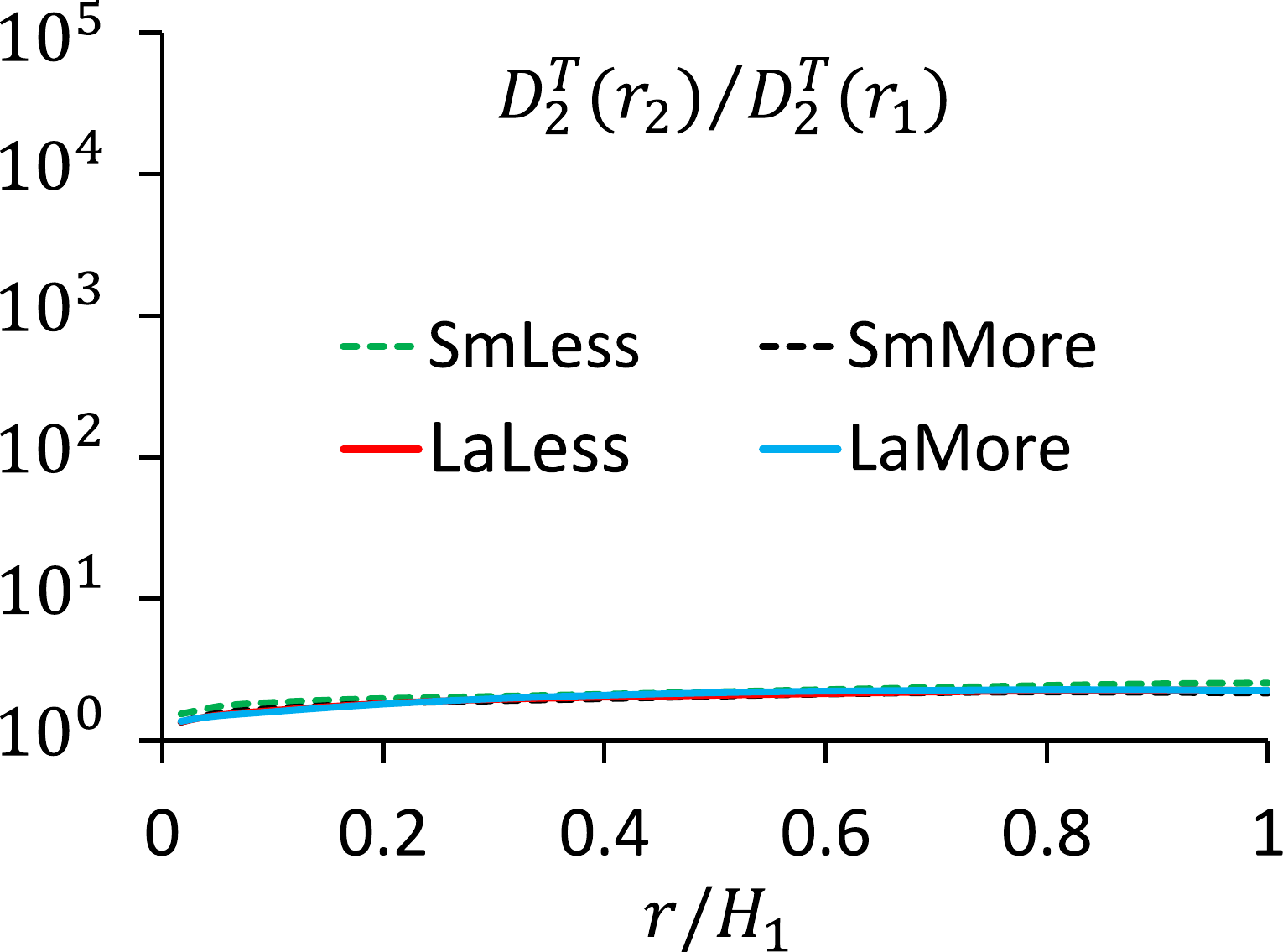}}
		\end{minipage}
		\begin{minipage}[b]{0.33\linewidth}
			\centering
			\makebox[-2.1em][l]{\raisebox{-\height}{(\textit{b})}}%
			\raisebox{-\height}{\includegraphics[height=3cm]{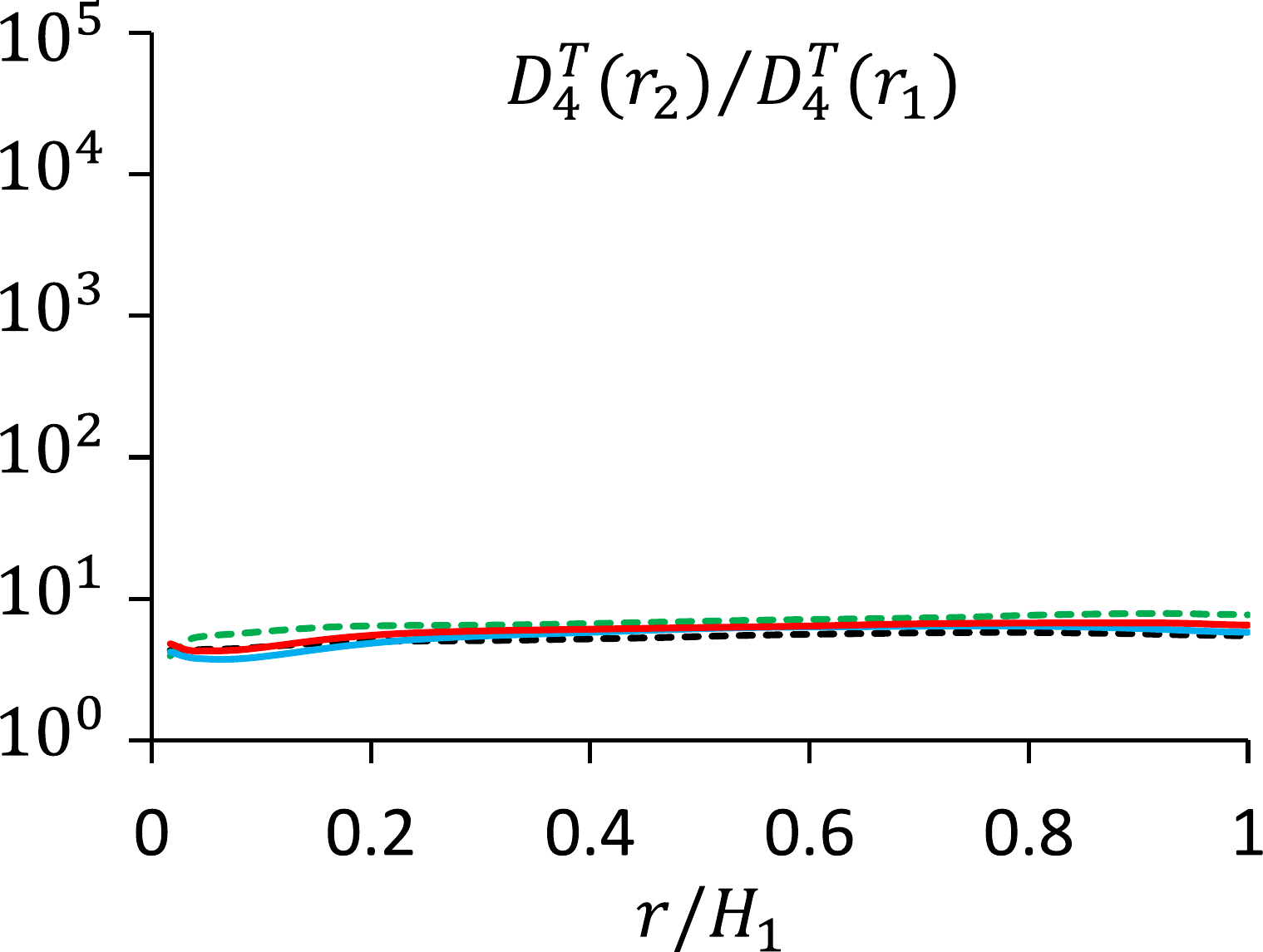}}
		\end{minipage}
		\begin{minipage}[b]{0.33\linewidth}
			\centering
			\makebox[-2.2em][l]{\raisebox{-\height}{(\textit{c})}}%
			\raisebox{-\height}{\includegraphics[height=3cm]{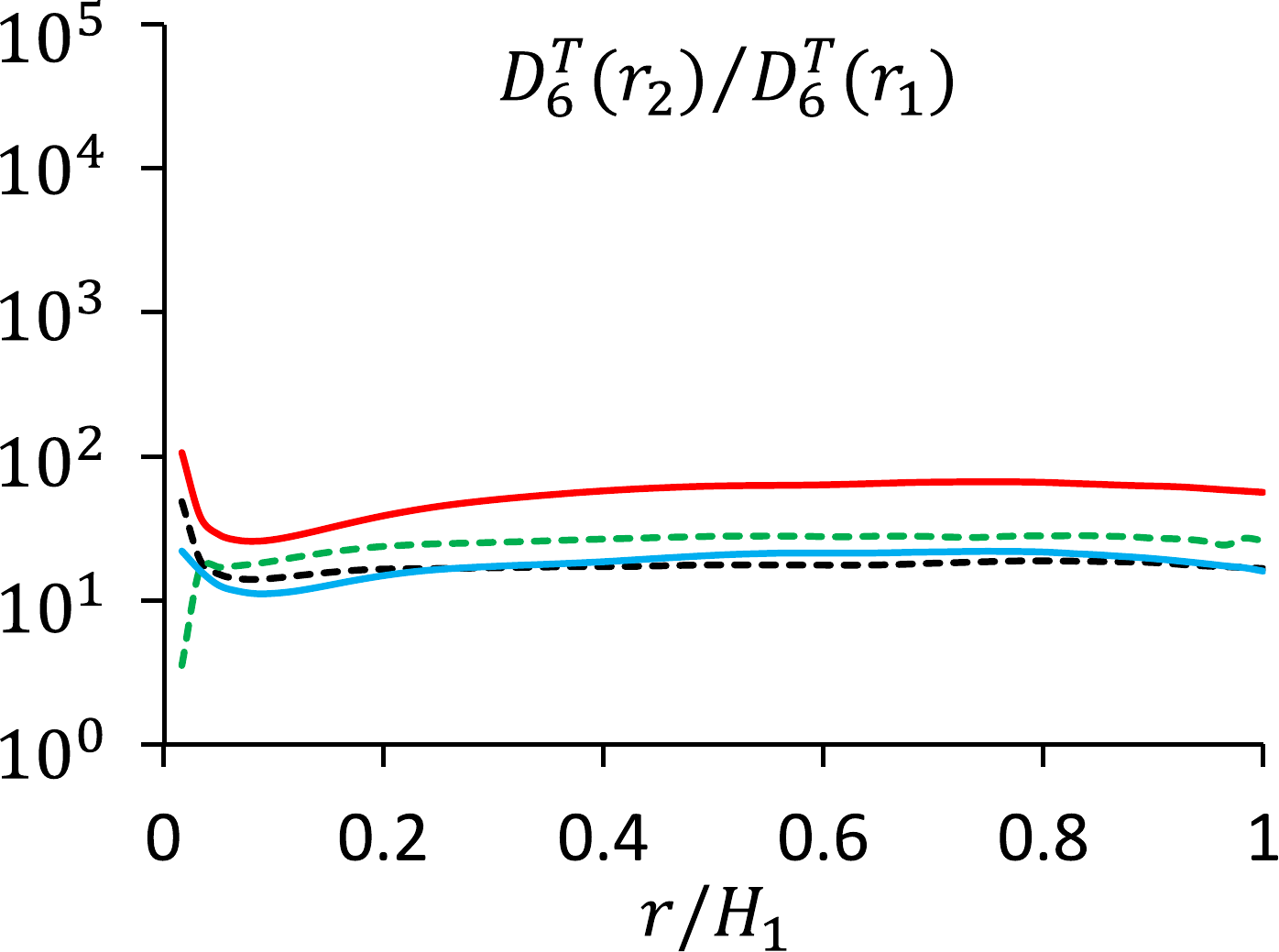}}
		\end{minipage}
	\end{minipage}
	\begin{minipage}[b]{1.0\linewidth}
		\vspace{2mm}
		\begin{minipage}[b]{0.33\linewidth}
			\centering
			\makebox[-2.2em][l]{\raisebox{-\height}{(\textit{d})}}%
			\raisebox{-\height}{\includegraphics[height=3cm]{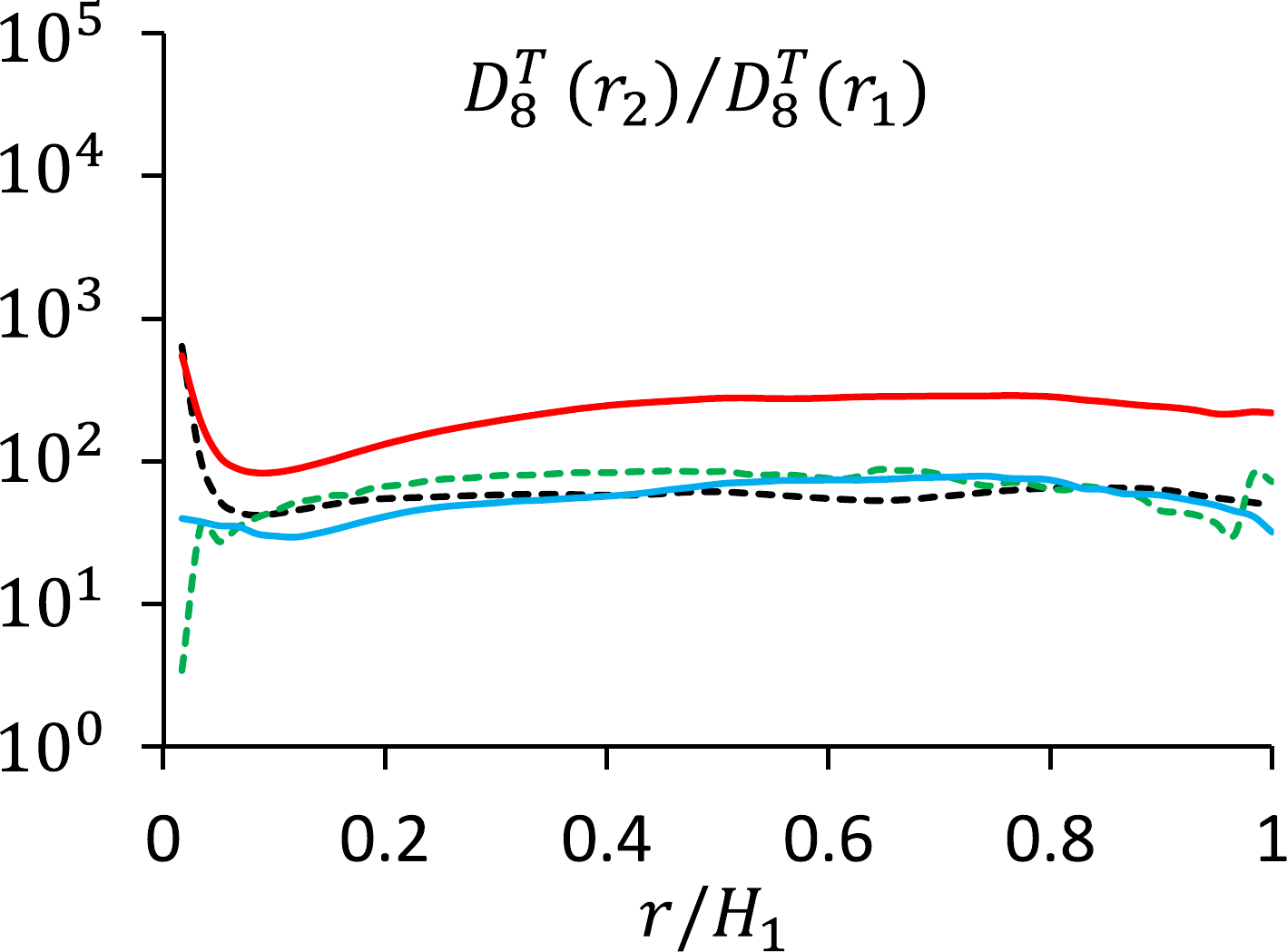}}
		\end{minipage}
		\begin{minipage}[b]{0.33\linewidth}
			\centering
			\makebox[-2.2em][l]{\raisebox{-\height}{(\textit{e})}}%
			\raisebox{-\height}{\includegraphics[height=3cm]{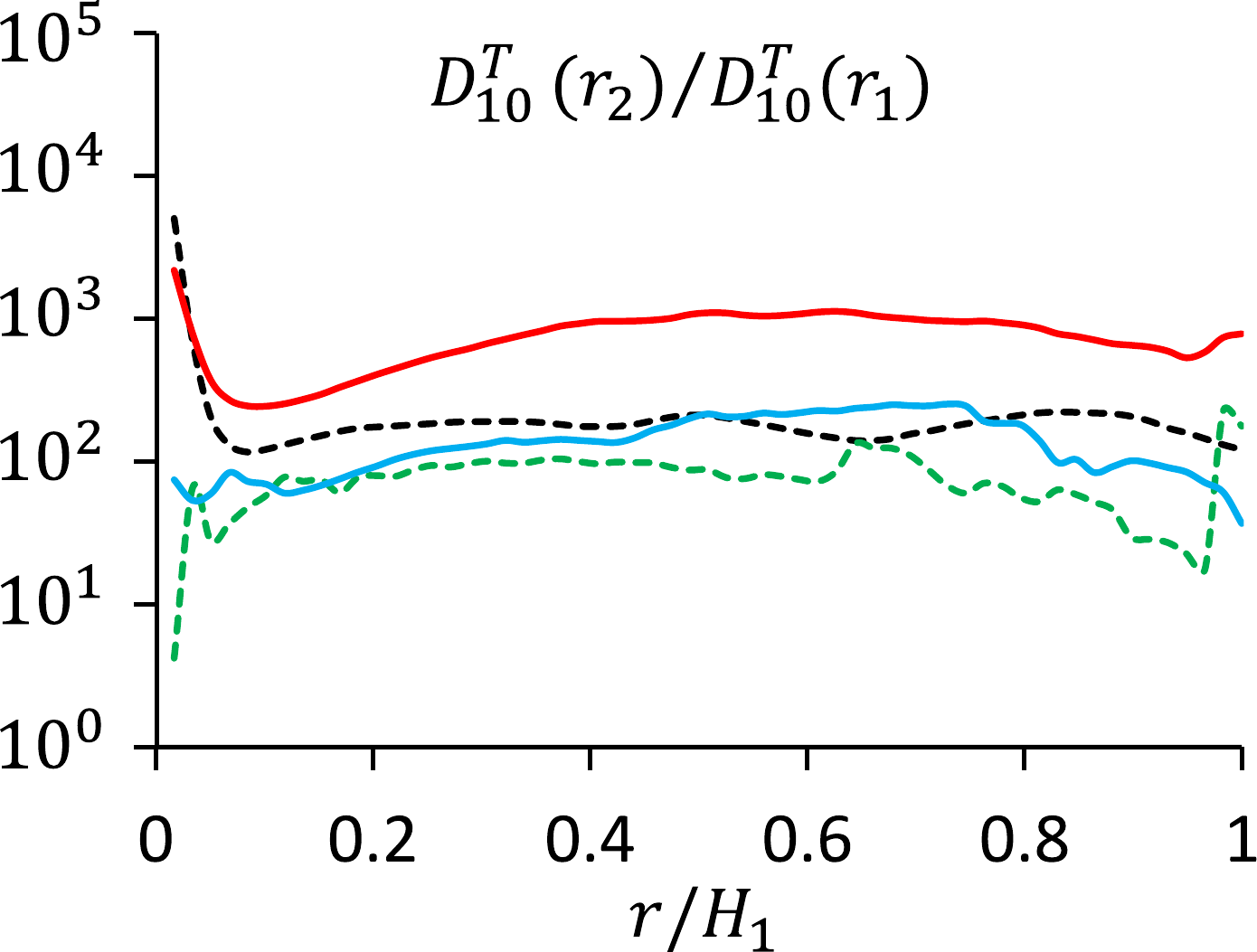}}
		\end{minipage}
		\begin{minipage}[b]{0.33\linewidth}
			\centering
			\makebox[-2.2em][l]{\raisebox{-\height}{(\textit{f})}}%
			\raisebox{-\height}{\includegraphics[height=3cm]{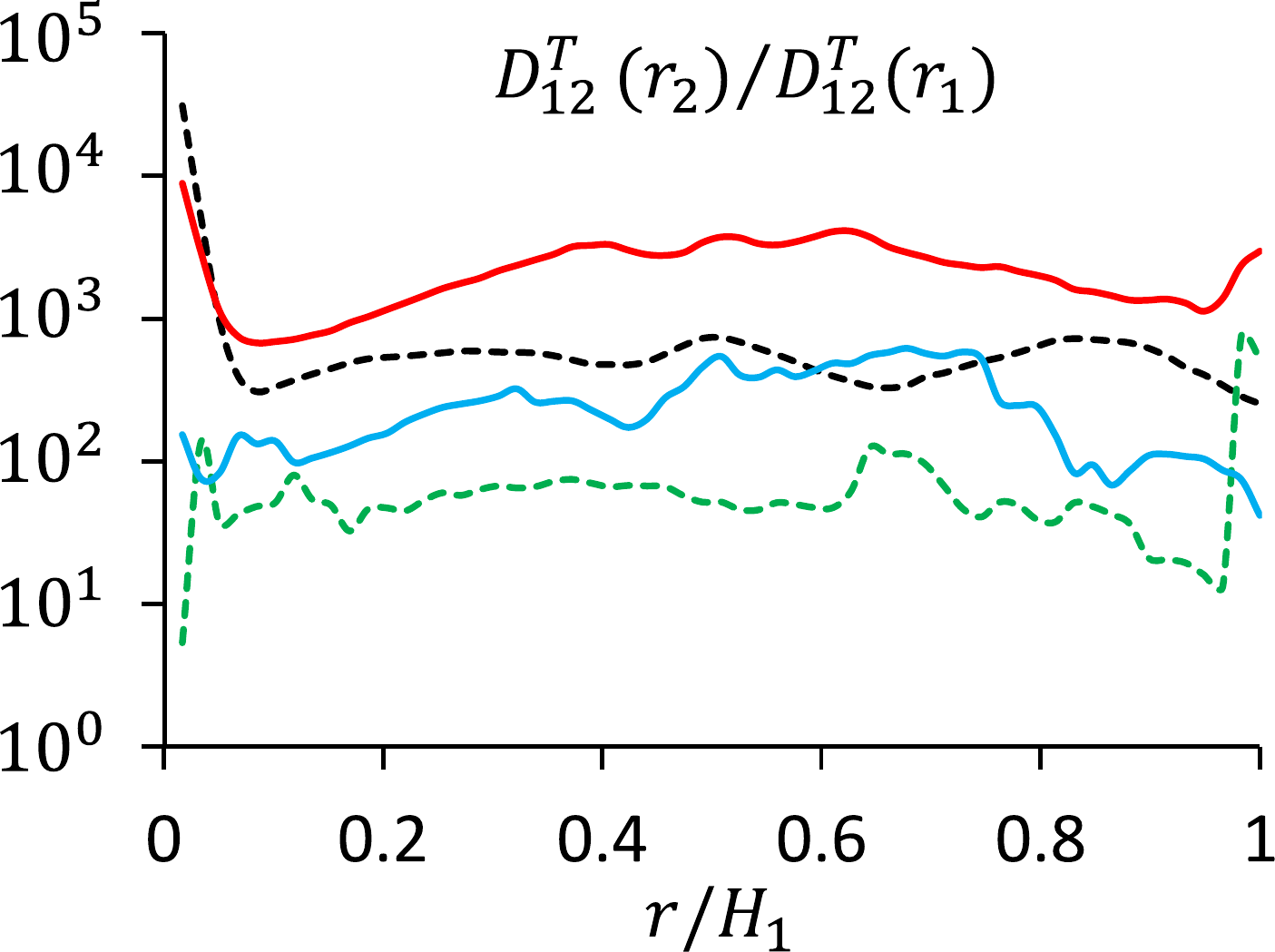}}
		\end{minipage}			
	\end{minipage}
	\caption{Ratio of $n$th-even-order transverse structure functions in different separation directions for all the cases ($n=2, 4, 6, 8, 10, 12$). Note that (\textit{a}) presents the same value as figure \ref{fig: D_LT_ratio}(\textit{b}), just with a different label of y-axis.} \label{fig: DT_ratio}
\end{figure}

Next, we consider measures of the anisotropy based on even-order structure functions up to order twelve by considering the ratios $D^L_n(r_1)/D^L_n(r_2)$ and $D^T_n(r_2)/D^T_n(r_1)$ which would be equal to unity for an isotropic flow. Using higher order structure functions allows for a characterization of how anisotropic the large fluctuations in the system are. The results in figures \ref{fig: DL_ratio} and \ref{fig: DT_ratio} show that the data is statistically well converged up to order 8. For order 10 and 12, although there is considerable noise, the general trend of the results seems clear. The results show that the ratios  $D^L_n(r_1)/D^L_n(r_2)$ and $D^T_n(r_2)/D^T_n(r_1)$ deviate increasingly strongly from unity as $n$ is increased, with the results for $n\geq 6$ reaching values very far from unity across most of the flow scales.  This shows that extreme fluctuations in the flow are more anisotropic than the ``typical'' fluctuations characterized by the $n=2$ results. Similar behavior has also been observed for single-phase turbulence, showing that higher-order moments are the most anisotropic \citep{2000_Kurien,2002_Warhaft}. For the longitudinal directions, the results show that just as for the $n=2$ results, the cases with smaller bubbles exhibit greater anisotropy that those with larger bubbles. However, for the transverse direction the \emph{LaLess} case is by far the most anisotropic. We are not aware of a good explanation for this differing dependence of the longitudinal and transverse anisotropy on the properties of the bubbles.

\section{Energy transfer}\label{sec: Energy transfer}

We now turn to consider the mean energy transfer in the flow, by analysing the third-order structure function as well as the related inter-scale energy transfer term \citep{2001_Hill,2018_Alexakis}
\begin{equation}
	\mathcal{F}(\boldsymbol{r})=\sum_{\gamma=1}^3\mathcal{F}_\gamma(\boldsymbol{r})\equiv \sum_{\gamma=1}^3 \sum_{\gamma=1}^3\frac{\partial}{\partial \xi_\gamma}D^{\gamma ii}_3(\boldsymbol{\xi})\Big\vert_{\boldsymbol{\xi}=\boldsymbol{r}} \;,\label{eq: F}
\end{equation}
that appears in the Karman-Howarth type equation governing $D^{ii}_2(\boldsymbol{r})$, where $D^{\gamma ii}_3(\boldsymbol{r})\equiv \langle\Delta {u_\gamma}(\boldsymbol{r}) \Delta u_i(\boldsymbol{r})\Delta u_i(\boldsymbol{r})\rangle$. Note that the divergence term in \eqref{eq: F} has been written using the variable $\boldsymbol{\xi}$ to emphasize that this term involves the divergence evaluated at $\boldsymbol{r}$.

The sign of $\mathcal{F}$ indicates the direction of the non-liner energy transfer between the scales of the flow, and in three-dimensional single-phase turbulence, we have $\mathcal{F}<0$ in the inertial and dissipation ranges, corresponding to a downscale transfer of energy from large to small scales on average. With our 2D PSV data we are not able to construct the full quantity $\mathcal{F}$, however, we are able to consider some of the contributions $\mathcal{F}_\gamma$.
\begin{figure}
	\begin{minipage}[b]{1.0\linewidth}
		\begin{minipage}[b]{0.5\linewidth}
			\centering
			\makebox[1.2em][l]{\raisebox{-\height}{(\textit{a})}}%
			\raisebox{-\height}{\includegraphics[height=4cm]{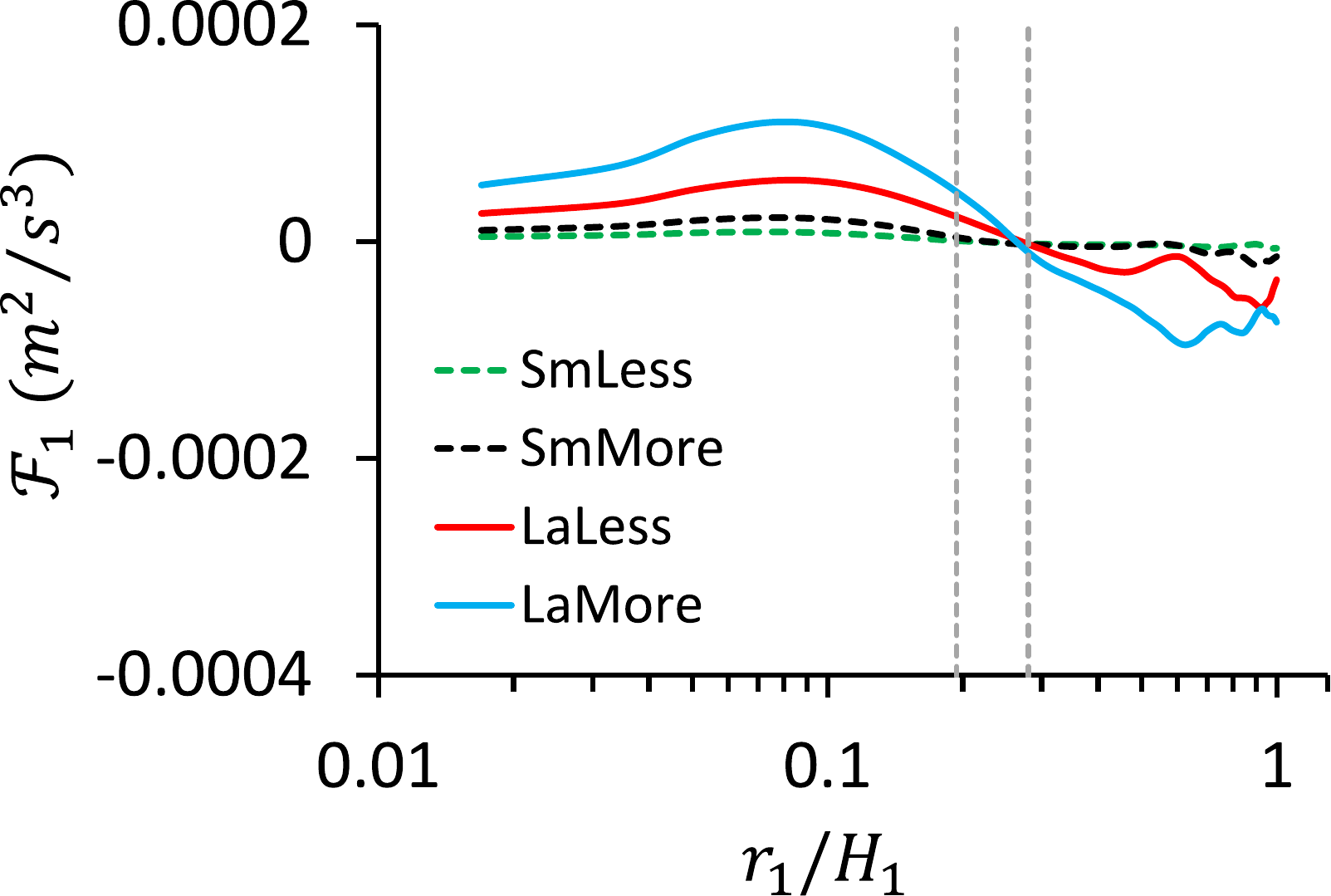}}
		\end{minipage}
		\begin{minipage}[b]{0.5\linewidth}
			\centering
			\makebox[1.2em][l]{\raisebox{-\height}{(\textit{b})}}%
			\raisebox{-\height}{\includegraphics[height=4cm]{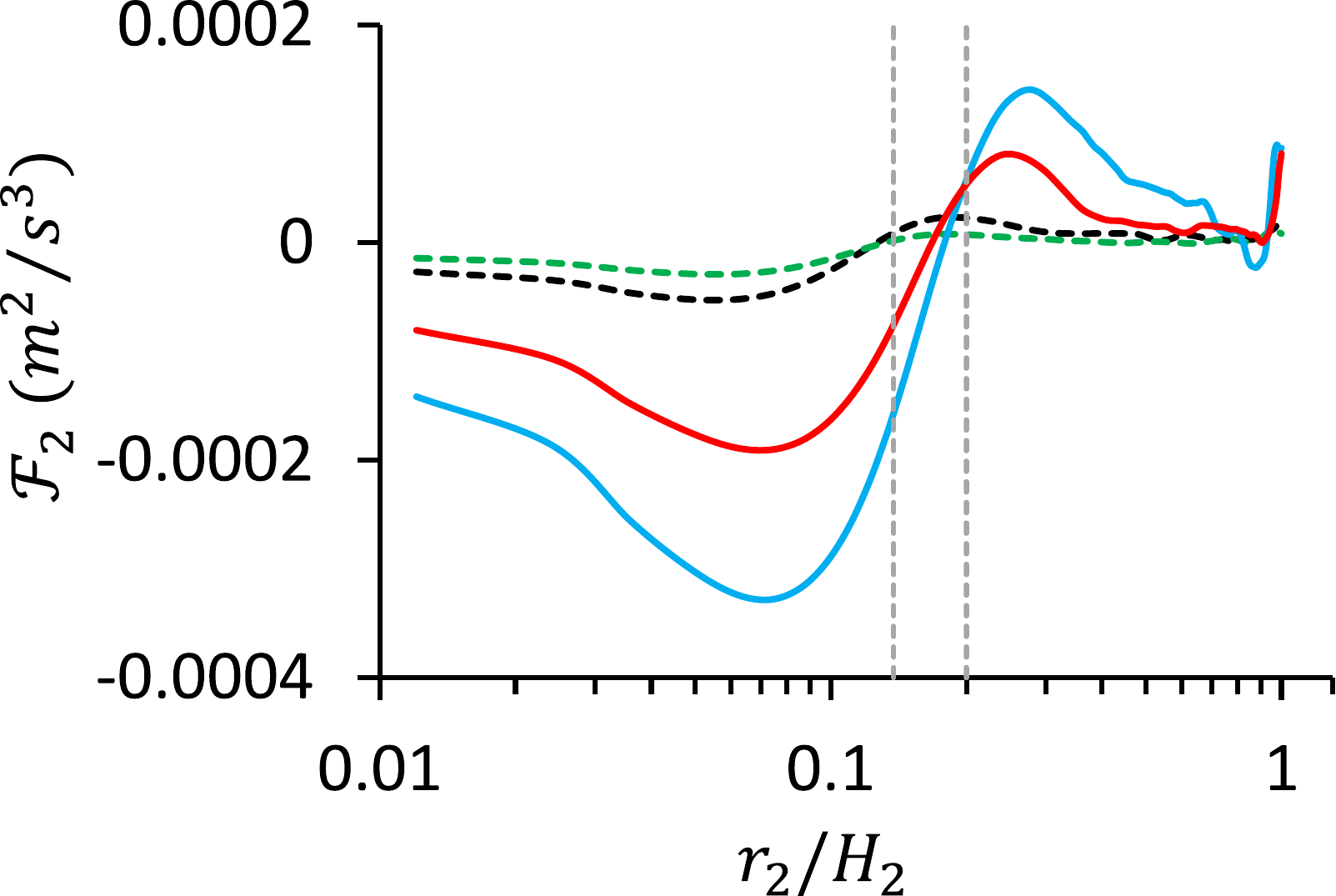}}
		\end{minipage}
	\end{minipage}
	\caption{Contribution $\mathcal{F}_1$ (\textit{a}) and $\mathcal{F}_2$  (\textit{b}) to the nonlinear energy transfer term for all the cases. In (\textit{a,b}) the two vertical dashed lines show $r=d_p$ for smaller and larger bubbles, respectively.} \label{fig: F1_F2}
\end{figure}

In figure \ref{fig: F1_F2} we plot the components $\mathcal{F}_1(r_1)=d(D^{111}_3+D^{122}_3)/dr_1$ and $\mathcal{F}_2(r_2)=d(D^{222}_3+D^{211}_3)/dr_2$. The peaks in the magnitudes of both $\mathcal{F}_1$ and $\mathcal{F}_2$ increase in the sequence \textit{SmLess, SmMore, LaLess}, and \textit{LaMore}, which corresponds to higher $Re_{H_2}$, $\alpha_p$, and/or $Re_p$. This is in part simply due to the fact that the kinetic energy in the flow increases in this sequence also, and with more kinetic energy at the large scales, there is more available to be passed downscale. Concerning $\mathcal{F}_1$, the results show that at scales $r>O(d_p)$, $\mathcal{F}_1<0$ corresponding to energy moving downscale, while at scales $r<O(d_p)$, $\mathcal{F}_1>0$, corresponding to energy moving upscale. In contrast to $\mathcal{F}_1$, the contribution $\mathcal{F}_2$ shows an opposite trend, with $\mathcal{F}_2<0$ for at smaller scales indicating a downscale energy transfer. The magnitude of $\mathcal{F}_2$ is, however, in general much larger than $\mathcal{F}_1$, implying a higher energy transfer rate in the horizontal direction of the flow. At the low Reynolds number of the flow, there is no inertial range over which either $\mathcal{F}_1$ or $\mathcal{F}_2$ are constant, indicating that the nonlinear energy transfer is not in the form of a constant flux cascade.

Our previous study \citep{2021_Ma} explored bubble-laden turbulent channel flows and also observed that there is a downscale energy transfer for $\mathcal{F}_3(r_3)$ (where in that paper the 3 direction corresponded to the spanwise direction of the channel flow). However, in that study we could not compute either $\mathcal{F}_1$ (streamwise) or $\mathcal{F}_2$ (wall-normal) due to the one-dimensional data set being used. Although the type of flow presently under consideration is considerably different from a pressure driven bubble-laden turbulent channel flow, the results nevertheless indicate that the nonlinear energy transfer behaviour may be very different in the different directions of the flow.

These results therefore show that the direction of the energy flux can differ from one direction to another, with $\mathcal{F}_1$ revealing an upscale energy transfer at smaller scales, while $\mathcal{F}_2$ reveals the opposite. Therefore, although the total energy transfer may be downscale in bubble-turbulent flows \citep{2020_Pandey}, some directions of the flow can exhibit an upscale transfer.
\begin{figure}
	\begin{minipage}[b]{1.0\linewidth}
		\begin{minipage}[b]{0.5\linewidth}
			\centering
			\makebox[1.5em][l]{\raisebox{-\height}{(\textit{a})}}%
			\raisebox{-\height}{\includegraphics[height=3.9cm]{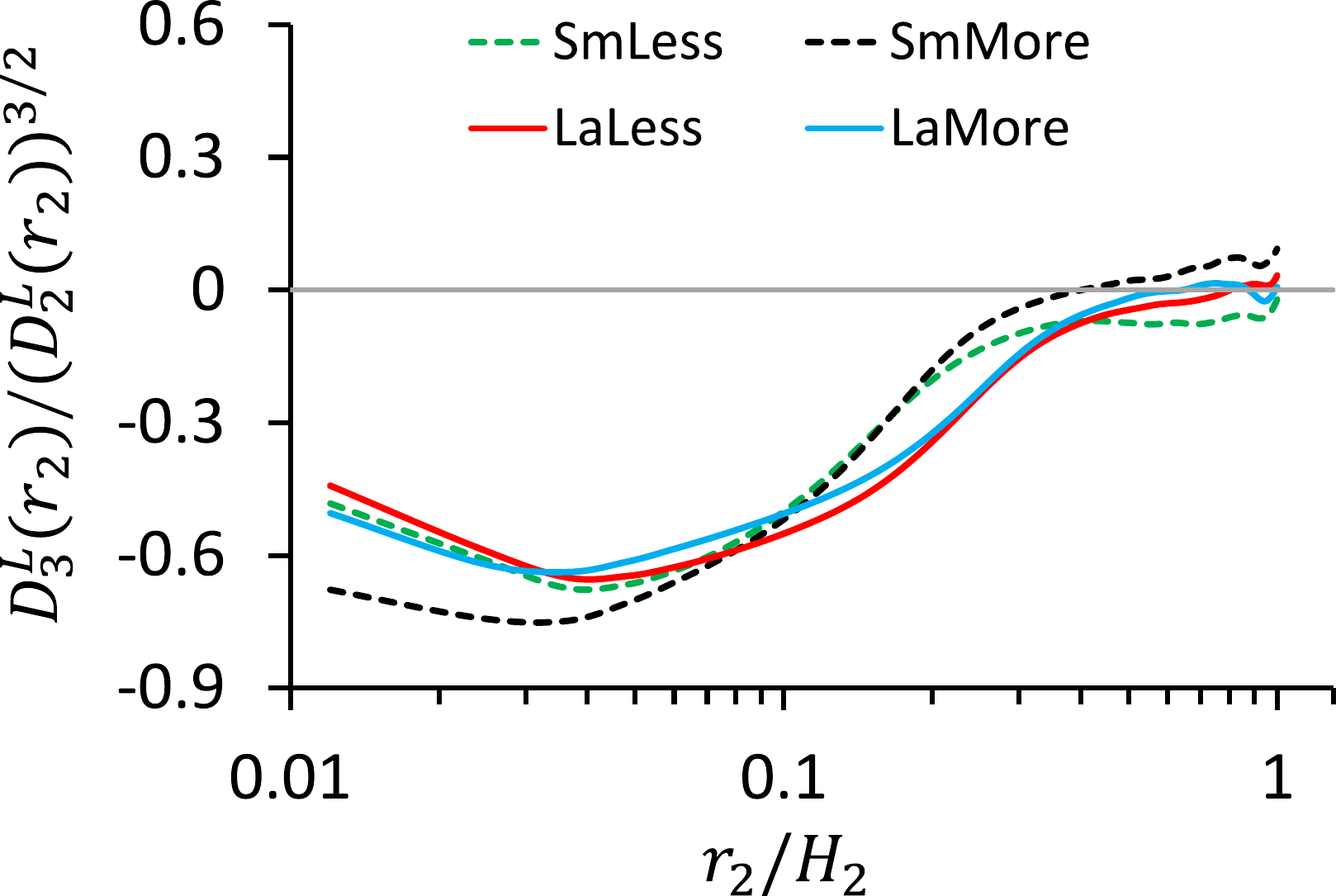}}
		\end{minipage}
		\begin{minipage}[b]{0.5\linewidth}
			\centering
			\makebox[1.5em][l]{\raisebox{-\height}{(\textit{b})}}%
			\raisebox{-\height}{\includegraphics[height=3.9cm]{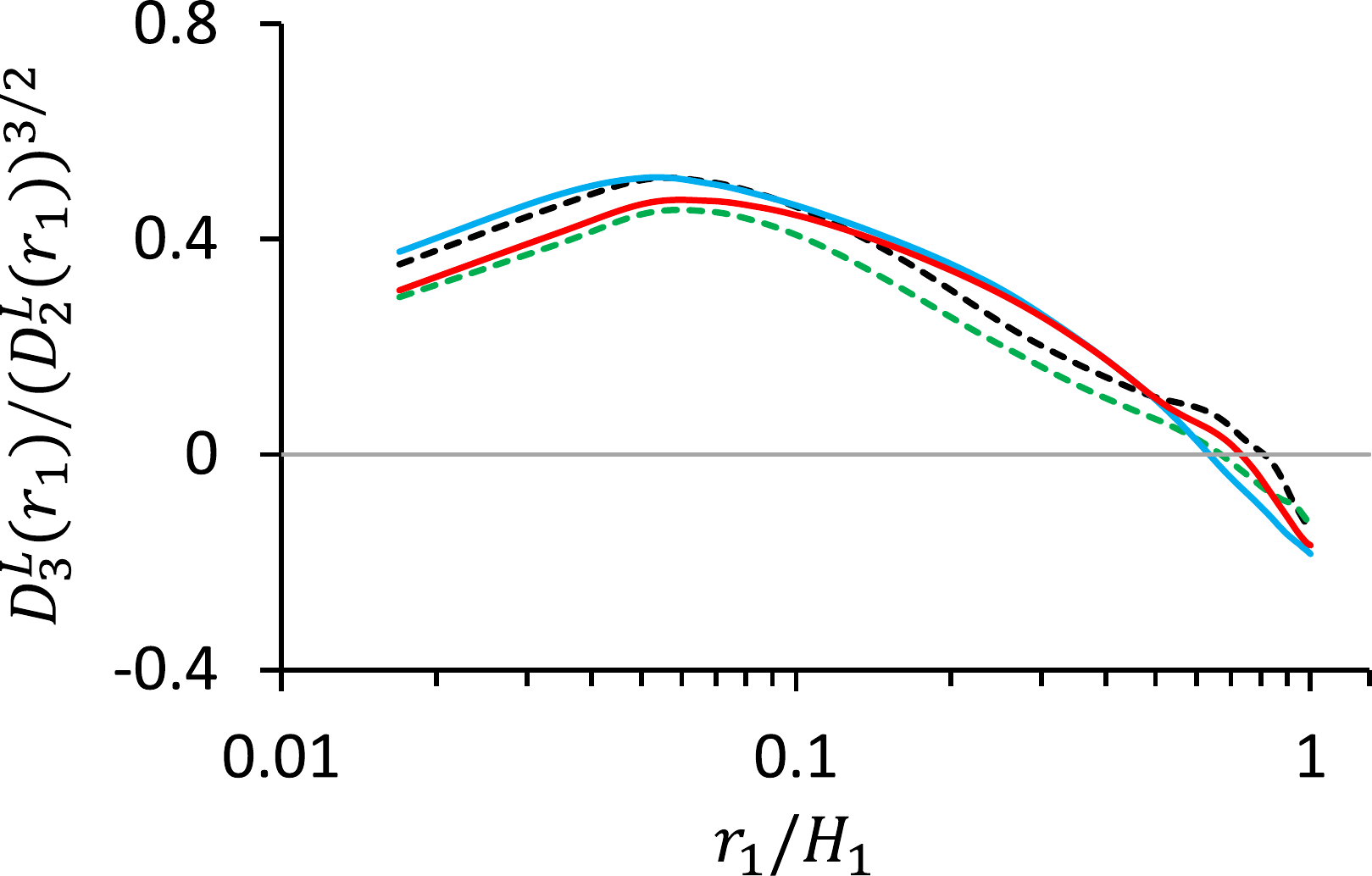}}
		\end{minipage}
	\end{minipage}
\caption{Normalized third-order longitudinal structure functions (skewness), with separations along the horizontal (\textit{a}) and the vertical (\textit{b}) directions. The horizontal lines in (\textit{a,b}) indicate the value of zero.} \label{fig: DLLL}
\end{figure}

The energy transfer term $\mathcal{F}$ depends on the third-order structure function, and its behavior may be further understood by considering the associated skewness of the velocity increments. Figure \ref{fig: DLLL} shows $D^L_3/(D^L_2)^{3/2}$ (the skewness of the longitudinal velocity increment) for the two separations $r_1$ and $r_2$. The results show that when $D^L_3(r_2)$ is plotted in this normalized form, the results for all of the cases almost collapse, and $D^L_3(r_2)/(D^L_2(r_2))^{3/2}$ is not far from the value $-0.4$ that occurs for single-phase isotropic turbulence \citep{2015_Davidson} for $r\rightarrow0$. The negative longitudinal skewness along the separation perpendicular to the bubble motion direction was also found in \cite{2021_Ma} for some of the bubble-laden cases considered, however for other cases this skewness was actually positive. This is very different from our results where all the cases considered have very similar negative values of the skewness. It is not clear as to whether this is due to the different kinds of flows being studied here and in \cite{2021_Ma}, or perhaps also because of differences in the parameter space of the bubbles being explored. For vertical separations $r_1$, $D^L_3(r_1)/(D^L_2(r_1))^{3/2}$ also almost collapses for all the cases but is positive at the small scales with a similar magnitude $\approx0.4$ to the results for $D^L_3(r_2)/(D^L_2(r_2))^{3/2}$. This again indicates strong anisotropy for the odd-order structure functions, just as for the even-order structure functions discussed in \S\,\ref{sec: Anisotropy}. 
  
Similar plots are obtained for the transverse structure functions $D^T_3(r_2)$ and $D^T_3(r_1)$. The corresponding skewness values were found to be close to zero in both separations directions for all the cases (not shown here).

\section{Extreme fluctuations in the flow}\label{sec: Extreme velocity increment}

Having considered the flow anisotropy and nonlinear energy transfer, we now turn to consider the extreme fluctuations in the flow. Single-phase turbulence at high Reynolds number exhibits fluctuations in space and time of its small-scale quantities that are orders of magnitude larger than the average values, a phenomenon referred to as small-scale intermittency \citep{1995_Frisch,1997_Sreenivasan}. Such extreme fluctuations are critical for many processes in both nature and engineering \citep{2021_Sapsis}, and their investigation continues to be an area of active research \citep{2015_Yeung,2019_Buaria}. However, very few studies have explored the topic of intermittency in bubble-laden turbulent flows (see the related works introduced in \S\,\ref{sec: introduction}). In what follows, we focus on extreme fluctuations of the velocity increments $\Delta\boldsymbol{u}$ in the bubble-laden flow we are considering.

\subsection{Probability density functions}\label{subsec: pdf}

\begin{figure}
	\begin{minipage}[b]{1.0\linewidth}
		\begin{minipage}[b]{0.5\linewidth}
			\centering
			\makebox[1em][l]{\raisebox{-\height}{(\textit{a})}}%
			\raisebox{-\height}{\includegraphics[height=4cm]{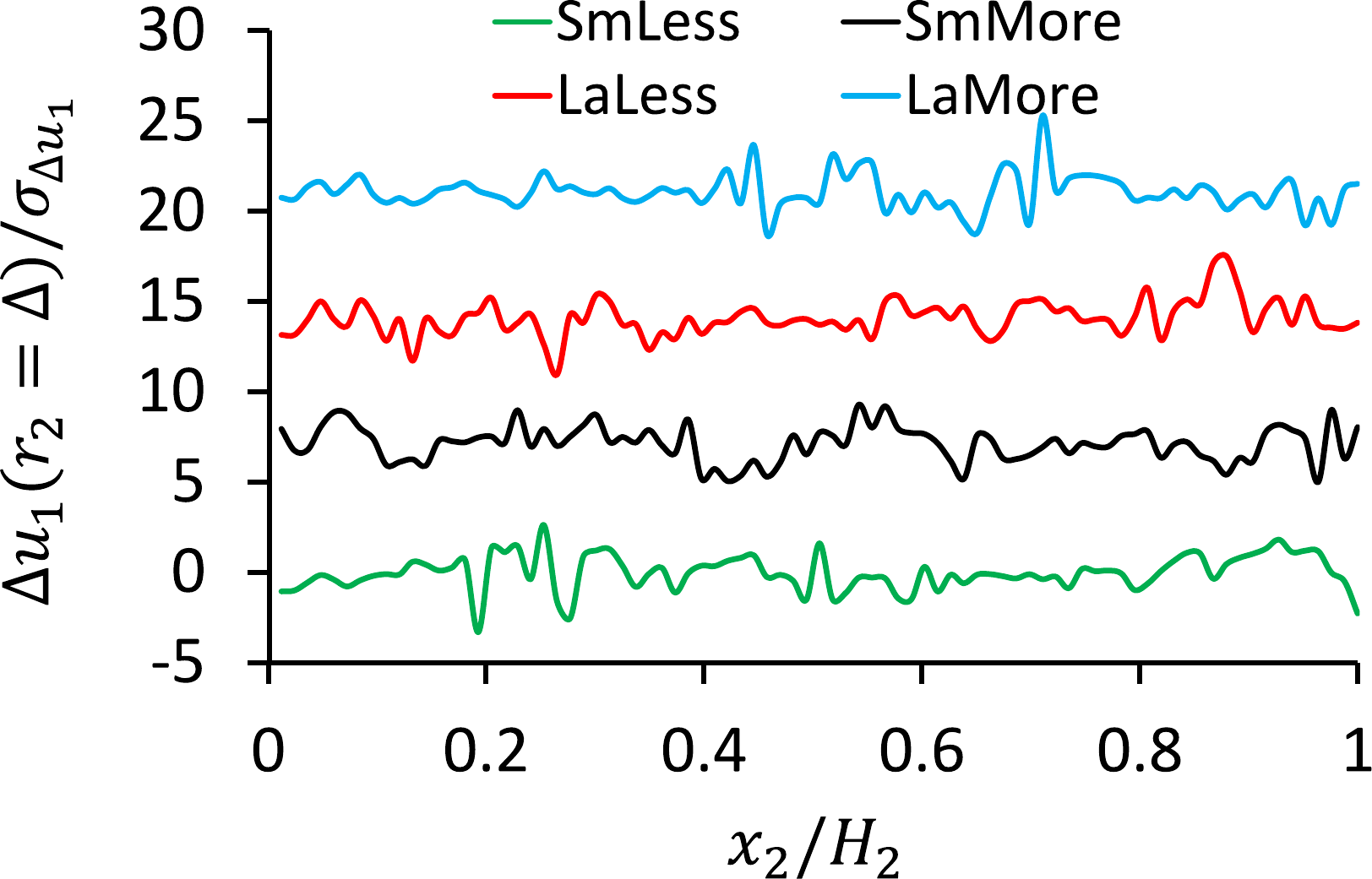}}
		\end{minipage}
		\begin{minipage}[b]{0.5\linewidth}
			\centering
			\makebox[1em][l]{\raisebox{-\height}{(\textit{b})}}%
			\raisebox{-\height}{\includegraphics[height=4cm]{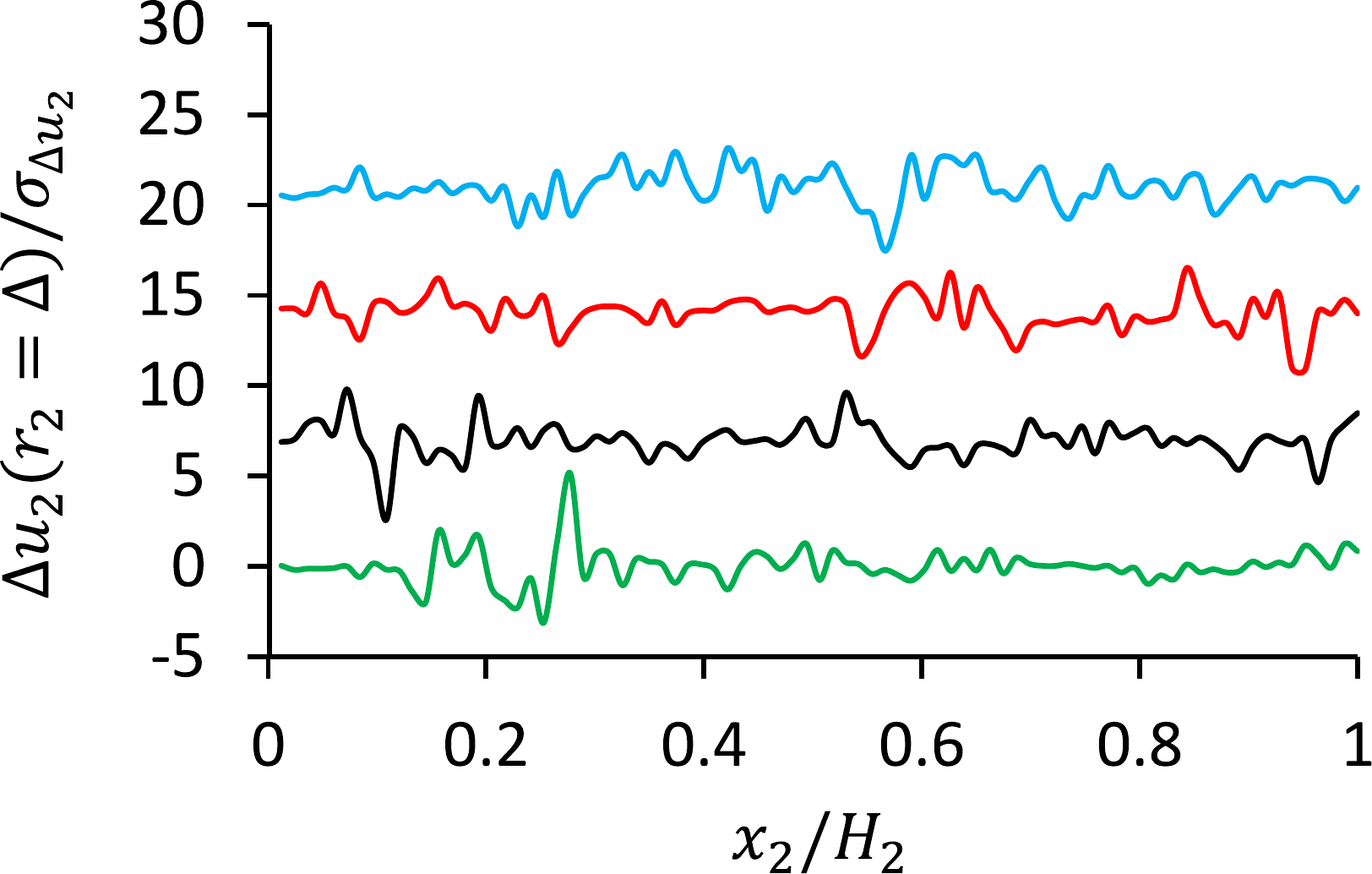}}
		\end{minipage}
	\end{minipage}
\caption{Instantaneous velocity increments normalized by standard deviations: (\textit{a}) $\Delta u_1(r_2=\Delta)/\sigma_{\Delta u_1}$ and (\textit{b}) $\Delta u_2(r_2=\Delta)/\sigma_{\Delta u_2}$, along an arbitrary horizontal line for an instant, when the line is free of bubbles for each considered case. In (\textit{a,b}) the curves for \textit{SmMore}, \textit{LaLess}, and \textit{LaMore} are shifted upward by $7, 14$, and $21$, respectively, for a better visual representation.} \label{fig: transient_delta_u}
\end{figure}

Figure \ref{fig: transient_delta_u} shows the instantaneous velocity increment $\Delta u_i(r_2=\Delta)$, normalized by their standard deviation $\sigma_{\Delta u_i}$ along a line (without bubbles) in the direction of $x_2$ within the FOV. The plots reveal intermittent fluctuations of $\Delta u_i(r_2=\Delta)$, with relatively small regions where $\Delta u_i(r_2=\Delta)/\sigma_{\Delta u_i}$ takes on large values, between which there are relatively small fluctuations of $\Delta u_i(r_2=\Delta)/\sigma_{\Delta u_i}$. We will return later to consider how these intermittent large fluctuations of $\Delta u_i(r_2=\Delta)/\sigma_{\Delta u_i}$ are associated with the bubble wakes in the flow.

\begin{figure}
	\begin{minipage}[b]{1.0\linewidth}
		\begin{minipage}[b]{0.5\linewidth}
			\centering
			\makebox[0.5em][l]{\raisebox{-\height}{(\textit{a})}}%
			\raisebox{-\height}{\includegraphics[height=4cm]{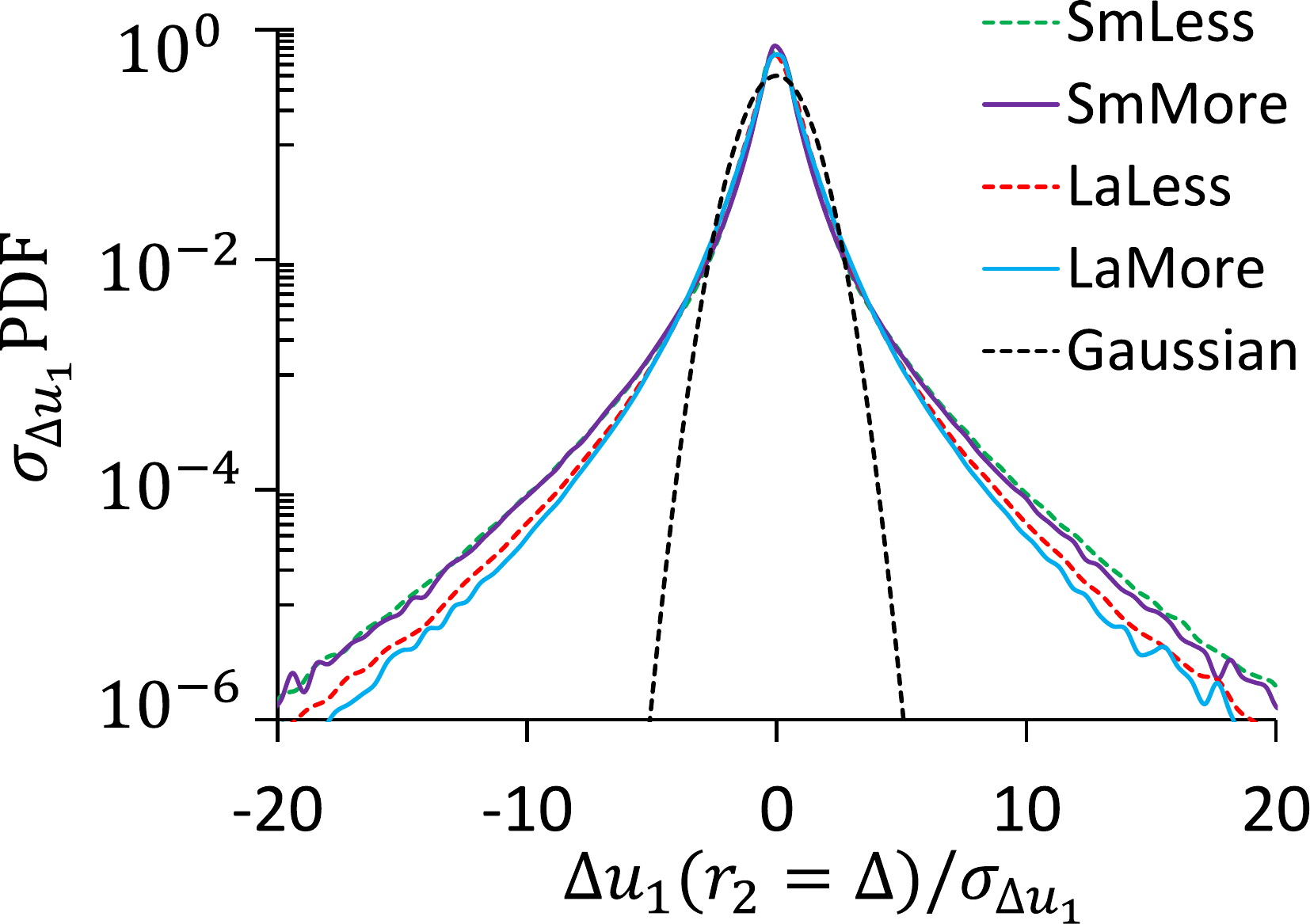}}
		\end{minipage}
		\begin{minipage}[b]{0.5\linewidth}
			\centering
			\makebox[0.5em][l]{\raisebox{-\height}{(\textit{b})}}%
			\raisebox{-\height}{\includegraphics[height=4cm]{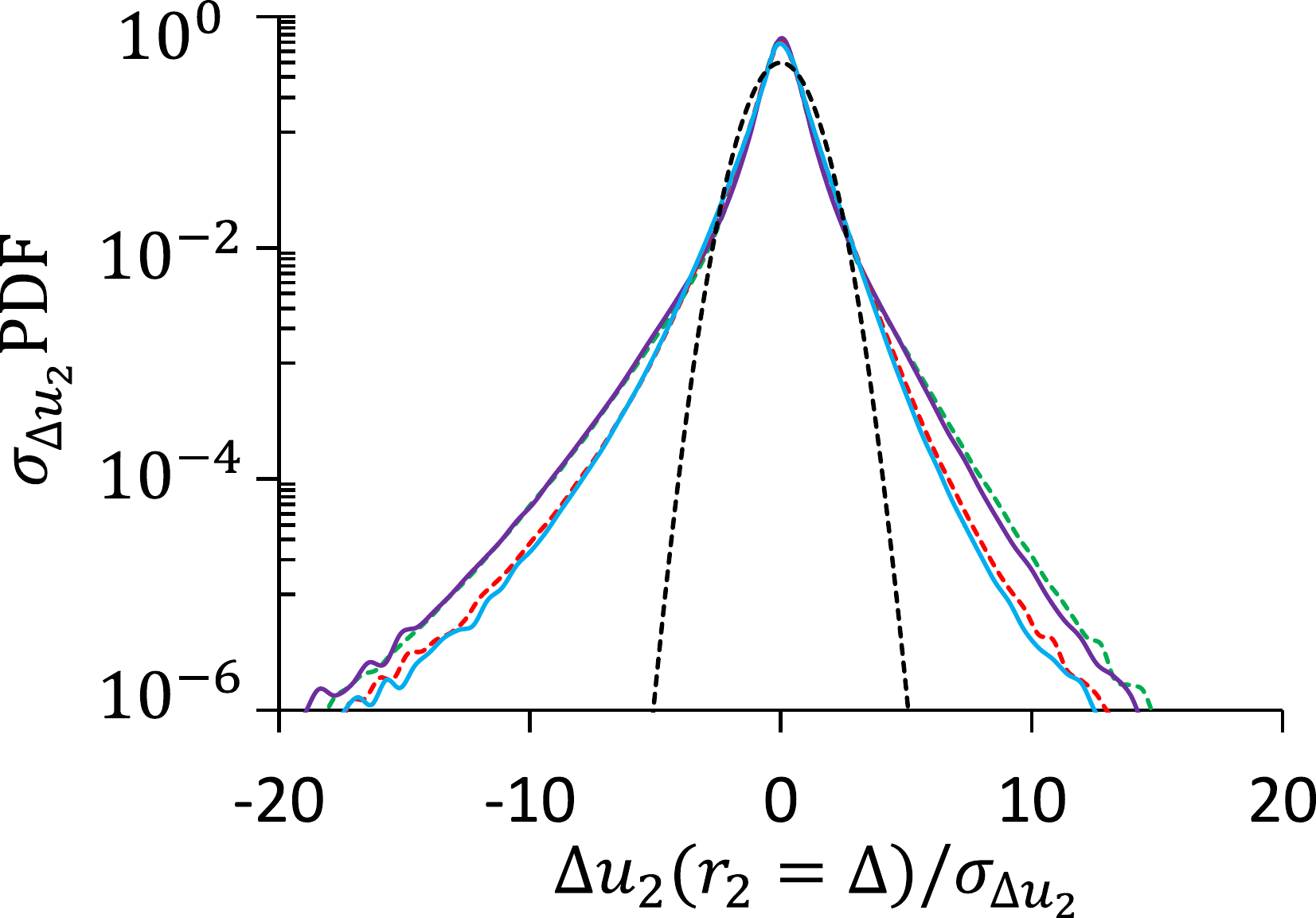}}
		\end{minipage}
	\end{minipage}
	\begin{minipage}[b]{1.0\linewidth}
		\vspace{3mm}
		\begin{minipage}[b]{0.5\linewidth}
			\centering
			\makebox[0.5em][l]{\raisebox{-\height}{(\textit{c})}}%
			\raisebox{-\height}{\includegraphics[height=4cm]{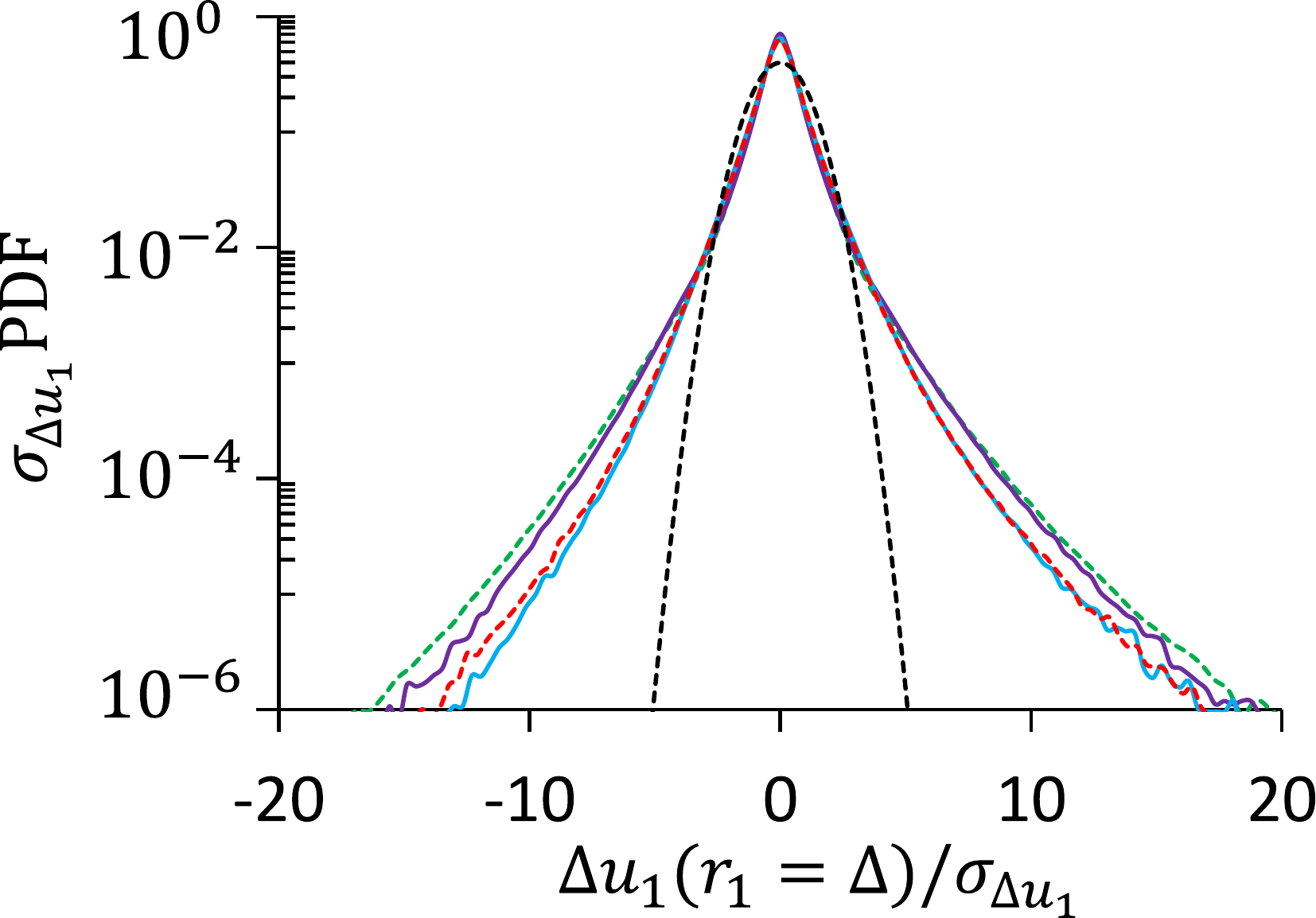}}
		\end{minipage}
		\begin{minipage}[b]{0.5\linewidth}
			\centering
			\makebox[0.5em][l]{\raisebox{-\height}{(\textit{d})}}%
			\raisebox{-\height}{\includegraphics[height=4cm]{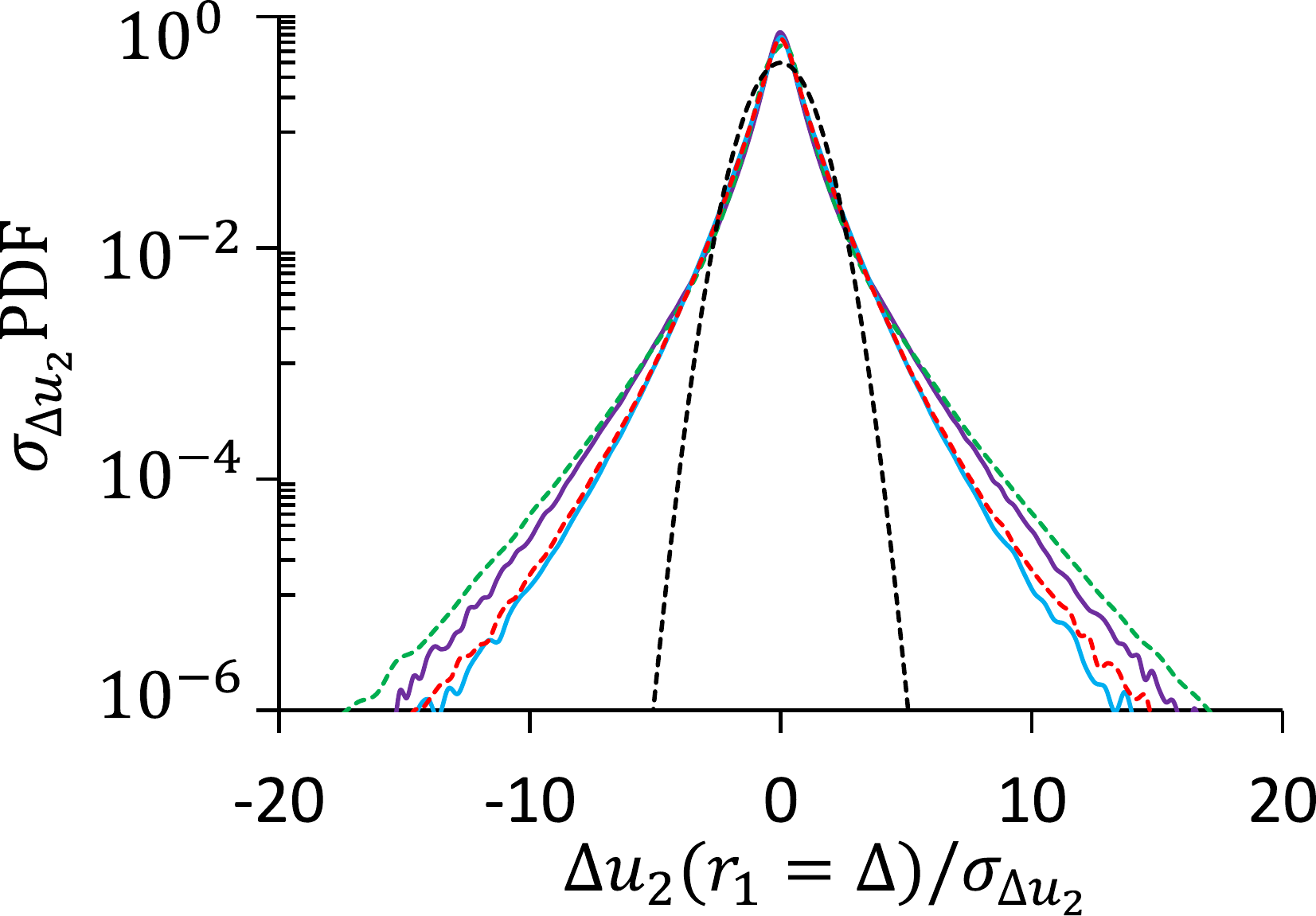}}
		\end{minipage}
	\end{minipage}
\caption{Normalized probability density functions of the transverse (\textit{a,d}) and longitudinal (\textit{b,c}) velocity increments, with the separation $r=\Delta$ along the horizontal (\textit{a,b}) and the vertical (\textit{c,d}) directions for all four cases.} \label{fig: pdf 4cases}
\end{figure}

To statistically quantify the extreme events in the flow, the PDFs of the velocity increments can be considered, and these are shown in figure \ref{fig: pdf 4cases}, normalized by the standard deviation, for separations equal to one PSV grid $(r=\Delta)$. All of the results in figure \ref{fig: pdf 4cases} show that the velocity increments have strongly non-Gaussian PDFs, consistent with the observations based on figure \ref{fig: transient_delta_u}. While the PDFs of the transverse velocity increments in plots (\textit{a,d}) are almost symmetric, the PDFs of the longitudinal velocity increment are negatively skewed for horizontal separations $r_2$, plot (\textit{b}), and positively skewed for vertical separations $r_1$, plot (\textit{c}), consistent with the skewness results in figure \ref{fig: DLLL}. The results also show that the PDFs become increasingly non-Gaussian in the order of \textit{LaMore, LaLess, SmMore, SmLess}, which corresponds to the order of decreasing Reynolds number. Therefore, surprisingly, while non-Gaussianity of the PDFs of the velocity increments becomes stronger as Reynolds number is increased in single-phase turbulence \citep{1995_Frisch}, the opposite occurs for the bubble-laden turbulent flow considered here. 

\begin{figure}
	\begin{minipage}[b]{1.0\linewidth}
		\begin{minipage}[b]{0.5\linewidth}
			\centering
			\makebox[0.5em][l]{\raisebox{-\height}{(\textit{a})}}%
			\raisebox{-\height}{\includegraphics[height=4cm]{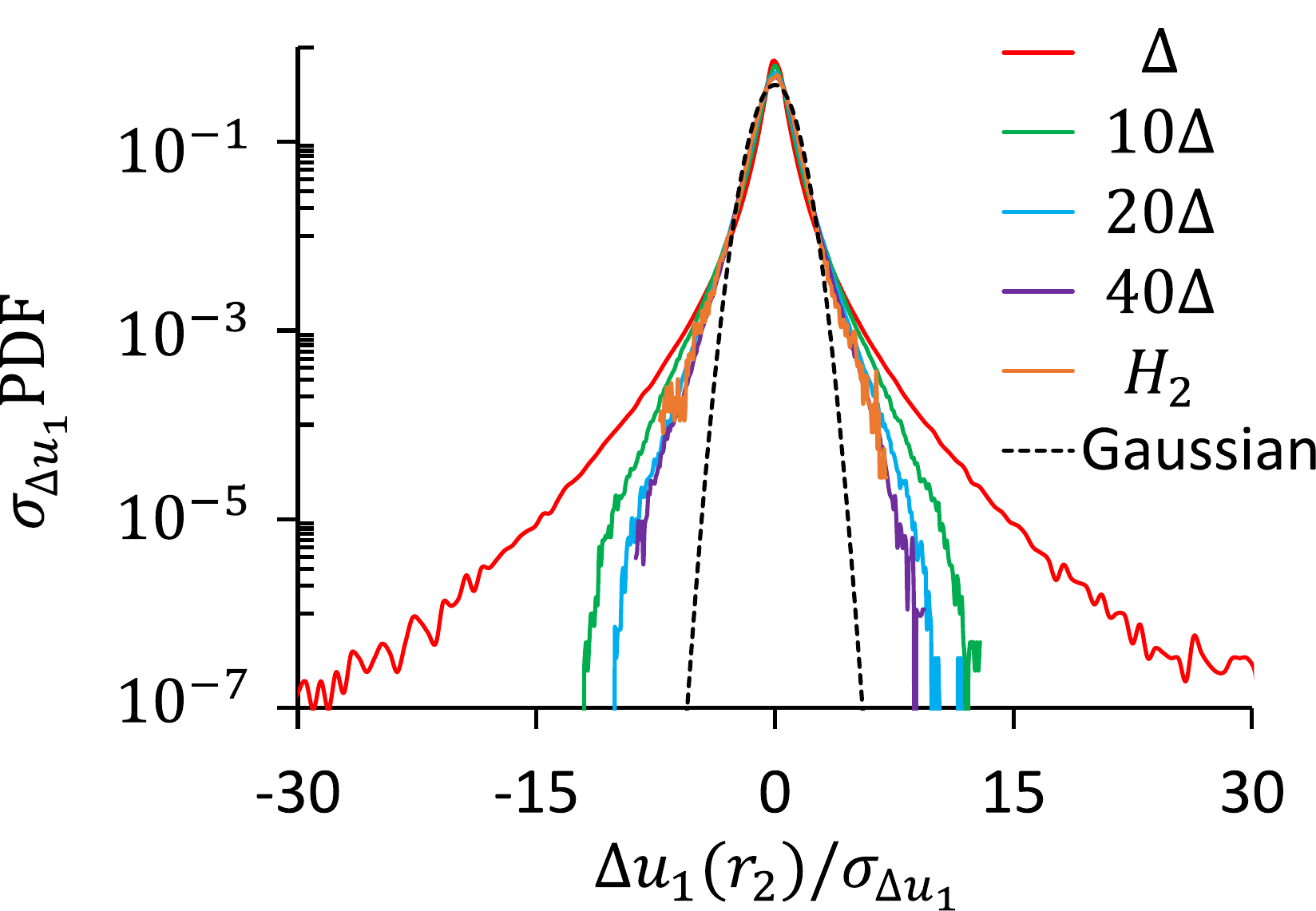}}
		\end{minipage}
		\begin{minipage}[b]{0.5\linewidth}
			\centering
			\makebox[0.5em][l]{\raisebox{-\height}{(\textit{b})}}%
			\raisebox{-\height}{\includegraphics[height=4cm]{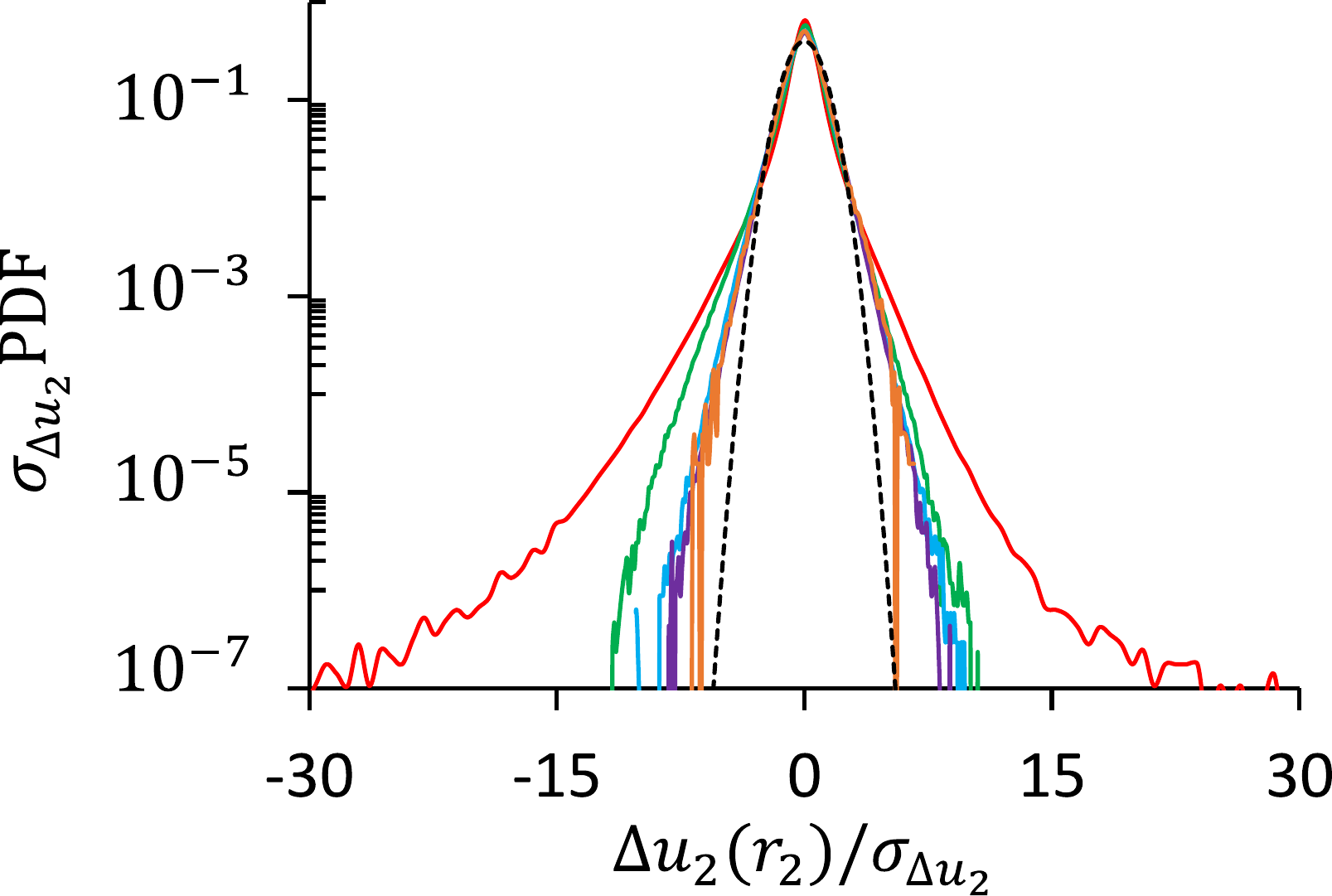}}
		\end{minipage}
	\end{minipage}
	\begin{minipage}[b]{1.0\linewidth}
		\vspace{3mm}
		\begin{minipage}[b]{0.5\linewidth}
			\centering
			\makebox[0.5em][l]{\raisebox{-\height}{(\textit{c})}}%
			\raisebox{-\height}{\includegraphics[height=4cm]{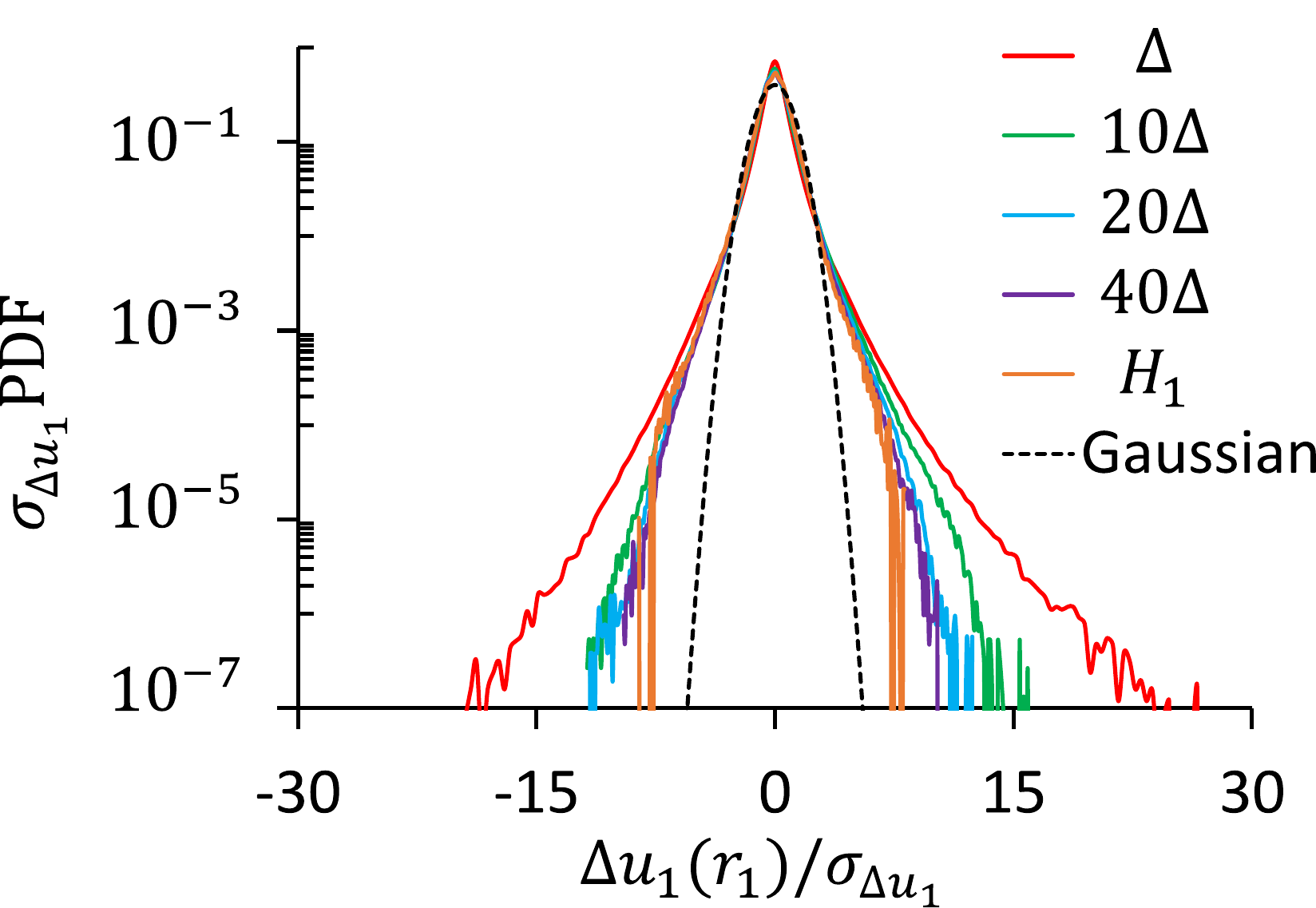}}
		\end{minipage}
		\begin{minipage}[b]{0.5\linewidth}
			\centering
			\makebox[0.5em][l]{\raisebox{-\height}{(\textit{d})}}%
			\raisebox{-\height}{\includegraphics[height=4cm]{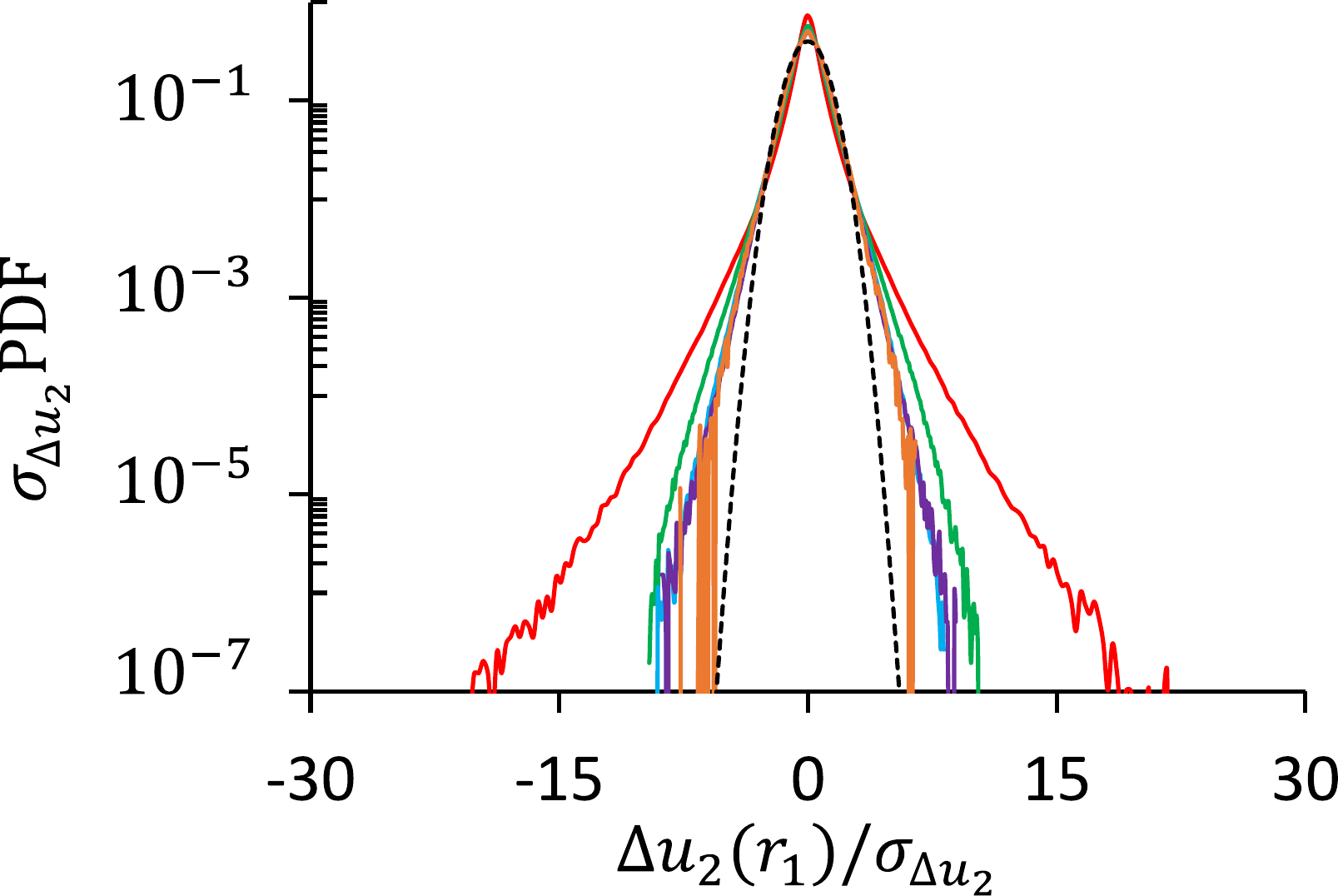}}
		\end{minipage}
	\end{minipage}
	\caption{Normalized probability density functions of the transverse (\textit{a,d}) and longitudinal (\textit{b,c}) velocity increments in the case \textit{SmMore} for various separation distances $r_i=\Delta, 10\Delta,20\Delta, 40\Delta, H_i$, where $\Delta$ is one PSV grid. (\textit{a,b}) are separations along the horizontal direction and (\textit{c,d}) are separations along the vertical  direction.} \label{fig: pdf SmMore}
\end{figure}

In turbulent flows, intermittency is characterized not simply by non-Gaussianity of the small-scale flow quantities, but also by the non-Gaussianity depending on scale (as well as Reynolds number) \citep{1995_Frisch}. To consider intermittency in the flow we therefore consider how the normalized PDFs of the velocity increments change with scale. Figure \ref{fig: pdf SmMore} shows the PDFs of the transverse (\textit{a,d}) and longitudinal (\textit{b,c}) velocity increments for various separation distances $r_\gamma=\Delta, 10\Delta,20\Delta, 40\Delta, H_\gamma$ and for the representative case \textit{SmMore}. The general behavior is very similar to that observed for single-phase turbulence \citep{2009_Ishihara}: as the scale is decreased, the PDFs of both the transverse (\textit{a,d}) and longitudinal (\textit{b,c}) velocity increments deviate more and more from a Gaussian distribution with the tails becoming increasingly heavy, and the PDFs becoming increasingly peaked. For the PDFs of longitudinal velocity increments are negatively skewed for $\Delta u_2(r_2)$ and positively skewed for $\Delta u_1(r_1)$, but the skewness becomes progressively weaker as the scale increases, and for the largest scale captured in the FOV, i.e. $r_\gamma=H_\gamma$, we find almost symmetric PDFs. 

We note that while the PDFs of the velocity increments should asymptote to the single-point PDFs of the velocity in the limit $r\rightarrow\infty$, our data for the PDF of $\Delta u_1(r_\gamma=H_\gamma)$ is very different (compare the left tails of the PDFs) from single-point PDF of $u_1$ shown in figure \ref{fig: pdf u and v}(\textit{a}). This is probably simply due to the fact that, as already noted earlier the size of the FOV is not large enough for this asymptotic regime to be observed. The aforementioned PDFs are however so different that the flow behavior must undergo an interesting transition where negative fluctuations of the vertical velocity begin to be suppressed (but not positive fluctuations) as the scale continues to be increased. This point will be explored in a future work by employing a larger FOV to observe this asymptotic regime.

\subsection{Flatness of velocity increment}\label{subsec: flatness}

\begin{figure}
	\begin{minipage}[b]{1.0\linewidth}
		\begin{minipage}[b]{0.5\linewidth}
			\centering
			\makebox[1.2em][l]{\raisebox{-\height}{(\textit{a})}}%
			\raisebox{-\height}{\includegraphics[height=3.6cm]{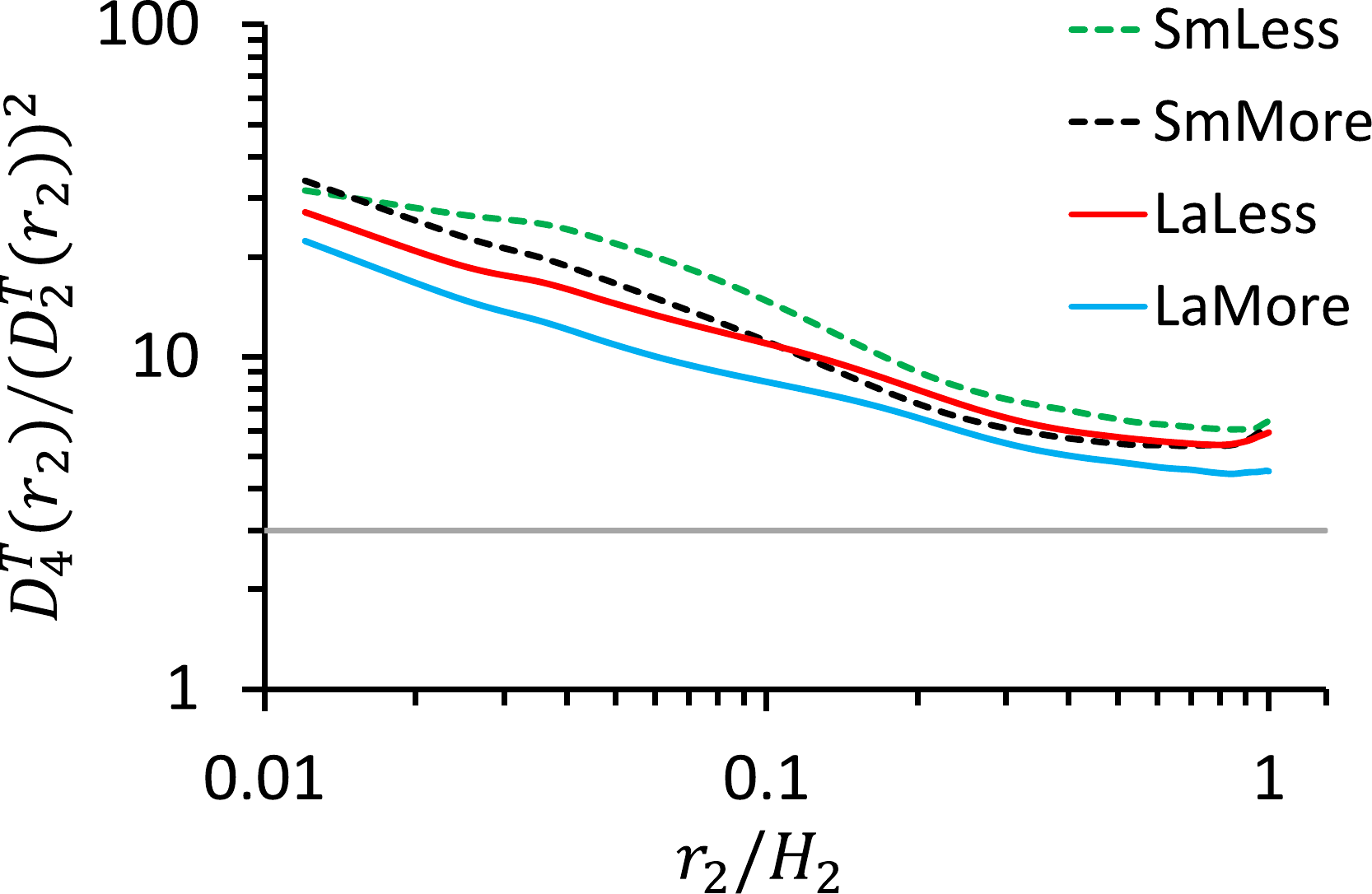}}
		\end{minipage}
		\begin{minipage}[b]{0.5\linewidth}
			\centering
			\makebox[1.2em][l]{\raisebox{-\height}{(\textit{b})}}%
			\raisebox{-\height}{\includegraphics[height=3.6cm]{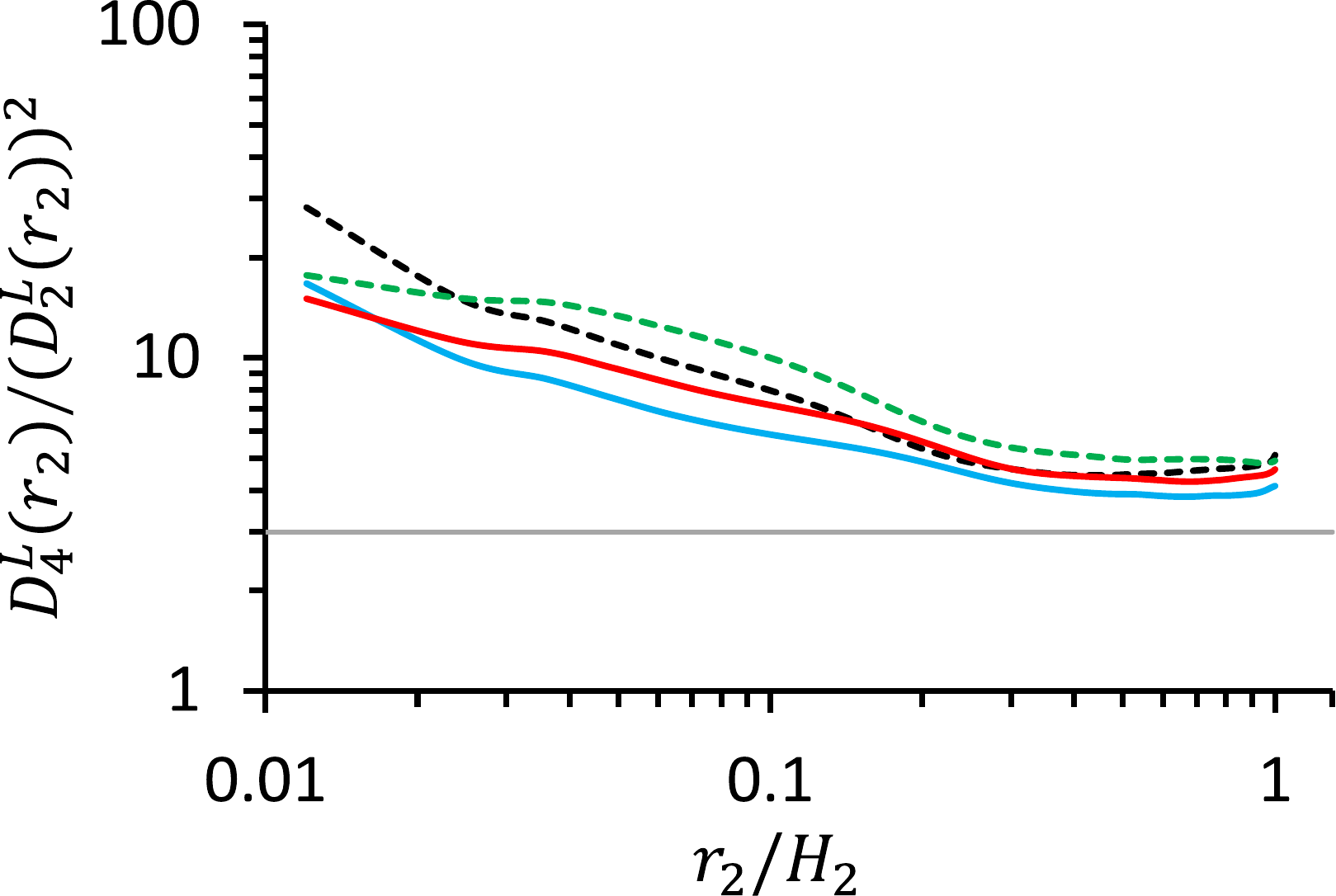}}
		\end{minipage}
	\end{minipage}
	\begin{minipage}[b]{1.0\linewidth}
		\vspace{3mm}
		\begin{minipage}[b]{0.5\linewidth}
			\centering
			\makebox[1.2em][l]{\raisebox{-\height}{(\textit{c})}}%
			\raisebox{-\height}{\includegraphics[height=3.6cm]{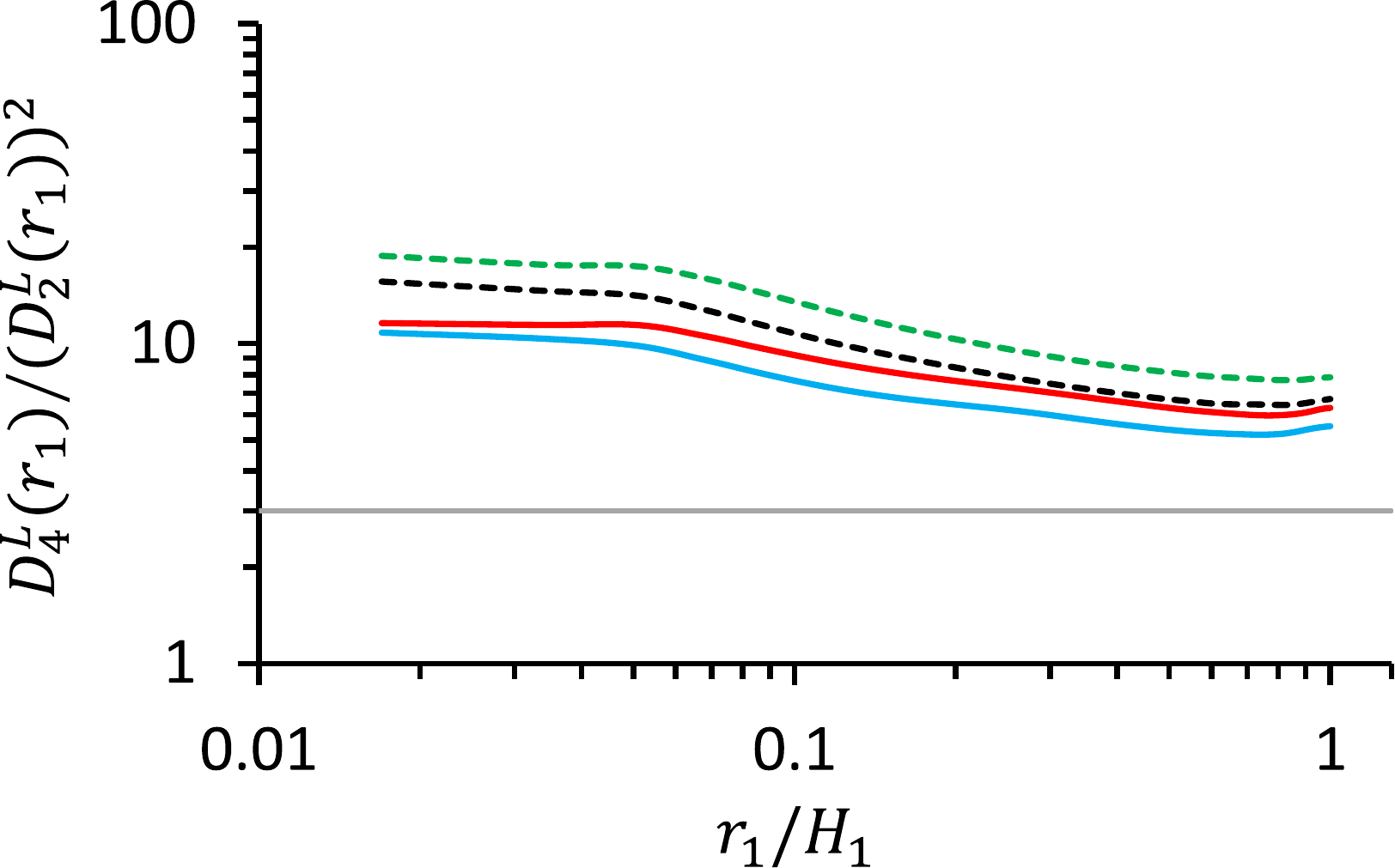}}
		\end{minipage}
		\begin{minipage}[b]{0.5\linewidth}
			\centering
			\makebox[1.2em][l]{\raisebox{-\height}{(\textit{d})}}%
			\raisebox{-\height}{\includegraphics[height=3.6cm]{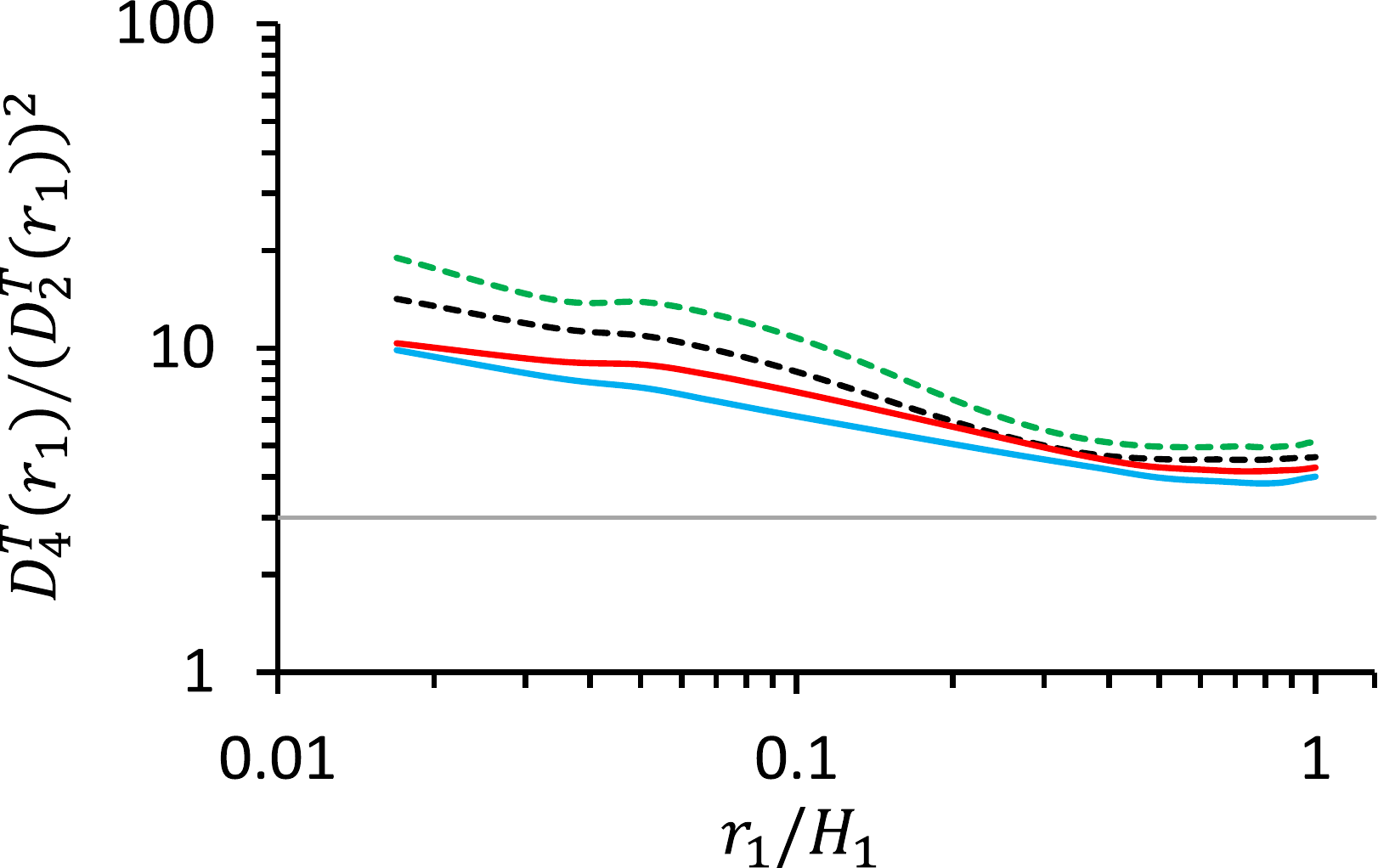}}
		\end{minipage}
	\end{minipage}
	\caption{Normalized fourth-order transverse (\textit{a},\textit{d}) and longitudinal (\textit{b},\textit{c}) structure functions, corresponding to the flatness of the velocity increments. The horizontal lines in (\textit{a,b,c,d}) indicate to the Gaussian value of $3$ for the flatness.} \label{fig: flatness}
\end{figure}

The intermittency and non-Gaussianity can be quantified using the flatness measures defined as $D^T_4(r_\gamma)/(D^T_2(r_\gamma))^2$ (figure \ref{fig: flatness}\textit{a,d}) and $D^L_4(r_\gamma)/(D^L_2(r_\gamma))^2$ (figure \ref{fig: flatness}\textit{b,c}). The first observation from the results in figure \ref{fig: flatness} is that the all four bubble-laden cases show a rather similar behaviour as $r$ increases, slowly reducing from values of up to 40 towards the Gaussian value of 3. However, there is still considerable deviation from 3 even at the largest scale $r_\gamma/H_\gamma=1$, consistent with the PDF results for $\Delta\boldsymbol{u}(r_\gamma=H_\gamma)$ in figure \ref{fig: pdf SmMore}. We also find that for both the transverse or longitudinal flatness measures and along either separations directions, the flatness is largest for the case with smaller $Re_{H_2}$, reflecting the same trend as the PDFs of the velocity increments at the single scale $\Delta$ (figure \ref{fig: pdf 4cases}). The values and behavior of the flatness are similar to those observed in \cite{2021_Ma} for the three bubble-laden cases with $\alpha_p=2.14\%$, although for the present flow the flatness values are larger than those in \cite{2021_Ma}. For separations $r_2$, the transverse flatness values are larger than the longitudinal ones, similar to what is observed in isotropic turbulence \citep{2009_Ishihara,2005_Li}. By contrast, for separations $r_1$, the transverse flatness values are slightly smaller than the longitudinal ones.

\subsection{Flow structures associated with extreme fluctuations}\label{subsec: Visualization}

\begin{figure}
	\begin{minipage}[b]{1.0\linewidth}
		\begin{minipage}[b]{0.5\linewidth}
			\centering
			\makebox[1.5em][l]{\raisebox{-\height}{(\textit{a})}}%
			\raisebox{-\height}{\includegraphics[height=4.5cm]{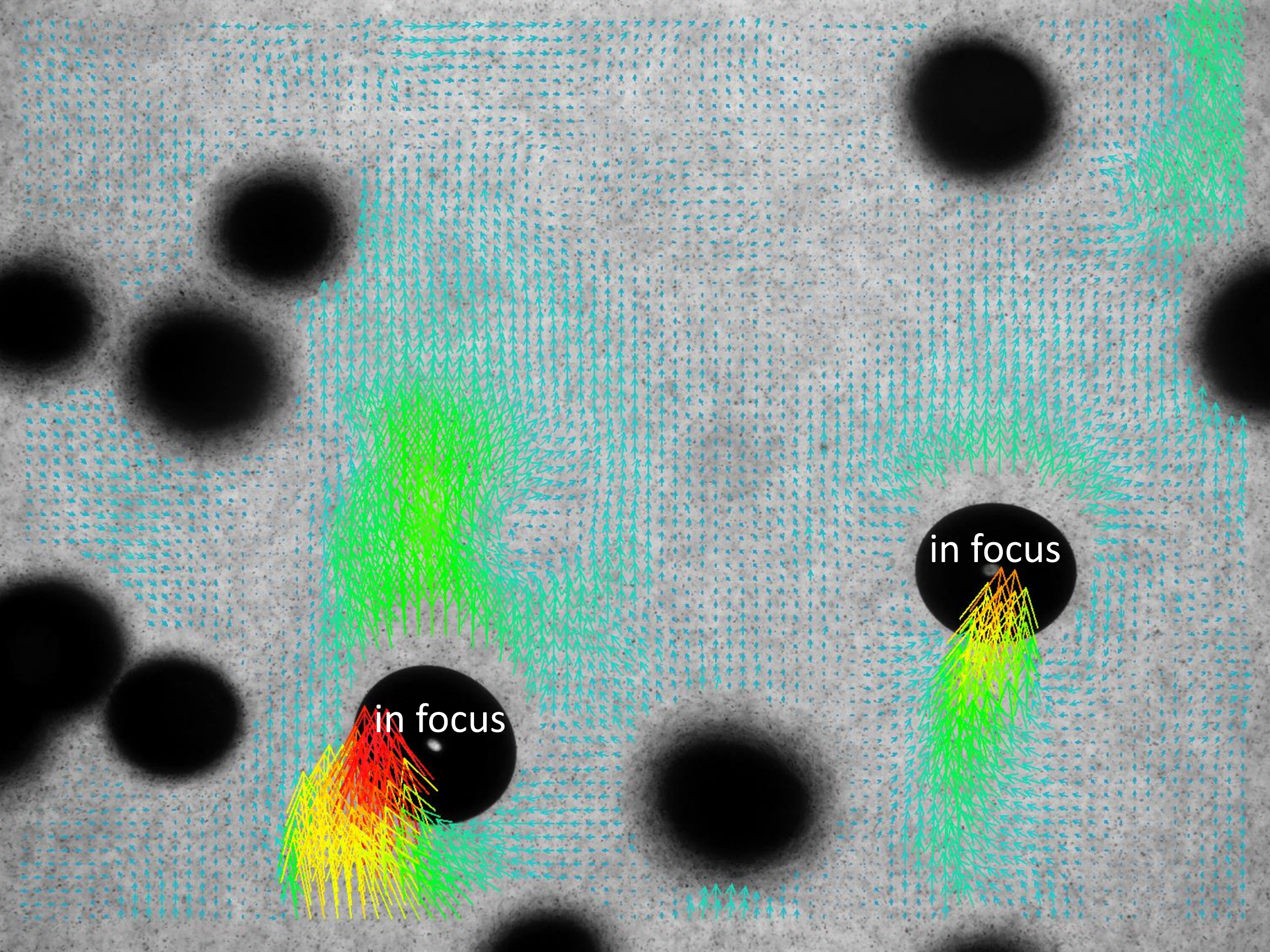}}
		\end{minipage}
		\begin{minipage}[b]{0.5\linewidth}
			\centering
			\makebox[1.5em][l]{\raisebox{-\height}{(\textit{b})}}%
			\raisebox{-\height}{\includegraphics[height=4.5cm]{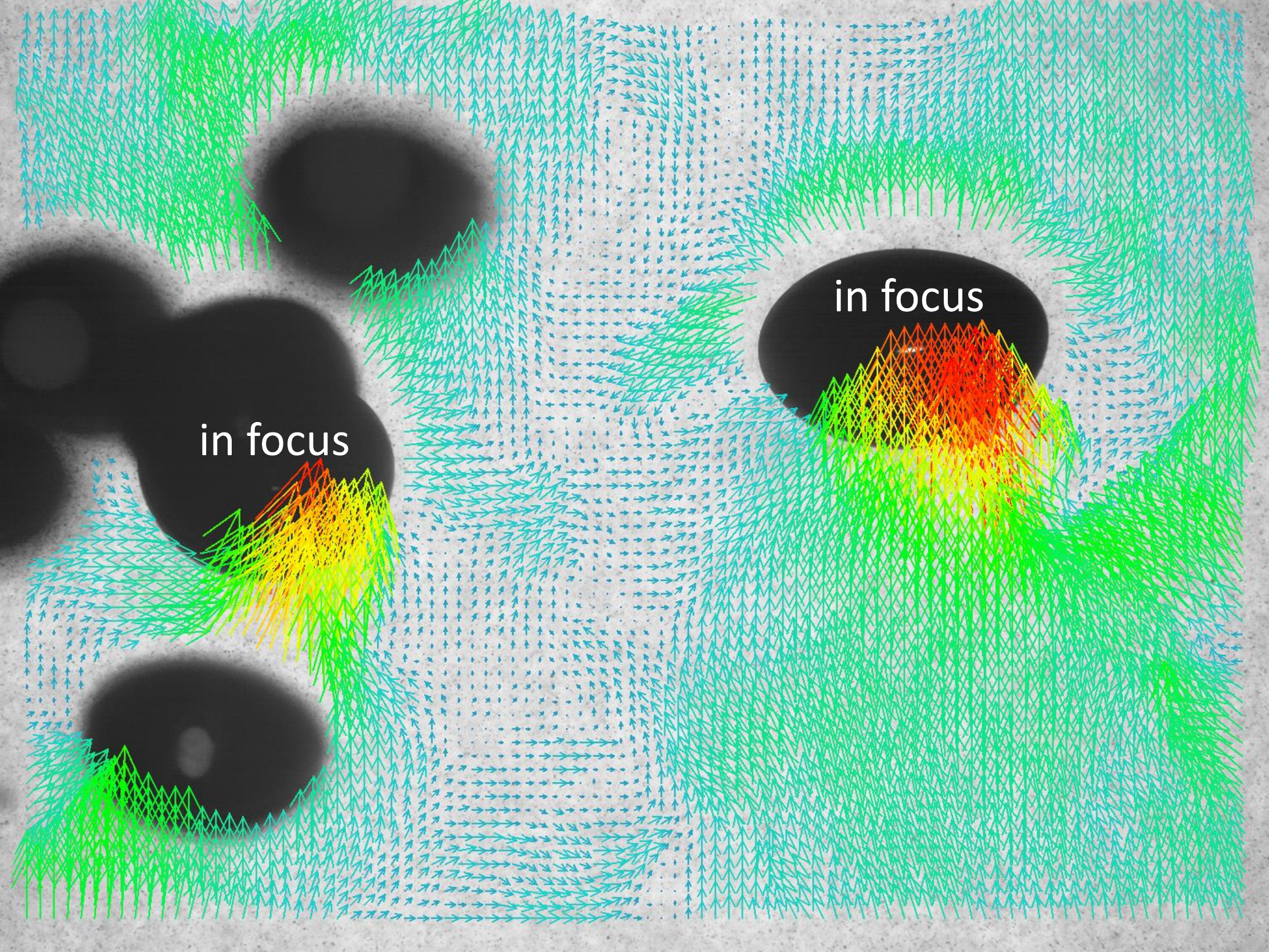}}
		\end{minipage}
	\end{minipage}
	\begin{minipage}[b]{1.0\linewidth}
		\vspace{3mm}
		\begin{minipage}[b]{0.5\linewidth}
			\centering
			\makebox[1.5em][l]{\raisebox{-\height}{(\textit{c})}}%
			\raisebox{-\height}{\includegraphics[height=4cm]{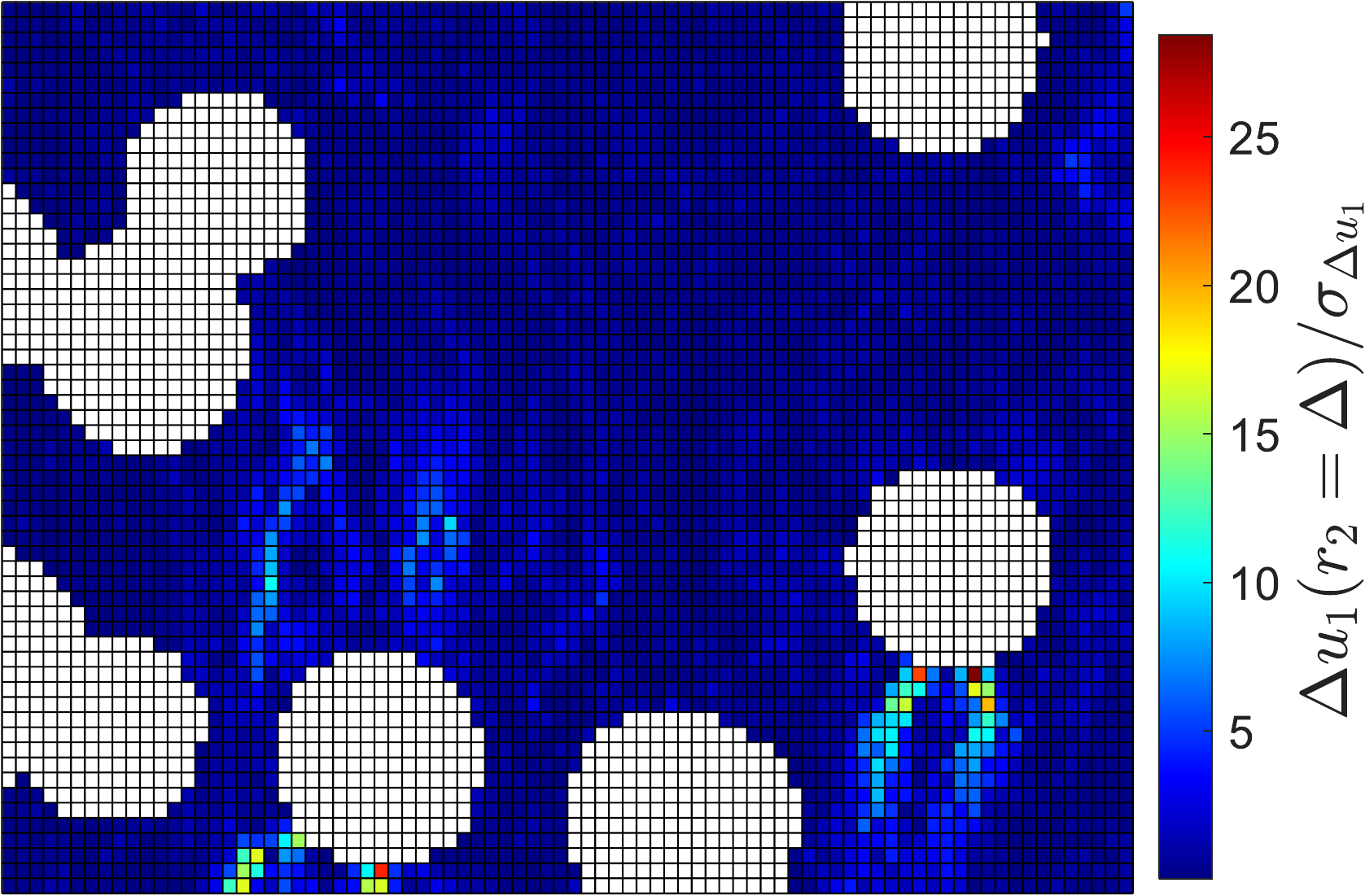}}
		\end{minipage}
		\begin{minipage}[b]{0.5\linewidth}
			\centering
			\makebox[1.5em][l]{\raisebox{-\height}{(\textit{d})}}%
			\raisebox{-\height}{\includegraphics[height=4cm]{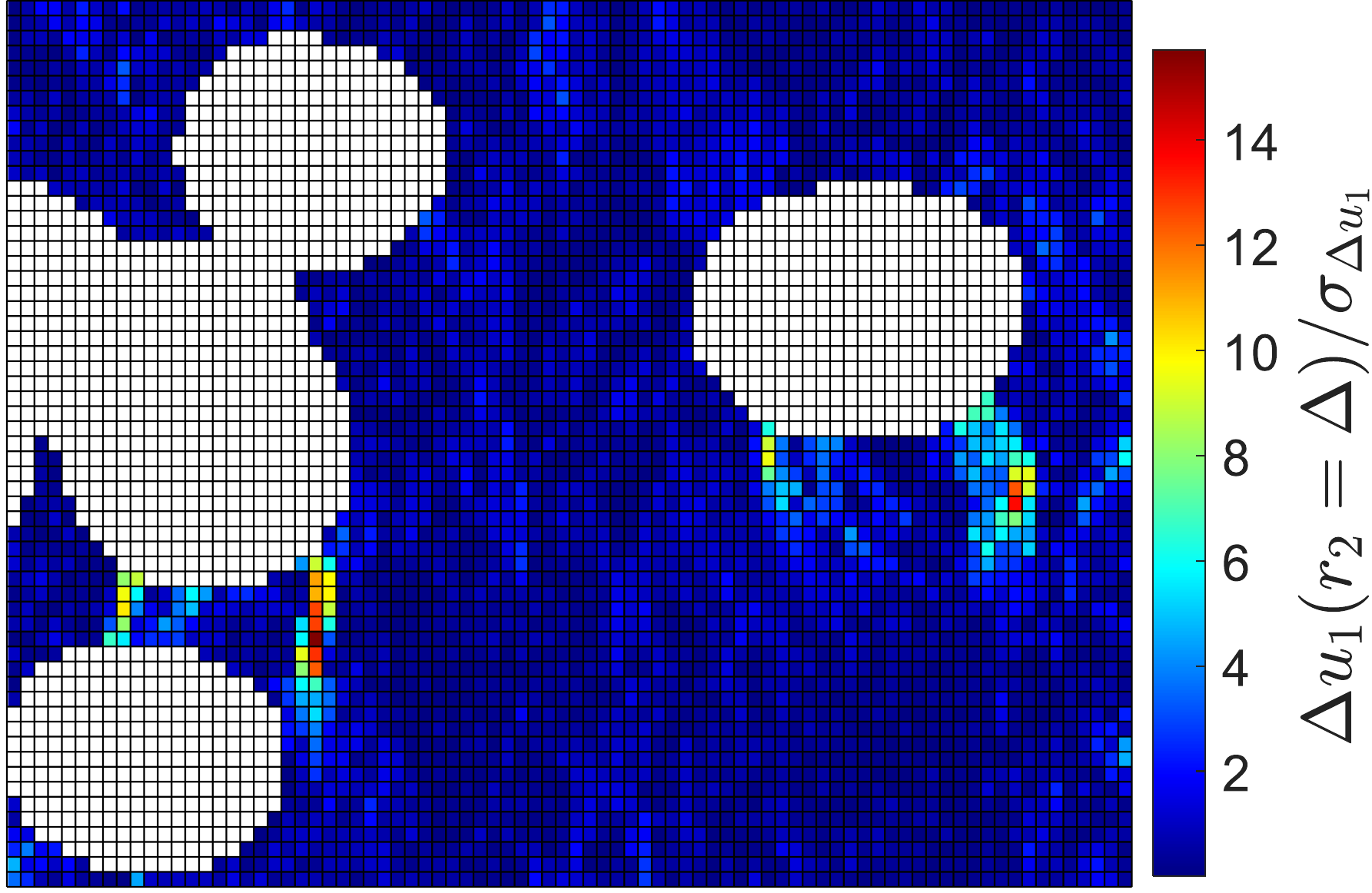}}
		\end{minipage}
	\end{minipage}
	\caption{Snapshot of the original velocity vector (\textit{a,b}) and the intensity distributions of normalized velocity increment $\Delta u_1(r_2=\Delta)/\sigma_{\Delta u_1}$ (\textit{c,d}). Here, (\textit{a,c}) are from the same instant based on the \textit{SmMore} case and (\textit{b,d}) are from the same instant based on the \textit{LaMore} case. The in-focus bubbles are denoted in (\textit{a,b}).} \label{fig: visualization}
\end{figure}

We now use flow visualization to understand which regions of the flow are associated with the extreme velocity increments. Figure \ref{fig: visualization}(\textit{c,d}) give two examples of such flow visualization extracted from a randomly chosen (but representative) FOV result for $\Delta u_1(r_2=\Delta)/\sigma_{u_1}$ for the cases \textit{SmMore} and \textit{LaMore}, respectively. Each grid in the figure corresponds to an interrogation window of the present PSV. For convenience, we also plot in (\textit{a,b}) the corresponding original velocity vector fields and denote in each case the in-focus bubbles. The images show that for both cases the regions where the velocity increment takes on extreme values are located in the outer-most region of the individual bubble wakes, while the fluctuations inside the wakes are relatively weak (see \textit{c,d}). At a result of this it appears that the small-scale intermittency in the present flow already characterized is quite different from the small-scale intermittency that occurs in single-phase turbulence that arises due to nonlinear self amplification of the velocity gradients \citep{2001_Tsinober,2005_Li,2019_Buaria,2020_Buaria}. In particular, the intermittency in the present flow is more similar to what is often called external intermittency \citep{1949_Townsend}, which arises near the interface between two regions of the flow with significantly differing behavior, e.g. such as at the turbulent/non-turbulent interface in turbulent jets \citep{2014_Silva,2021_Gauding}. In the bubble-laden flow we are considering, the intermittency arises at the interface between the edge of the bubble wake and the surrounding relatively quiescent flow.

\begin{figure}
	\begin{minipage}[b]{1.0\linewidth}
		\vspace{3mm}
		\begin{minipage}[b]{0.5\linewidth}
			\centering
			\makebox[1.5em][l]{\raisebox{-\height}{(\textit{a})}}%
			\raisebox{-\height}{\includegraphics[height=4.5cm]{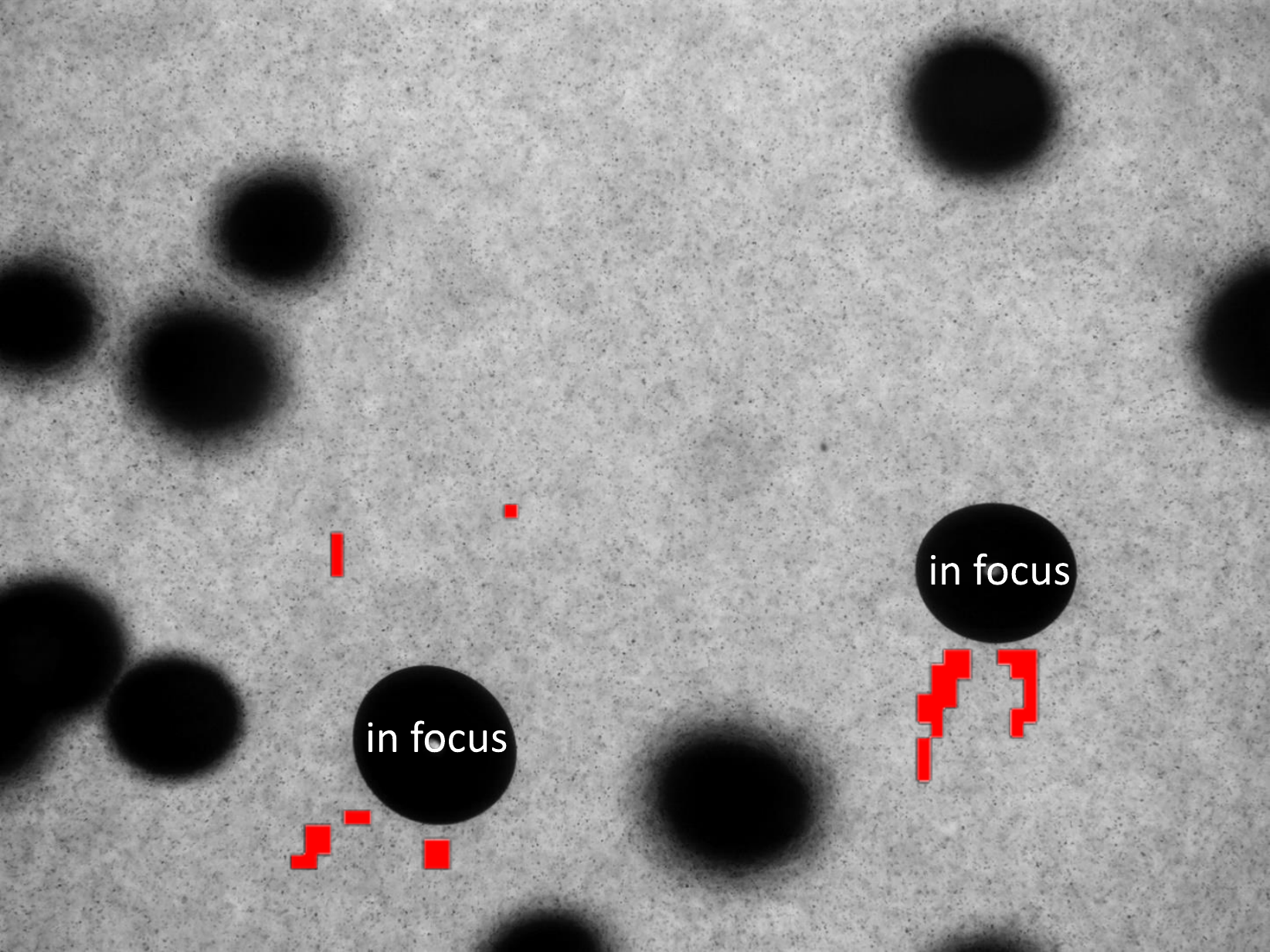}}
		\end{minipage}
		\begin{minipage}[b]{0.5\linewidth}
			\centering
			\makebox[1.5em][l]{\raisebox{-\height}{(\textit{b})}}%
			\raisebox{-\height}{\includegraphics[height=4.5cm]{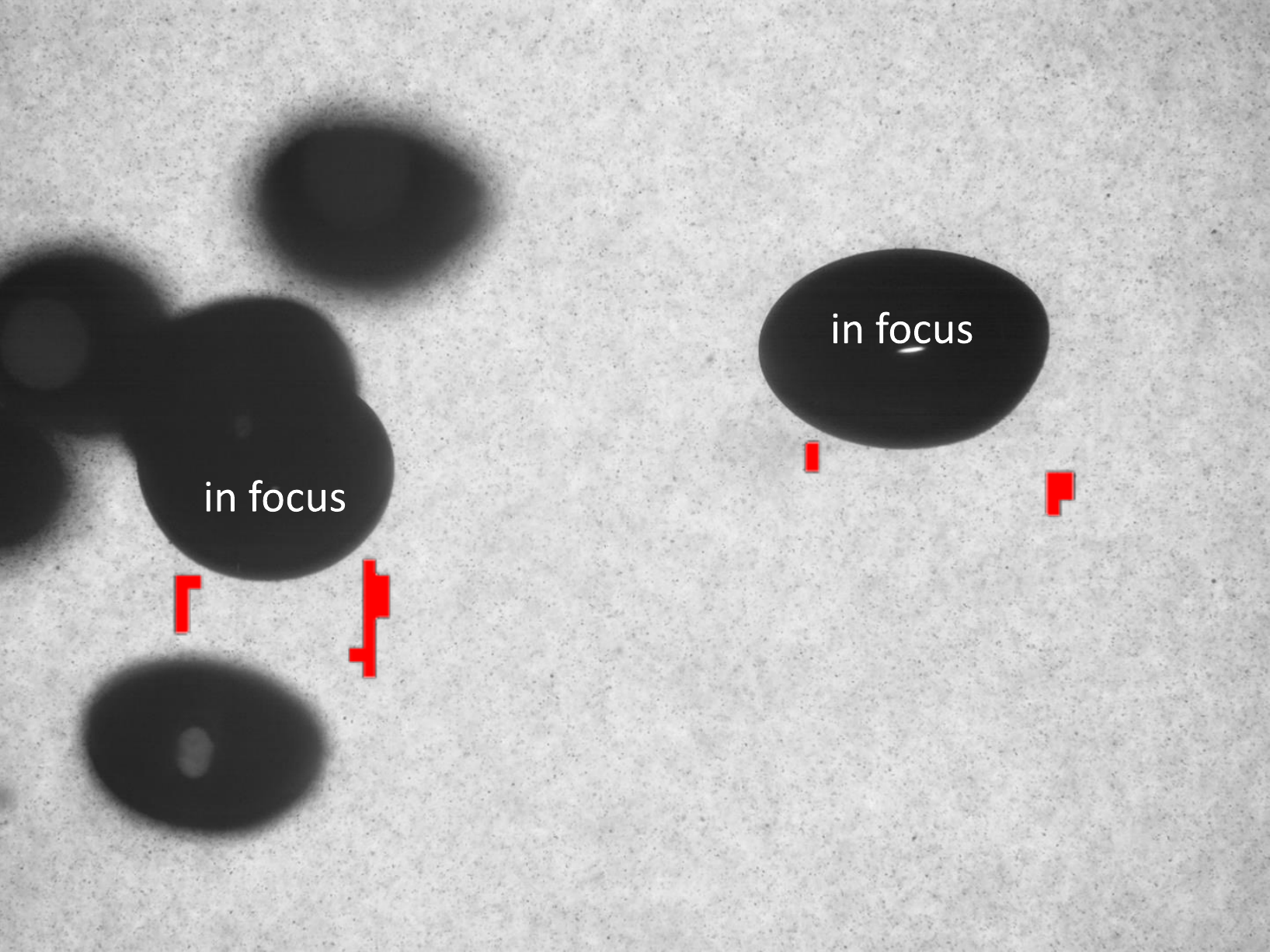}}
		\end{minipage}
	\end{minipage}
	\begin{minipage}[b]{1.0\linewidth}
		\vspace{3mm}
		\begin{minipage}[b]{0.5\linewidth}
			\centering
			\makebox[1.5em][l]{\raisebox{-\height}{(\textit{c})}}%
			\raisebox{-\height}{\includegraphics[height=4.5cm]{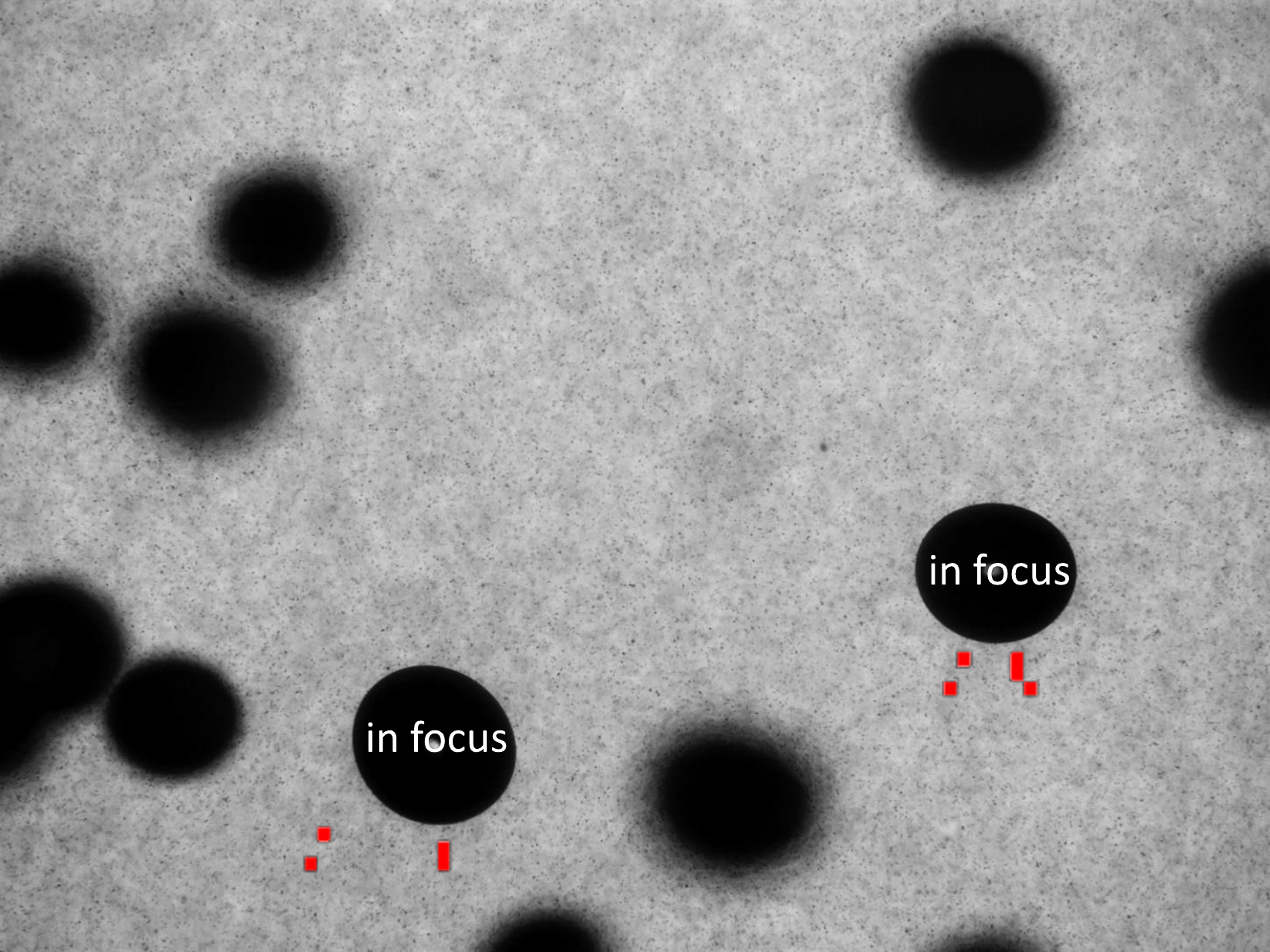}}
		\end{minipage}
		\begin{minipage}[b]{0.5\linewidth}
			\centering
			\makebox[1.5em][l]{\raisebox{-\height}{(\textit{d})}}%
			\raisebox{-\height}{\includegraphics[height=4.5cm]{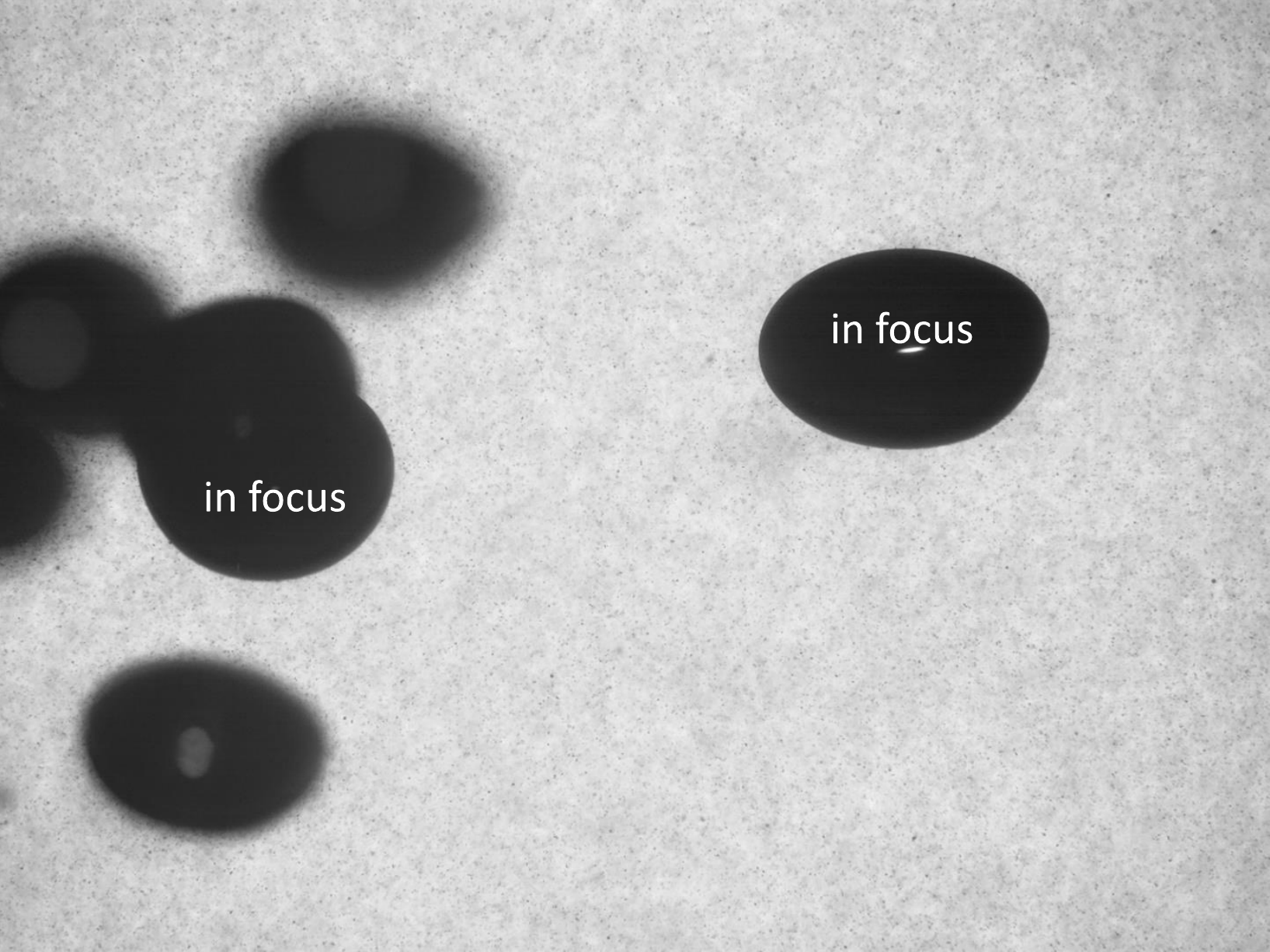}}
		\end{minipage}
	\end{minipage} 
	\caption{Snapshots, indicating the regions with extreme values, where $\left |\Delta u_1(r_2=\Delta)\right |\geq8\sigma_{\Delta u_1}$ (\textit{a,b}) and $\left |\Delta u_1(r_2=\Delta)\right |\geq16\sigma_{\Delta u_1}$ (\textit{c,d}) highlighted in red. Panels (\textit{a,c}) are from the same instant from \textit{SmMore} and (\textit{b,d}) are from the same instant from \textit{LaMore}. The in-focus bubbles are denoted in each panel.} \label{fig: extreme value}
\end{figure}

One further observation on figure \ref{fig: visualization}(\textit{c,d}) is that the extreme values are larger for the smaller bubble wakes than the larger bubble wakes (note the different legend scales in \textit{c,d}). This point is made clearer in figure \ref{fig: extreme value}(\textit{a,b}) where instead $\left |\Delta u_1(r_2=\Delta)\right |\geq8\sigma_{\Delta u_1}$ is plotted. Looking at the in-focus bubble on the right-hand side for both cases (where a more complete wake structure is visible than for the left in-focus bubble), we find that the regions of extreme values are larger for the \textit{SmMore} case. By increasing the threshold to $\left |\Delta u_1(r_2=\Delta)\right |\geq16\sigma_{\Delta u_1}$ (figure \ref{fig: extreme value}\textit{c,d}), we see that there are still some locations with extreme values satisfying this threshold for the \textit{SmMore}, while there are none for the \textit{LaMore} case. Although this observation is only for one snapshot, we have conducted the same analysis for many different snapshots and virtually all of them show the same behaviour. These observations are consistent with the PDF results in figure \ref{fig: pdf 4cases} that show that cases with smaller bubbles (lower $Re_{H_2}$) have heavier tails than the cases with larger bubbles (higher $Re_{H_2}$).

An explanation for why the intermittency is largest for the cases with smaller bubbles and lower volume fractions was given in \cite{2021_Ma}. When the volume fraction is not too large (but still large enough for the bubbles to make a statistically significant contribution to the flow properties), there are relatively few regions in the flow where turbulence is produced due to the bubble wakes, meaning that turbulence in the flow is very patchy and therefore intermittent. As the volume fraction increases, the regions of the flow influenced by the bubble wakes becomes less rare, and hence intermittency in the flow reduces. This then explains why the \textit{SmLess} case is the most intermittent, and the \textit{LaMore} case the least intermittent. Moreover, since the flow we are considering has no background turbulence (i.e. the only turbulence in the flow is the result of the bubble wakes), then this may explain why the intermittency we see is higher than that in \cite{2021_Ma} where the flow considered had background turbulence, so that turbulent activity in the flow was not restricted to regions close to the bubbles.

\section{Conclusions}

In this paper we have presented an analysis of the multiscale properties of a bubble-laden turbulent flow, based on experimental data of a flow in a vertical column with bubble swarms rising in water. The experiment takes advantage of a recently developed PSV technique for particle/bubble flows and provides the first comprehensive data set for computing multipoint measurements (without having to invoke Taylor's hypothesis, in contrast to, e.g. \cite{2005_Rensen}) of flows laden with finite-sized bubbles. By acquiring a very large number of snapshots of the velocity field we are able to compute structure functions of up to twelfth orders, and PDFs that resolve the heavy tails associated with extreme fluctuations in the flow.

The results show that the level of anisotropy in the flow produced by the rising bubbles is strong in general, and not negligible at any scale in the flow. Moreover, the results show that (i) the differing behaviour of the second-order longitudinal and transverse structure functions when measured for separations in different directions shows that both velocity components and separation directions of the 2 dimensional data need to be considered in order to fully characterize the anisotropy of the flow; (ii) the bubble size and void fraction are both important parameters determining the amount of anisotropy in the flow; (iii) higher-order structure functions reveal greater anisotropy across the scales of the flow, such that extreme fluctuations in the flow are the most anisotropic. 

Since the PSV data captures two-dimensional data, we were able to consider the energy transfer between scales for two separation directions in the flow. The results revealed a downscale energy transfer on average for horizontal separations, but an upscale energy transfer on average in the vertical direction. However, the horizontal energy transfer was much stronger than that in the vertical direction.

We also investigated extreme events in the flow by considering the normalized probability density functions of the velocity increments in the flow. The results showed that the probability of extreme fluctuations increases with decreasing scale, just as in single-phase turbulence. However, the results also showed that for a given scale the probability of extreme events decreased with increasing Reynolds number, contrary to what occurs in single-phase turbulence. To explore the origin of these extreme fluctuations in the bubble-laden flows, we visualized regions of extreme small-scale velocity increments in the FOV and observed that they are typically located at the boundary of the wakes produced by the bubbles. For the cases with smaller bubbles and lower void fractions, vast regions outside of the bubbles wakes exhibit weak fluctuations, and so this combined with the extreme fluctuations at the bubble wake boundaries leads to strong intermittency. For larger bubbles which produce larger flow Reynolds numbers, and with larger void fractions, the wake regions become less rare in the flow and hence the flow is less intermittent than the former case, even though the Reynolds number is higher. Furthermore, the extreme values were also observed to reach larger values (compared to the standard deviation) for the smaller bubbles, which again causes the smaller bubble cases to exhibit greater intermittency than the larger bubble cases, in addition to the effect arising from the fraction of the flow modified by the bubble wakes.

Finally, it is worth pointing out two limitations with the present PSV data. First, the FOV is not large enough to resolve the integral length scales of the flow, and therefore as discussed in \S\,\ref{subsec: pdf} we were not able to fully resolve the transition in the properties of the bubble-laden flow as the scale approaches the intergral length scale. Second, the bubbles considered here all have approximately fixed shape, mimicking in nature small bubbles in a contaminated flow. We are currently performing new experiments with deformable bubbles and differing FOVs in order to address these two points and will report the results in future work.

\section*{Acknowledgements}
The authors would like to acknowledge Thomas Ziegenhein for providing some routines for calculating bubble statistics. We also thank Ronald Franz and Uwe Hampel for providing the high-speed camera used in the present study. 

\section*{Funding}
T.M. acknowledges funding by Deutsche Forschungsgemeinschaft (DFG, German Research Foundation) under Grants MA 8408/1-1 and MA 8408/2-1.

\section*{Declaration of Interests}

The authors report no conflict of interest.

\section*{Data availability}
The data that support the findings of this study are available from the first author T.M. on request.\\

\bibliographystyle{jfm}
\bibliography{BIT_exp_JFM}

\end{document}